\documentclass[twoside,12pt]{article}
\usepackage{epsfig}

\newcommand{\be}{\begin{equation}}
\newcommand{\ee}{\end{equation}}
\newcommand{\bea}{\begin{eqnarray}}
\newcommand{\eea}{\end{eqnarray}}

\usepackage[dvipsnames]{xcolor}
\usepackage[colorlinks=true,urlcolor=Blue,linkcolor=Blue]{hyperref}
\usepackage[all]{hypcap}
\usepackage{cite}

\begin{document}

\title{Heavy-Ion Storage Rings and Their Use in Precision Experiments with Highly Charged Ions}
\author{Markus Steck and Yuri A. Litvinov\\
GSI Helmholtzzentrum f{\"u}r Schwerionenforschung, 64291 Darmstadt, Germany}

\maketitle
\begin{abstract}
Storage rings have been employed over three decades in various kinds of nuclear and atomic physics experiments with highly charged ions.
Storage ring operation and precision physics experiments benefit from the availability of beam cooling which is common to nearly all facilities. 
The basic aspects of the storage ring components and the operation of the ring in various ion-optical modes as well as the achievable beam conditions are described. 
Ion storage rings offer unparalleled capabilities for high precision experiments with stable and radioactive beams.
The versatile techniques and methods for beam manipulations allow for preparing beams of highest quality at any energy of interest.
The rings are therefore part of the experiment .
Recent experiments conducted in a wide energy range and with various experimental installations are discussed. 
An overview of active and planned facilities, new experimental set-ups and proposed physics experiments  completes this review.   

\end{abstract}
\tableofcontents

\section{Introduction}
\label{s:intro}

Over the last three decades storage rings were successfully employed for physics research. 
They enable storage of freshly produced exotic ions, which are either stable ions in a defined high atomic charge state or 
exotic radioactive nuclides.
This turns to be a straightforward way to achieve the efficient use of such rare species.
Furthermore, it offers unparalleled possibilities for precision experiments in the realm of atomic and nuclear physics 
as well as for experiments of astrophysical relevance and for testing fundamental interactions. 
Since the storage ring itself is an indispensable part of the experiment, its parameters are decisive for the experiments to be conducted.

The design of the storage rings was governed by the accelerator facilities which were available in the different laboratories and 
were planned to serve as injectors to provide fast ion beams for storage. 
The injectors define the available beam species and the injection energy. 
The production of specific secondary particles often requires specific energies, which might not be directly suited for the experiment.
The beam energy can be varied by operation of the storage ring in a synchrotron mode.
Both acceleration and deceleration of the beam after injection is feasible. 
In addition to experiments aiming at maximum achievable energies, given by the maximum rigidity of the magnetic system, 
a range of new experimental ideas requires specific stored exotic beams at very low energies, which is one of the challenges today.
The particles can be stored as a bunched or coasting beam. 
The beam quality in all the storage rings can be improved by beam cooling systems which also allow various non-Liouvillean methods of beam manipulations.
As a result, cooled secondary beams with small transverse size and high momentum definition are available for precision experiments.
The cooling capability is the biggest advantage of conducting experiments in storage rings.
The stored beam can be brought in collisions with electron, photon and atomic targets. 
Proposals for using neutron and ionic targets are discussed in the frameworks of future projects.
The use of internal targets in combination with cooling offers the platform for well-defined experimental conditions. 
Such internal targets are very thin but, thanks to high revolution frequencies in the ring, high luminosities can be achieved.
Beam cooling is essential for accumulation of low intensity beams and it also supports efficient deceleration of beam species which require high energy for their production. 
In accordance with the beam manipulation methods and the beam quality of the stored beam, specific experimental methods were devised. 
The rings are operated at ultra-high vacuum to achieve an efficient beam storage.
On the one hand this allows one to preserve high atomic charge state of the ions, but on the other hand this sets severe constraints on any experimental equipment to be brought into the ring vacuum environment. The experiments are based on particle detection systems, many of which are specifically developed for storage rings.

This work is organized as follows.
It starts with the discussion of the major machine properties essential for making high-precision experiments possible.
A brief review of the present status of physics research at heavy-ion storage rings is followed. 
The work is restricted to experiments with highly-charged heavy ions that are conducted at the presently operated facilities.
The new-generation storage ring projects and the corresponding envisioned research programs are outlined at the end.

\section{Historical Remarks}

Many storage ring were designed in the middle of the 1980s and commissioned around 1990. 
Even earlier storage rings were built to test the accelerator concepts employed in the operation of storage rings.
Two particular storage rings were constructed to test the concept of beam cooling. The NAP-M ring at Novosibirsk demonstrated electron cooling with an electron beam device, which was later used in modified versions in many other storage rings. Stochastic cooling was developed in the ICE ring at CERN to a degree which later allowed its application in storage rings for cooling and accumulation of antiprotons. 
Among many others, these developments were essential for the success of ``the large project, which led to the discovery of the field particles W and Z, communicators of weak interaction'' which was awarded with the 1984 Nobel Prize to Carlo Rubbia and Simon van der Meer \cite{NobelPrize1984}.
For more details the reader is referred to an overview article by H. Koziol and D. M\"ohl  and references therein \cite{koziol&moehl}.
The concepts for the various rings used in physics experiments borrowed techniques which were developed at the Low Energy Antiproton Ring (LEAR) \cite{lear} and even earlier at the Intersecting Storage Rings (ISR) \cite{isr} both operated at CERN. An overview of those machines can be found in a review article by R.E. Pollock \cite{pollockrev}. They were all built at existing accelerator laboratories profiting from the infrastructure and employing the existing accelerators as injectors. 
	They can be categorized as light-ion storage rings for experiments using protons and light ions at medium energy in the range of a few hundred MeV to a few GeV, mainly for nuclear and particle physics. The heavy-ion storage rings cover an energy range from very few MeV/u to several hundred MeV/u and provide ion beams from protons up to fully-ionized uranium. They consequently found a broad range of applications in atomic and nuclear physics. Some of the advanced technologies and beam manipulation methods which were pioneered at LEAR were included in the concept of these storage rings and later expedited during their operation. These are aspects as beam cooling, accumulation of beam particles, deceleration, ultra-high vacuum technology and many others. Furthermore, methods which are common to storage rings and synchrotrons were optimized for the requirements of storage rings, e.g. radio-frequency (rf) manipulations to produce beams with a special bunch structure, injection and extraction, and beam diagnostics. 
	
The light-ion storage rings used a similar concept and operation mode. They all started from a proton cyclotron as an injector which provided a continuous proton beam with moderate intensity. In order to boost the stored particle number, schemes to fill the transverse acceptance of the storage ring were employed. The increased transverse emittance was tempered by cooling the beam after injection and the usual adiabatic damping of the emittance during acceleration when operating the storage ring as a synchrotron. The variable energy of the beam after acceleration in the ring provided the conditions for experiments with internal gas, pellet or solid fiber targets, preferably unpolarized and polarized hydrogen targets were used, see Section \ref{s:targets}. The different facilities had options to operate with other light ions and with polarized protons and deuterons. The main focus were experiments for studying  hadronic interaction between light fast beam particles and light target material which allowed the production and study of mesons, baryons and strange particles, spectroscopy of hadronic states and the investigation of symmetry breaking in hadronic interactions. For more details the reader is referred to dedicated conference proceedings, e.g. \cite{STORI96, STORI99}.
	 
The heavy-ion storage rings performed a much more diversified experimental program which required ions over the full range of ion masses with different charge states of the stored ions. The low-energy heavy-ion rings used beams from existing low-energy facilities, such as electrostatic accelerators, linear rf accelerators and cyclotrons. The intensity of the stored beam was also maximized by transverse injection schemes in combination with cooling. Depending on the injection energy, the experiments used beams with and without acceleration in the storage ring.  
The  medium-energy ion storage rings are coupled to synchrotrons, which provide a bunched beam for single turn injection into the storage ring at high energy. Such a scheme allows the injection of highly charged ions with intermediate stripping and the injection of rare isotopes which are produced in a target and separated in-flight in a dedicated separator. As high energy is a prerequisite to generate the required ion species, the energy can be matched to the experiment requirement by acceleration or deceleration of the beam directly in the storage ring.

The experiments in the low-energy rings were strongly supported by the availability of  beam cooling, which allowed various high precision atomic physics measurements. Low primary beam energies limit the range of secondary ions that can be produced. Typically, multiple-charged heavy ions were stored. The interaction of the ion beam with internal targets, with the free electrons of the electron cooler or with laser light are favoured methods for studying predominantly atomic physics aspects. 
One of the examples of the low-energy storage rings is the Test Storage Ring (TSR) \cite{Kramer-1990} which was in operation in Heidelberg from 1988 until 2012. 
The medium-energy rings extended the range of experimental conditions to higher energy and to beams of ions with very high atomic charge states for precision atomic physics studies. In addition, the storage of rare isotopes produced by projectile fragmentation and/or in-flight fission of a primary heavy-ion beam became possible which enabled experiments in nuclear structure and nuclear astrophysics.

Several of the storage rings of the first phase have been closed down. 
Only the medium-energy Experimental Storage Ring (ESR) in Darmstadt \cite{esrmachine} and the high-energy COoler SYnchrotron (COSY) in J{\"u}lich \cite{cosymachine} are still in operation. 
In 2007, the HIRFL-CSR complex in Lanzhou has been taken into operation. 
There, highly charged heavy ions are used for experiments in the medium-energy experimental Cooler-Storage Ring (CSRe) \cite{csremachine}. 
In 2018, the first nuclear physics experiments were performed in the medium-energy Rare-Ion Ring (R3) in Tokyo \cite{rareriring}.
The R3 ring in RIKEN is a very special machine which follows a cyclotron driver accelerator instead of a heavy-ion synchrotron like in the case of the ESR and the CSRe.
Furthermore, R3 is the only heavy-ion storage ring in operation, where no beam cooling is applied.
An overview of the ring parameters of the three active heavy ion storage rings is given in Table 1.

\begin{table*}[htbp]
	\caption{Active heavy ion storage rings.}
	
	\begin{tabular*}{0.9\textwidth}{lcccc}
		\\
		\hline
		~~~~~~~~~~~~~~~~~~~     & ~~~~~~~~ESR~~~~~~~~   &  ~~~~~~~~CSRe~~~~~~~~ &  ~~~~~~~~R3~~~~~~~~    \\
		
		\hline  
		circumference   &      108.4 m         &  128.8 m     &  60.4 m      \\
		maximum magnetic rigidity   &  10 Tm      &   9.4 Tm    & 6.0 Tm    \\	
		maximum beam energy U$^{92+}$   &  550 MeV/u      &   500 MeV/u    & 230 MeV/u   \\	
		minimum beam energy U$^{92+}$   &  3 MeV/u      &   5 MeV/u    & -   \\	
		beam cooling   &  stochastic, electron      &   electron    & no    \\
		internal target   &  gas jet      &   gas jet     & no    \\
		standard mode \\
		tunes $Q_x/Q_y$  & 2.2/2.4 & 2.53/2.57 & -  & \\
		acceptance $ (A_x/A_y) [\mu m]/ (\delta p/p) [\%] $ & 450/150/$\pm$3.6 & 200/150/$\pm$2.6& -& \\  
		isochronous mode  \\
		tunes $Q_x/Q_y$  & 2.26/2.45 & 1.70/2.72 & 1.22/0.88 & \\
		acceptance $ (A_x/A_y) [\mu m]/ (\delta p/p) [\%] $ & 20/20/$\pm$0.3 & 20/20/$\pm$0.25 & 20/10/$\pm$0.5 & \\  
		transition energy $\gamma _t$ & 1.4 & 1.4  & 1.21 & \\
		
		\hline		
	\end{tabular*}
	\label{tablemass}
\end{table*}

Some of the storage ring facilities have been or will be modified as part of new projects. A more detailed description of those facilities will be given in Section \ref{s:rings}.

There is also a group of storage rings which is part of an accelerator chain and used for specific beam manipulations, particularly beam accumulation and beam cooling. As they are not hosting experiments, they are only briefly mentioned here for the sake of completeness. The Low Energy Ion Ring (LEIR)~\cite {leir} is a part of the heavy ion chain at CERN and is used to deliver high intensity heavy ion bunches for collision experiments in the Large Hadron Collider (LHC). Also at CERN, the combination of Antiproton Decelerator AD~\cite{ad} and the Extra Low ENergy Antiproton deceleration ring (ELENA)~\cite{elena} provides intense antiproton beams for experiments down to 0.1 MeV. 

There was notable progress in recent years in constructing electrostatic trapping devices, like, e.g., Penning traps and dedicated ultra-low energy storage rings.
An example of the latter is the Cryogenic Storage Ring (CSR) \cite{Novotny-2019} in Heidelberg which is operated at a temperature well below 10~K and was used in first experiments  in 2015.
Although medium- and highly-charged ions can be used in the CSR for atomic, molecular, and cluster physics experiments, we restrict the scope of this work to magnetic storage rings.

	\section{Storage Ring Operation}
 	  \subsection{Beam Cooling}
	\label{s:bcool}
	
	The outstanding feature for performing high precision storage ring experiments is the availability of beam cooling, which is even more powerful in combination with internal experimental installations like thin targets. Beam cooling reduces the uncertainty caused by the finite emittance and energy spread of the beam by a non-Liouvillean process. Depending on the initial beam quality, cooling can reduce the emittance and energy spread of the stored beam by orders of magnitude. This is most important for secondary beams which are produced by the interaction of the primary beam with solid target material resulting in an increased emittance and energy spread of the beam, due to the nuclear reaction process and the interaction with matter. Beam cooling can also compensate the heating caused by the passage of the circulating beam through matter for experiments using internal targets and the interaction with residual gas. Cooled beams are well defined in the absolute energy and in their spatial location. Therefore, the uncertainty caused by the finite energy spread and emittance is reduced, which is a prerequisite for achieving high precision measurements. The lifetime of the stored ion beam is also affected by cooling. On the one hand, energy loss and multiple scattering in the target and the residual gas are compensated and do not contribute to the loss rate. On the other hand, electron cooling results in losses by the recombination of ions and electrons as a result of the interaction process which provides the cooling.
	
	Stochastic cooling and electron cooling in storage rings have different beneficial regimes, they are often used in a complimentary way. Stochastic cooling which is based on detection of the deviation of sub-ensemble of beam particles from the reference orbit and nominal energy and the generation of a correction signal is an advanced large bandwidth feedback technique. Stochastic cooling benefits from large deviations from the reference particle as this results in large correction kicks and consequently is most powerful for beams of poor quality.  Electron cooling acts through Coulomb collisions of the ions in a merged cold electron beam which results in the beam frame in a transfer of unwanted momentum of the individual ions to the electron beam. The Coulomb interaction in the electron cooling process is strongest for small momentum deviations and the cooling rate increases with the reduction of the ion beam emittance and energy spread. Therefore stochastic cooling is often used as pre-cooling for electron cooling. Another feature of the two cooling methods  is the dependence on beam energy. Stochastic cooling depends on the ion optical properties of the storage ring, particularly on the change of revolution time with energy, and on high impedance electrode systems which are used for signal detection and correction kicks. Among other aspects, this makes stochastic cooling preferential at higher energies, typically at relativistic energies with a Lorentz factor $ \beta=v/c $ exceeding 0.7, with $v$ being the particle velocities and $c$ the speed of light in vacuum. Electron cooling suffers from technical difficulties in accelerating a cold electron beam to very high energy and the relativistic reduction of the electron density which makes it less powerful at relativistic energies. This is just a general guideline and the cooling concept in storage rings depends on details of the storage ring design, such as ion optical layout and available space. Therefore it is advantageous to include the implementation of powerful beam cooling into the storage ring concept from the beginning. 
	
	There exists a considerable interest to force the ion beam to ultimate beam quality \cite{Schiffer-1985}. The regime of extreme beam quality is labeled beam crystallization and has been studied theoretically and experimentally, for electron and laser cooling. Different from ion traps, where 3-dimensional crystallization is readily achieved \cite{Pallace}, in storage rings the dispersion of the particle orbits in the magnetic structure of the storage ring results in additional heating which hampers 3-dimensional ordering. One-dimensional ordering, however could be demonstrated \cite{prlorder}.
	
	\subsubsection{Electron Cooling}
	\label{s:cooling}
	Electron cooling is achieved by merging the ion beam with an intense cold electron beam of the same velocity. The thermal energy of the ions is transferred by Coulomb interaction to the electron beam which is continuously renewed and resembles a reservoir of cold electrons. An extensive introduction into electron cooling is given in \cite{poth-ec,meshkov-ec,meshkov-ec2}.
	The cooling rate depends on charge $ q = Qe $ and the mass $ M \approx Am_0 $ of the ion, with $A$ the mass number and $m_0$ the average nucleon mass (see Eq.~(\ref{mass})), the electron density $ n_e $ of the electron beam and the ratio $\eta _e = L/C$ of the length $ L $ of the cooling section to the ring circumference $ C $. The cooling rate is particularly sensitive to velocity deviations  $ \delta v $ of the ion from the average electron velocity in all degrees of freedom. For the regime of tempered and only moderately pre-cooled beam the cooling rate $\tau $ scales as
	
	\be  \tau ^{-1 } \propto \frac{Q^2}{A} n_e \eta _e ({\delta v})^{-3}   \ee
	
	The cooling rate is small for hot ion beams with large relative velocities and increases to the maximum value at small relative velocities $\delta v$, which is determined by the temperature of the electron beam. Consequently, electron cooling is most powerful for ion beams which are pre-cooled or already have small energy spread and emittance.
	For powerful cooling the electron beam should have low temperature, large electron density and a long interaction section with the ion beam. The ion optical properties of the storage ring in the cooling section influence the cooling time. A large ion optical beta function in the cooling section reduces the transverse velocities of the ions for a given emittance, but results in a large ion beam size which may not be matched to the electron beam size. The value of the dispersion function in the cooling section also affects the speed of cooling. Therefore the choice of the ion optical functions is a compromise in the quest for the achievement of fast cooling and the compliance with requirements of storage ring operation and experiments.

	In the traditional electron cooling systems the electrons are emitted from a thermal gun which is immersed in a longitudinal magnetic guiding field. It was experimentally and theoretically investigated that the magnetic field affects the cooling power and inceases the cooling force experienced by ions~\cite{magcool}. A quantitative description of the increase of the cooling force in an analytical approach turned out to be difficult and did not match experimental observations well. An empirical formula, however, derived from systematic experimental investigations and named after the inventor V.V.~Parkhomchuk  matches many experiments and turned out to be a good approximation~\cite{parkhomchuk}. The formula is frequently used for predictions of the cooling force and is included in several simulation tools employed in the design of new cooling systems. The magnetic field of the cooling device extends from the electron gun to the electron collector. The guiding field of the electron beam also covers the interaction region between ions and electrons and therefore affects the ions which are circulating in the storage ring. In a system with constant magnetic guiding field strength the temperature of the cathode determines the transverse temperature of the electron beam. A method to reduce the transverse electron temperature was conceived by decreasing the magnetic guiding field between electron gun and the cooling section. The electrons are bound to the magnetic field lines, the electron beam size is expanded in the decreasing magnetic guiding field and the temperature is reduced as the quantity $ p^e_\perp /B $, with the transverse momentum $ p^e_\perp $ of the electron and the magnetic field strength $ B $, is conserved. This has been demonstrated to greatly improve the resolution in ion-electron recombination experiments which are sensitive to the transverse electron temperature. The electron cooling force is also increased for the expanded beam, but the increased force is foiled by the reduction of the electron density of the expanded beam~\cite{transexphakan}.

\begin{figure*}[t!]
\centering\includegraphics[angle=-0,width=0.8\textwidth]{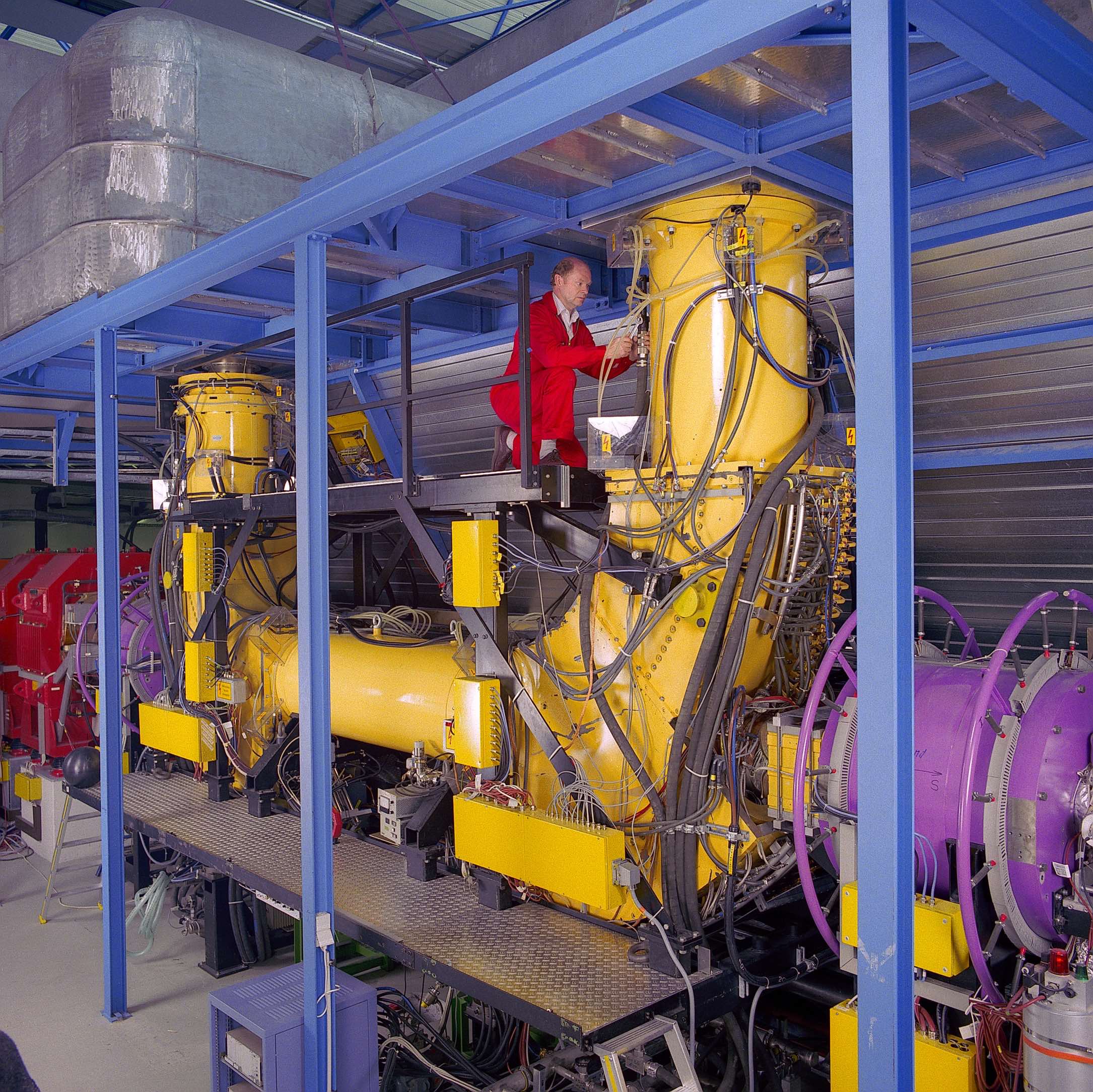}
\caption{(Colour online) The ESR electron cooling system is a typical low and medium energy electron cooling system. A magnetic guiding system provides a longitudinal magnetic field in which the electrons move from the electron gun (located in the front vertical part of the system) to the electron collector (in the rear vertical part). The magnetic guiding field is provided by solenoid and toroid magnets. The electron and ion beams are merged in the horizontal 2.5 m long cooling section. Photo: A. Zschau, GSI, Darmstadt.}

\label{esrecool}
\end{figure*}

	The cooling force is proportional to the electron density, consequently the extraction of a high electron current from the gun is beneficial for powerful electron cooling. According to $ n_e= j_e/v $, the electron density for a beam of fixed current density $ j_e $ is reduced when the electron beam is accelerated to higher velocity $ v $. The current density at the extraction from the cathode is limited and cooling at higher energy suffers from the reduced electron density.   
	In the majority of the existing electron cooling systems the acceleration of the electrons was performed in electrostatic acceleration structures. Electron energies up to 300~keV are readily available with standard high voltage systems. Figure \ref{esrecool} shows the electron cooling system of the ESR as a typical electron cooling device. The parameter range of the ESR electron cooling system in operation is listed in Table 2.  Higher energies require special design of the high voltage system. Only recently the use of a bunched electron beam accelerated in a linear rf accelerator resulted in electron cooling of a relativistic heavy ion beam in a collider \cite{bnllerec}.

\begin{table}[h]
\caption{Parameter range of the ESR electron cooling system employed in cooling experiments.}
\begin{center}
\vspace*{2mm}
\begin{tabular*}{0.6\textwidth}{ll}
\hline
electron energy & 1.6~-~230~keV \\
electron current for cooling & 0.001~-~1.0~A \\
maximum electron current & 3 A\\
relative current loss to ground & $ \le 2 \times 10^{-4} $ \\ 
gun perveance & 1.95~$ \mu $ P \\
cathode temperature & $\sim$ 1000 K \\
electron beam diameter & 50.8~mm \\
magnetic field strength & 0.01~-~0.11~T\\
maximum magnetic field strength & 0.2 T \\
length of cooling section & 2.5~m\\
basic vacuum pressure & $1\times 10^{-11}$~mbar \\
vacuum pressure with electron beam & $\le~5\times 10^{-11}$~mbar \\
\hline
\end{tabular*}
\end{center}
\end{table}

	\subsubsection{Stochastic Cooling}
	
	Stochastic cooling was developed for the reduction of the phase space volume of antiprotons, which were produced in a thick target and which were needed for collision experiments in large colliders at CERN and FNAL. The concept is similar to feedback systems which detect the deviation of particles from the nominal value and apply a correction signal to reduce the deviation. An overview of the status of the stochastic cooling technique can be found in \cite{moehl-sc} and references therein.
	The process itself is stochastic in the sense that a signal is derived from the fluctuations of the beam noise which is picked up by high frequency electrode systems, typically operating in the frequency range of a few GHz. The correction signal is sent from the pick-up electrode across the storage ring to another electrode used as kicker to correct the deviations. Since the applied electric field affects only the direction of the particle motion, the kicker electrodes are placed on the opposite side of the storage ring at a location where, due to the betatron motion, the measured position deviations result in angular deviations. The correction signal has to arrive in proper time and phase at the kicker in order to damp the intrinsic particle motion. The electronic chain between pick-up and kicker has to be adjusted to provide the correction signal in time and phase with the circulating particles over the full bandwidth of the system. The cooling rate of stochastic cooling is described by
	
	\be  \tau ^{-1} \propto \frac{2W}{N}  \ee
with the particle number $ N $ and the bandwidth of the electronic feedback circuit $ W $.
	
	Various aspects contribute to the performance of a stochastic cooling system. An important aspect is the ion optical properties of the storage ring, particularly the transition energy $ \gamma_t $ which determines the variation of revolution time with particle momentum. Concepts to minimize the change of the ion inter-particle distance along the orbit, the so-called unwanted mixing, between pick-up and kicker and to maximize the wanted mixing between kicker and pick-up have been worked out and considered for implementation in dedicated cooling rings. 
	
	For efficient transverse cooling the number of transverse particle oscillations due to the focusing elements of the ring, between pick-up and kicker, should be designed according to a phase advance $ \phi _{pu-k} = (n+\frac{1}{2}) \pi $, with an integer number $ n $. As the choice of the tune, the number of betatron oscillations around the ring, must avoid resonances which result in particle loss, the ion optical parameters of the storage ring have to be chosen in an early stage of design in order to achieve stable condition for beam storage and optimum conditions for stochastic cooling.
	
	The method of stochastic cooling, originally designed for antiprotons, has been extended to rare isotope beams~\cite{scesrnolden, scesrnolden2, Nolden-2004}, as another application to a hot secondary beam, but also to stable beams, ranging from protons to bare uranium. For highly charged ions stochastic cooling benefits from the increased signal strength, proportional to the charge $ (Qe)^2 $, of the detected signal and correspondingly increased kick strength of the correction signal. As the final beam quality in stochastic cooling is limited by electronic noise, for experiments with heavy ions and rare isotopes it is favorably combined with electron cooling which takes over when the initial fast stochastic cooling is approaching the noise limit. Despite of the stochastic nature of the cooling method stochastic cooling has also been applied to beams of very few ions, down to single ions, resulting in a fast reduction of the uncertainty in longitudinal momentum and transverse divergence originating from their production in a thick target, see Section \ref{s:t12}.
	
	\subsubsection{Laser Cooling}
	
	Laser cooling, originally developed in traps to damp the thermal particle motion, is considered as a method to provide fast cooling and to achieve smallest momentum spread for swift stored ion beams in storage rings~\cite{lasercooltsr}. Different from traps, an overlap between the ion beam and a collinear laser beam can be achieved in an efficient way to provide only longitudinal cooling. The laser light of frequency $ \Omega $ excites an atomic transition with frequency $\omega_{12}$ in an ion, which moves with a velocity corresponding to the Lorentz factors $\beta $ and $ \gamma $. The laser frequency has to be matched according to the formula describing the Doppler effect for the moving ions $ \Omega = \gamma \omega_{21} (1-\beta \cos{\theta} ) $, with the angle $\theta $ between the direction of the ion and the laser beam. Frequent repetition of the process of directed momentum transfer in the absorption process and the subsequent isotropic emission of a photon results in a directed momentum transfer which in combination with a counteracting force results in a reduction of longitudinal velocity spread. The counteracting force is usually provided by an external potential, e.g. by the ring rf system. The requirement of certain transitions between atomic states and the available powerful laser systems allows the application of laser cooling only to a few selected incompletely stripped ions \cite{Wen-2019}.
	So far $^7$Li$^+$, $^9$Be$^+$, $^ {24}$Mg$^+$, $^{12}$C$^{3+}$, $^{16}$O$^{5+}$ ions have been used in laser cooling experiments at storage rings, see, e.g. \cite{Wen-2019} and references cited therein. There are plans to apply laser cooling to heavier Li-like ions at relativistic energies which can profit from increased power or bandwidth in modern laser systems \cite{Winters-2015}.    
	
	\subsection{Internal Target}
	\label{s:targets}
	
	Internal targets in storage rings offer various advantages compared to traditional solid targets with a single passage of the beam. The high revolution frequency in the order of MHz results in a corresponding increase in the luminosity, which can balance the reduced thickness of the target. The interaction of reaction products with the target material is reduced proportional to the target thickness and allows reaction studies at lowest momentum transfer. Furthermore, the use of thin targets reduces unwanted multiple scattering and energy smearing. Cooled beams in a storage ring improve the definition of the interaction point and the particle energy and consequently the resolution.
	
	The operation of internal targets in a storage ring has to consider the requirements of ultra-high vacuum conditions which is indispensable particularly for low energy and heavy ion beams. The targets are designed for a good definition of the target volume and a minimum pressure rise in the vicinity of the target with a small increase of the average storage ring residual gas pressure.
	Various types of targets have been developed for the installation in storage rings~\cite{targetstori14}. The most common types are cluster jet, droplet beam and pellet targets. Cluster jet targets are very flexible in the operation with different materials, jet targets ranging from hydrogen over nitrogen to many noble gases have been operated. They provide particle densities in the range $10^{10}$ to $10^{15}$ atoms/cm$^{2}$. For the highest target densities the gas is forced by a high input pressure through a cooled nozzle. A system of skimmers shapes the gas jet to the size required by the experiment. The size of the nozzle through which the gas is expanded into the interaction region and the geometry of the skimmers which shape the jet and deflect scattered particles into the pumping system define the size of the target at the interaction point of a few millimeters.  Target densities as low as $10^{10}$ atoms/cm$^{2}$ are useful for experiments for heavy target materials and beams of highly charged heavy ions stored at low energy. The upper limit of the acceptable target density is either defined by the technical abilities to produce the highest densities or by the ability to compensate the energy loss and scattering of the beam particles by cooling and rf systems. Depending on the beam species and the beam energy, target densities exceeding $10^{14}-10^{15}$ hydrogen atoms/cm$^{2}$ have been shown to exceed the capacity of the beam cooling system and result in fast beam loss \cite{nimpetridis}.

\begin{figure}[hbtp]
	\centering
	\includegraphics[width=0.8\textwidth]{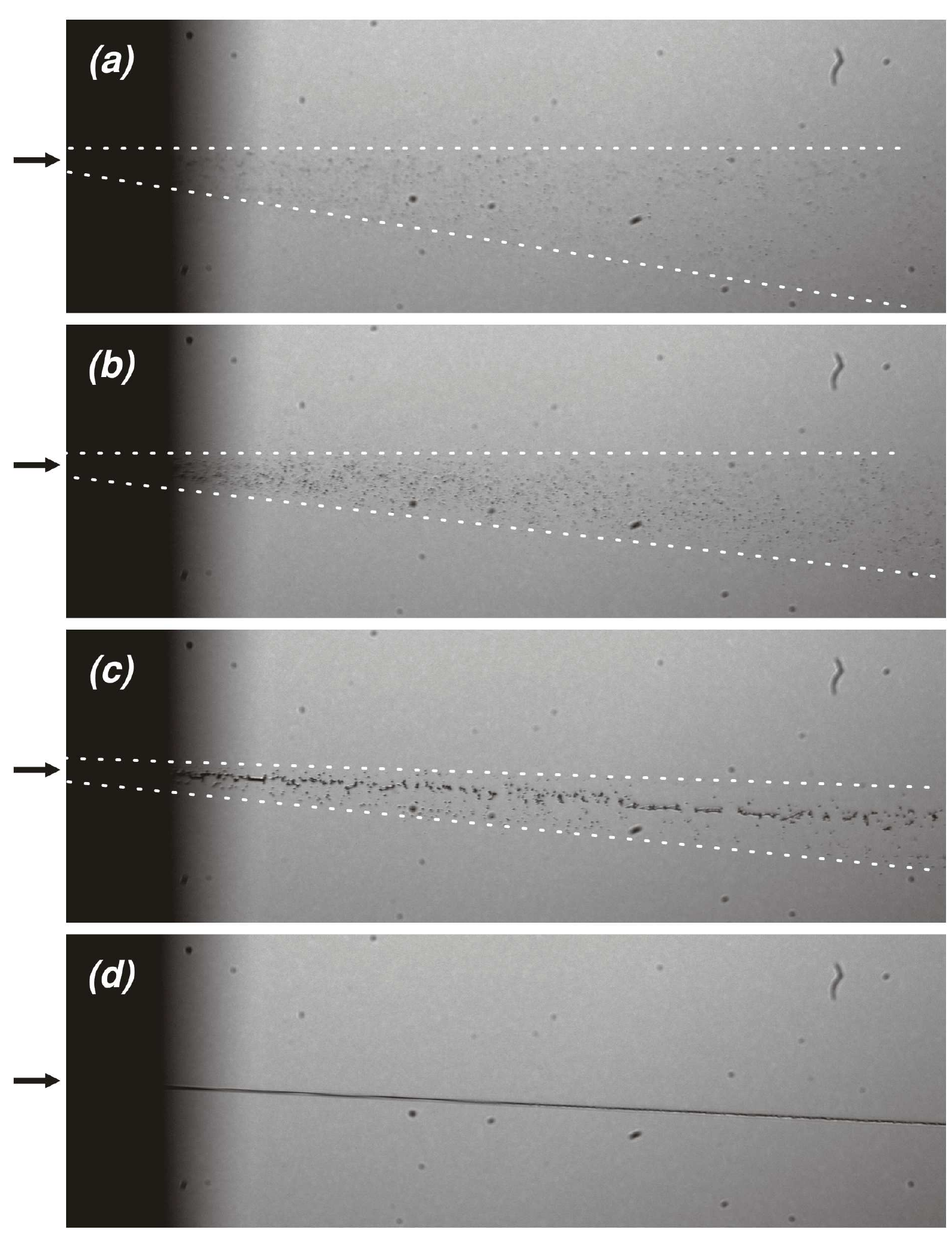}
		\caption{Transition from a cluster jet produced at a nozzle temperature of (a) 31 K to droplet beams at different temperatures of (b) 29 K and (c) 26 K. Note the reducing angular divergence of the beam (dotted lines) until eventually forming a hydrogen filament at (d) 23 K. The arrows indicate the nozzle position. Taken from \cite{Kuhnel-2009}.}
	\label{droplets}
	\end{figure}

	The production of a droplet target requires significantly lower nozzle temperatures, thus leading to liquefaction of the gas within the nozzle reservoir. The liquid expands into vacuum and undergoes cavitation induced fragmentation, forming a droplet beam with a broad droplet size distribution. This technique is particularly important for the realization of high density target beams utilizing very light target gases like helium and hydrogen~\cite{Kuhnel-2009}. The formation of a droplet beam has been observed by a shadow imaging system, see Figure.~\ref{droplets}.

	 Similar to droplet beam targets, pellet targets also use a cooled nozzle through which the liquid is expanded and pellets are shaped by means of a piezo-transducer. The individual pellets have a size on the order of micrometers and are separated by several millimeters. Thus a pellet target with a density of $10^{20}$ atoms/cm$^{2}$ in the pellet delivers an average target density on the order of $10^{15}$ atoms/cm$^{2}$. This is still in the regime where beam cooling can compensate the heating by the target.  An option to achieve higher target densities are storage cells, but at the expense of a long volume in which the beam particles interact with the target, as e.g. for a polarized target \cite{poltarget}.
	
	Even higher target densities are feasible using solid material, e.g. strip or fiber targets. The target density, however, significantly exceeds the limit which can be compensated by beam cooling. Solid targets can be positioned into the halo or the outer edge of the stored beam thus the targets interact only with a small fraction of the stored beam particles which are rapidly lost. Another method to reduce the interaction rate of the ion beam with a solid target is sweeping of the ion beam over the target. This allows to control the reaction rate and the lifetime of the stored beam. Solid targets however suffer from reduced resolution and increased background and to some extent sacrifice the virtues of the use of stored beams.
	
	The actual target density in an experiment can be determined by beam based measurements from the energy loss which follows the well-known Bethe-Bloch formula for the reduction of kinetic energy in a path length of matter

		\be  \frac{dE}{dx} = \frac{Z^2e^4}{4\pi \epsilon _0^2 m_e } \frac {n_t}{\beta ^2 c^2}~(ln (\frac{2m_ec^2 \beta ^2}{l(1-\beta^2)} )-\beta ^2) \label{BBloch}\ee

	with the electron mass $m_e$, the elementary charge $e$, the vacuum permittivity $\epsilon_0$, the electron density in the target $n_t$, the ion charge number $Z$, the mean ionization potential $l$ and the Lorentz factor of the beam  $\beta$. The energy loss can be measured precisely by observing the change of revolution frequency when the beam interacts with the target without any compensation by the beam cooling or rf system. The relative change of energy

		\be \frac{dE}{E} = \frac{1+\gamma}{\gamma} \frac{1}{\eta} \frac{df_0}{f_0} \ee

corresponds to the measured change of revolution frequency $f_0$ and is proportional to the momentum slip factor 
$ \eta = \frac{1}{\gamma^2 } - \frac{1}{\gamma^2_t}$ and the transition energy $\gamma _t$ of the storage ring. 
This method is also applicable to a beam based measurement of the average ring vacuum when no target interacts with the stored beam.
This average, however, does not take into account the composition of the residual gas.
	
	Other kinds of targets were employed in the storage rings as well. Free or quasi-free electrons are required for atomic physics experiments. The most common type of ion-electron interaction experiments are studies of dielectronic recombination, see Section \ref{s:dr}, which use the electron beam of an electron cooling device at non-zero relative velocity. The velocity of the electrons, unlike the condition for electron cooling, is detuned relative to the ion velocity. As the cooling by the electrons is absent, such a detuning is introduced only for a short time period, typically for some 10 ms. After the time of interaction at the detuned energy, cooling is re-established to compensate the emittance and momentum spread growth and a coherent energy shift by readjustment of the electron energy. Alternatively, a second electron cooling system can be used to maintain a cooled ion beam. The use of two electron beam systems, one operated as electron target, at different electron energies or cooling by stochastic cooling are options to operate at a continuous detuning without compromising the quality of the cooled beam \cite{Stohlker-2011}.
	Alternatively, transverse electron targets which use electrons not confined by a longitudinal magnetic field and cross the ion beam with very low energy have also been considered \cite{Brandau-2017e}. They have the advantage of a well defined interaction point in the longitudinal degree of freedom.
	
	Laser systems are also used in atomic physics experiments providing photons interacting with the ion beam \cite{laserspec}. The laser beam, overlapped in a straight section with the ion beam, excites atomic transitions in incompletely stripped ions for precision spectroscopy. As the interaction depends on the angle between ion and laser beam, a precise control of the direction of the two beams in the interaction region is needed. The precise control of the relative beam direction is achieved by the use of mechanical scrapers which allow an angular adjustment of the two beams with an accuracy of better than 0.1 mrad. This accuracy is sufficient as the internal angular spread due to the finite transverse emittance is at least of that order.
	
Very recently it was proposed to employ thermal neutrons as an internal target in storage rings. Either neutrons from a reactor \cite{Reifarth-2014}
 or thermalized in heavy water neutrons from a spallation reaction \cite{Reifarth-2017} could be used for this purpose. 
 The realization of such target is highly complicated and still requires extensive feasibility studies.
 However, if constructed, it will enable a wealth of physics cases related to neutron induced reactions on stored radioactive beams.

	\subsection{Beam Diagnostics}
\label{S:BeDi}
	The diagnostics of the stored beam in storage rings is preferentially non-destructive. Various diagnostic devices are common with synchrotrons~\cite{beamdiag}. Capacitive beam position pick-ups allow the measurement of the transverse position of a bunched beam at certain positions around the ring and consecutively the adjustment of the beam orbit. The beam current and thus the number of stored ions can be measured with fast current transformers for beams with a time structure imposed by the external rf fields. The coasting beams can be detected with slow current transformers. The standard current transformers are based on high permeability amorphous material which encircles the beam in a toroidal core, where the azimuthal field component changes the magnetic properties and allows the measurement of the beam current in the range of microAmperes to Amperes. For very low currents, e.g. for a low intensity or a low energy beam of ions or molecules with low charge, cryogenic current comparators were more recently developed. These comparators use a SQUID sensor to extend the measurement range to the regime of beam currents of nanoAmperes~\cite{ccc}.   Current monitors are particularly useful to measure the beam lifetime by monitoring the stored current as a function of time and for the normalisation of reaction rates.  
	
\begin{figure*}[t!]
\centering\includegraphics[angle=-0,width=0.8\textwidth]{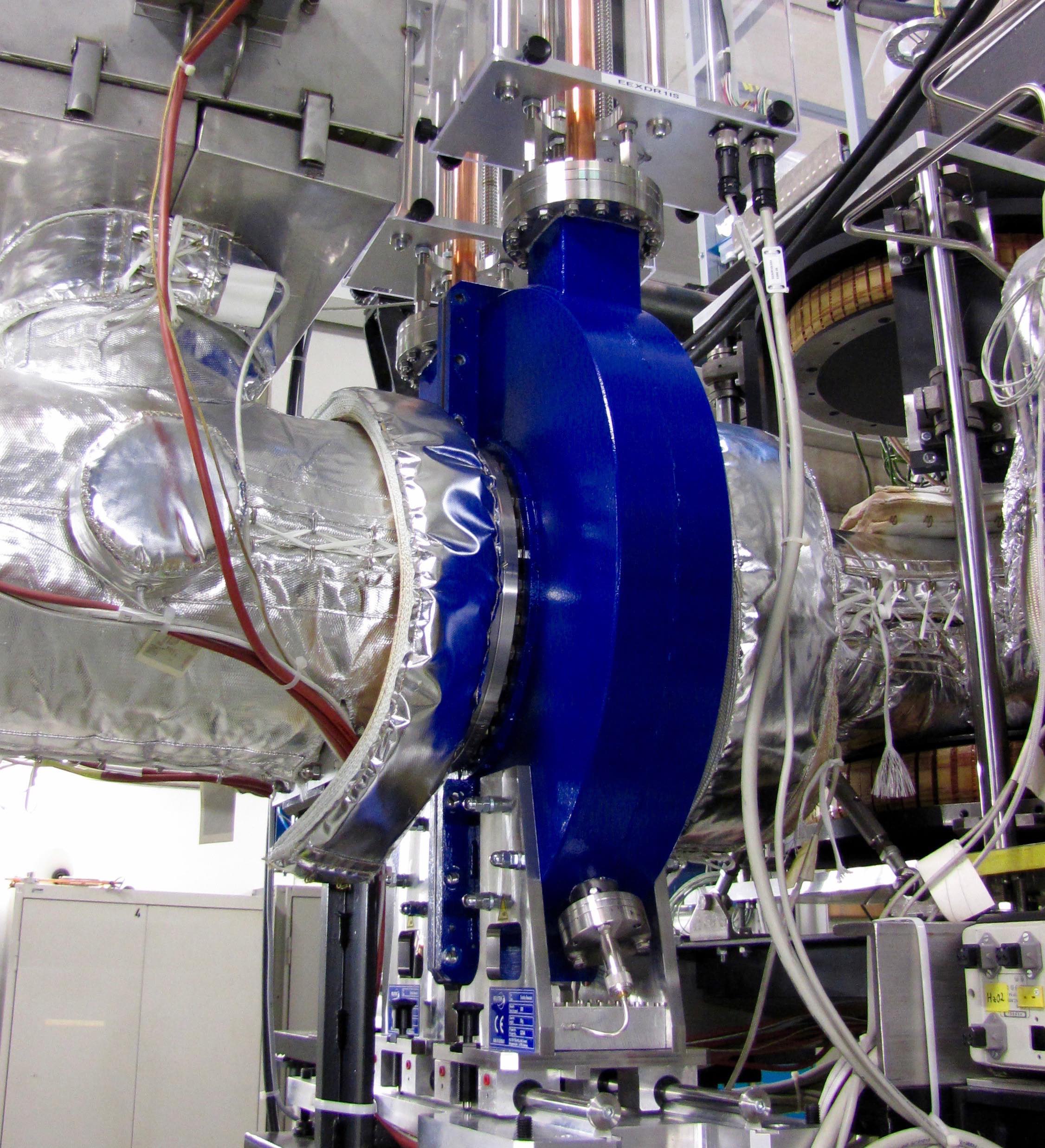}
\caption{(Colour online) Resonator for Schottky noise detection at a frequency of 245 MHz installed at the ESR storage ring. The system provides high sensitivity due to its large quality factor and consequently provides fast measurements with high time resolution by narrow band detection. Photo: M. S. Sanjari, GSI, Darmstadt.}

\label{esr-schottky}
\end{figure*}

	The availability of beam cooling demands devices that measure the longitudinal momentum spread and the transverse emittance with good resolution. Another requirement to monitor the cooling process is the ability to perform the measurements with good time resolution. Schottky noise  detection in circular accelerators ~\cite{schottky} is the appropriate technique to detect the distribution of revolution frequencies which is proportional to the momentum distribution of the beam, $ {\delta f}/{f} \propto \eta {\delta p}/{p} $, with the momentum slip factor $\eta $. The Schottky signal provides information on the momentum distribution which is reflected in the distribution of revolution frequencies and on the average beam momentum derived from the center of the frequency distribution. If either the beam velocity $\beta $ or the orbit length $C$ is known with good precision, the other quantity can be determined with the same precision, the revolution frequency can be easily determined with an accuracy on the order $ 10^{-6} $ and in special cases even more precisely. For a multi-component beam consisting of ions with mass $M$  and charge $Q$ the revolution frequency $f$ of the different ions stored simultaneously can be compared resulting in a precise measurement of nuclear masses according to formula

	\be \frac{\delta f}{f} = - \alpha _p \frac{\Delta (M/Q)}{(M/Q)} + \eta  \gamma ^2 \frac{\Delta \beta }{\beta } \label{eqx}\ee
with the momentum compaction factor $\alpha_p= {\delta L/L}/({\delta p/p})$ of the storage ring and the Lorentz factor $\gamma$.
Standard broadband system allow the detection of many Schottky signal harmonics at lower frequency, but are not sensitive enough to detect higher harmonics. A special resonant Schottky noise detection system was developed with an improved sensitivity which allows the detection of single ions and the measurement of the frequency distribution of the  Schottky signal with a time resolution of the order 10 ms \cite{esrschottkyreso}. It is based on a resonant cavity which couples in a narrow bandwidth to the electric field of the passing ions (Figure \ref{esr-schottky}). The high sensitivity detection is important for the detection of short-lived nuclei and the measurement of their decay with good time resolution. 

	For the measurement of the transverse beam emittance a measurement of the beam width $ \sigma_x $ is sufficient as the emittance $ \epsilon_x $ of a beam in a storage ring can be determined from the relation $ \epsilon_x = {\sigma_x}^2/\beta_x$. The ion-optical beta function $ \beta_x $ at the profile monitor is known sufficiently well from ion-optical calculations and can also be determined experimentally by tune response measurements. The time evolution of the beam distribution, due to cooling or interaction with an internal target, correspondingly requires a precise beam size measurement with appropriate time resolution.
	
	Detectors have been developed which measure ionized particles produced by the interaction of the stored beam with the residual gas. Position sensitive detectors measure the projection of the residual gas ions in a homogeneous transverse electric field onto the detector plane. Various concepts for such beam profile monitors installed in the ultra-high vacuum of the storage ring have been implemented. The sensitivity and time resolution is dependent on the vacuum conditions, but also on the beam species and the beam energy as the ionization of residual gas particles is proportional to $\beta $ and $Q^2$ of the beam.
	
	Other monitors of the transverse beam distribution make use of particles which change charge during their circulation in the ring due to interaction with the residual gas. At low energies, protons can capture an electron from the residual gas or an internal target resulting in neutral hydrogen atoms moving straight in the direction of the beam with the same velocity unaffected by the field of magnetic ring components. Neutral particle detectors at the end of the straight section provide an image of the stored beam. For heavy ions the down-charged ions generated by electron capture are separated from the stored beam in dispersive sections \cite{Klepper-1992, Klepper-2003}. Position sensitive detectors can to be placed close to the circulating beam, as the horizontal separation of charge states is in the range of millimeters to centimeters depending on the charge change of the ion and the value of the dispersion function of the ring. For low energy ions the detectors have to be installed in the vacuum, whereas for higher energies it is acceptable to install the detectors in vacuum pockets which separate the detector volume from the ring vacuum system by a thin entrance window.
	The detection of ionized beam particles for incompletely stripped stored ions is also possible with this technology. Particle detector systems are also powerful in coincidence measurements of particles emitted in the interaction of the stored beam with a target and of beam particles which have changed their charge. The efficiency to transport the particles into the detector is 100 \%, if the dispersion function and the acceptance defined by the vacuum chamber allow the transport from the interaction point to the detector. If the particle detectors have sufficient transverse position resolution, the size and consequently the transverse emittance of the stored beam can be measured.  
	
	Some of the detector systems employed in recent experiments are briefly discussed in the corresponding sections of Chapter \ref{s:masses}.
	
	\subsection{Beam Parameters}
	\label{s:bpar}
	
	 Powerful cooling systems are mandatory to prepare ion beams of highest phase space density for precision experiments in storage rings. However, the achievable beam quality is not only defined by the cooling system. The cooling to highest phase space density forces the beam into a regime when additional processes contribute to the beam quality. Coulomb forces between the ions in the dense beam result in heating processes. Intrabeam scattering (IBS) counteracts the achievement of high phase space density with a heating rate which is inversely proportional to the phase space volume occupied by the ion beam and increases linearly with the ion number \cite{ibs}.
In the course of the cooling process, the reduction of the phase space volume ceases when the heating rate due to IBS equals the cooling rate. Other limitations originate from the cooling process itself. The stochastic cooling rate depends on the ion number and the electron cooling rate depends on the density of the electron beam. Finally, the beam energy also affects the resulting beam quality due to the dependence of IBS on energy and the dependence of the electron cooling rate on the beam energy for a fixed electron current.
	 
\begin{figure*}[h!]
\centering\includegraphics[angle=-0,width=0.8\textwidth]{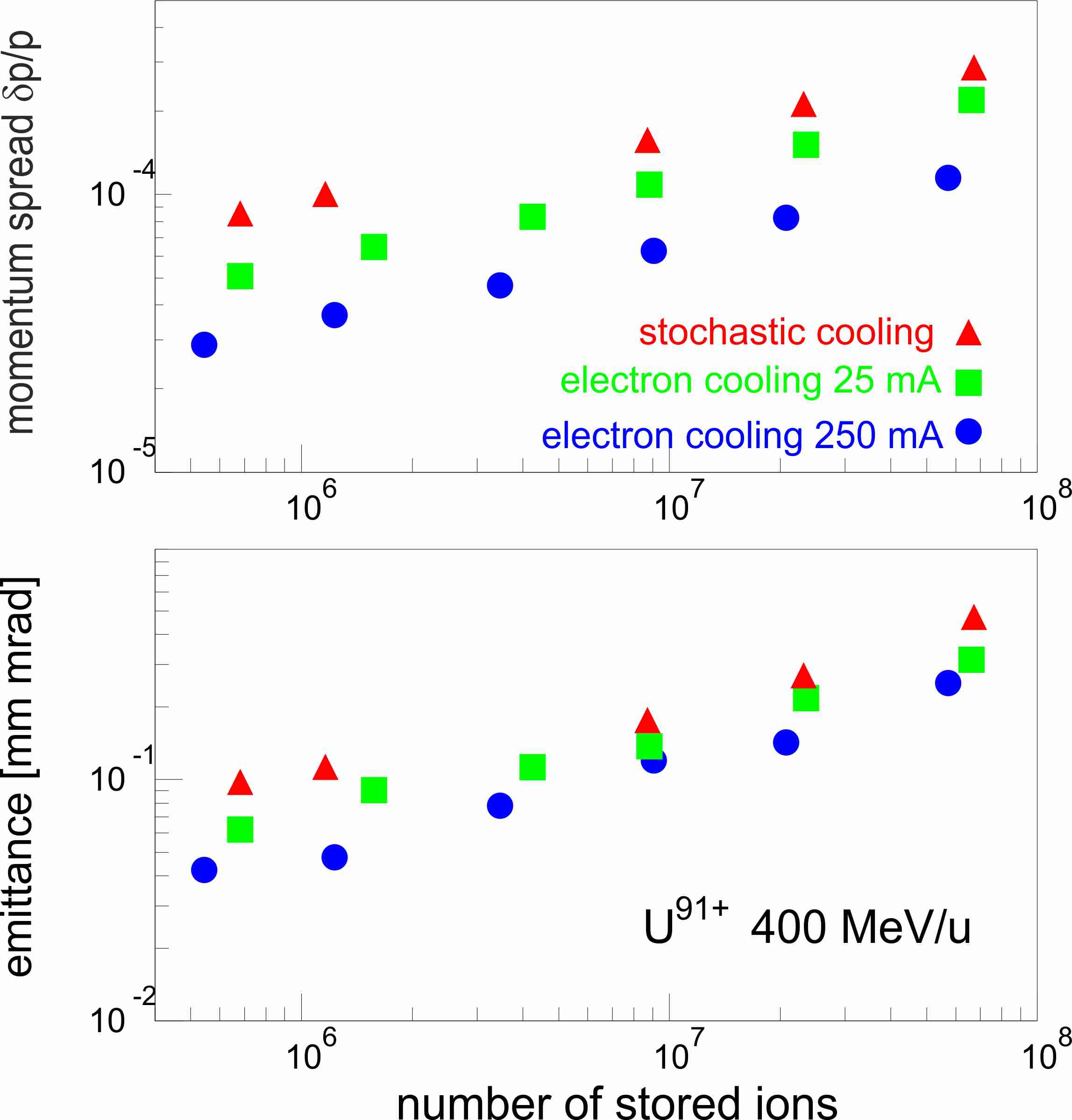}
\caption{(Colour online) Beam parameters (momentum spread, horizontal emittance) of a cooled beam of U$^{91+}$ at an energy of 400 MeV/u cooled by stochastic and electron cooling. The cooling rate of the electron beam was varied by application of two electron current (25, 250 mA).
The dependence on the ion number is caused by intrabeam scattering.}

\label{equi-ecsc}
\end{figure*}

	 Various simulation codes have been developed to study this intertwined interplay of beam parameters \cite{ibsmw, betacool}. Complementarily, a lot of information is available from beam experiments. A typical example of experimental observations in Figure~\ref{equi-ecsc} shows the effect of the cooling rate on the equilibrium momentum spread and transverse emittance. The measured values are shown with the cooling rate as a parameter for two different values of the electron current and in comparison to the equilibrium values for stochastic cooling. There is a clear trend of increased beam quality starting from stochastic cooling with a moderate beam quality and an improvement of the beam quality for the increased electron current resulting in an order of magnitude stronger cooling. Highest beam quality is clearly achieved with strongest electron cooling. However, a detrimental effect is the decrease of beam lifetime which is inversely proportional to the electron current.  

	 Another limitation of beam quality comes from space charge effects. The space charge forces between the ions result in defocusing which shifts the betatron tune and can result in a beam heating by resonances and corresponding particle loss. The effect of the ion beam space charge is more pronounced at low beam energy and is the strongest for highly charged ions. The tune shift $ \delta Q_{x,y} $ in the horizontal or vertical phase space due to space charge is known from theory 
	  \be \delta Q_{x,y} = -\frac{r_p N_ i Q^2 }{2 \pi  \beta ^2 \gamma ^3 A \epsilon _{x,y}} (\frac{2}{1+\sqrt{( \epsilon _{y,x} \beta _{y,x})/( \epsilon _{x,y} \beta _{x,y})}}) \ee
	  with the classical proton radius $r_p$, the number of ions $N_i$, charge and mass number of the ion $Q$ and $A$, the transverse emittance $\epsilon_{x,y}$ and the beta function $\beta_{x,y}$ in the respective plane.
	  
	  This is the most serious limitation for experiments with highly charged ions at low energy as well as for the process of ion deceleration to low energy, e.g. for beam energies below 5 MeV/u the particle number for cooled beams of bare heavy ions is limited by the space charge tune shift to some $ 10^7$. 
	  
	  As the IBS rate increases $\propto Q^4/A^2$, even for ions of low charge it is the limiting process for achieving best beam quality, to some extent also caused by lower cooling rates for ions with lower charge. This is known from the storage of protons and antiprotons and was also observed during the preparation of singly-charged ion beams for laser cooling studies. 
	  
	  The operation of an internal target causes additional heating of the stored ion beam. The interaction of the stored beam with the target results in a growth of the transverse emittance $\epsilon $ with a rate
	  \be \frac{d \epsilon}{dt}=\frac{f_0}{2} (\beta _t \theta_{rms}^2+\frac{D_t^2}{\beta _t} \delta_{loss}^2+ \beta _t D_t^{ \prime 2} \delta^2_{loss}) \ee
with the mean square scattering angle $\theta _{rms} ^2 $ and mean square momentum deviation $\delta _{loss} ^2$ per target traversal, the ion optical beta function $\beta _t$ at the target location and the dispersion function $D_t$ and its derivative $D_t^{ \prime}$ at the target location. 
	  An obvious conclusion from the emittance heating rate is the preference for a small beta function at the target location in order to minimize the emittance heating. A small dispersion function at the target supports the achievement of highest spatial resolution as the interaction point of the particles with the target is independent of the particular particle momentum and also reduces heating.
	  
	  For intermediate target densities in the range $10^{13}-10^{14}$ atoms/cm$^{2} $, the additional heating can be compensated by electron cooling with a moderate electron beam current without significant increase of beam emittance and momentum spread. 
The beam emittance and momentum spread is still determined by IBS and no significantly different beam equilibrium values with and without target are evidenced.
	  For the highest available target densities exceeding $10^{15}$ hydrogen atoms/cm$^2$, beam losses were observed in experiments due to the interaction with the internal target. If the energy loss in the target is too large for compensation by cooling it still can be compensated by the rf system of the ring. For a bunched beam the energy loss is compensated by the rf potential which provides a force which increases linearly with the energy deviation. In addition, the cooling system is needed to compensate the much weaker energy straggling due to the target. If the increased local particle number in a beam bunched by a conventional harmonic rf system constitutes a problem to stable beam operation, dedicated barrier bucket systems \cite{bbsystems} allow a flexible control of the longitudinal particle density and an operation with reduced bunch currents to provide the required energy loss compensation. 
	  
	  For high intensity beams the increase of beam emittance due to IBS results in a beam size at the target that can exceed the size of the target. As a result, the particles in the tails of the stored beam will not pass through the target and thus do not contribute to the luminosity any more. This introduced luminosity uncertainty has to be considered in experiments which are aiming at the measurement of absolute reaction cross sections. Dedicated luminosity monitors have to be considered, see, e.g., Section~\ref{s:pg}.
	  
\begin{figure*}[t!]
\centering\includegraphics[angle=-0,width=0.8\textwidth]{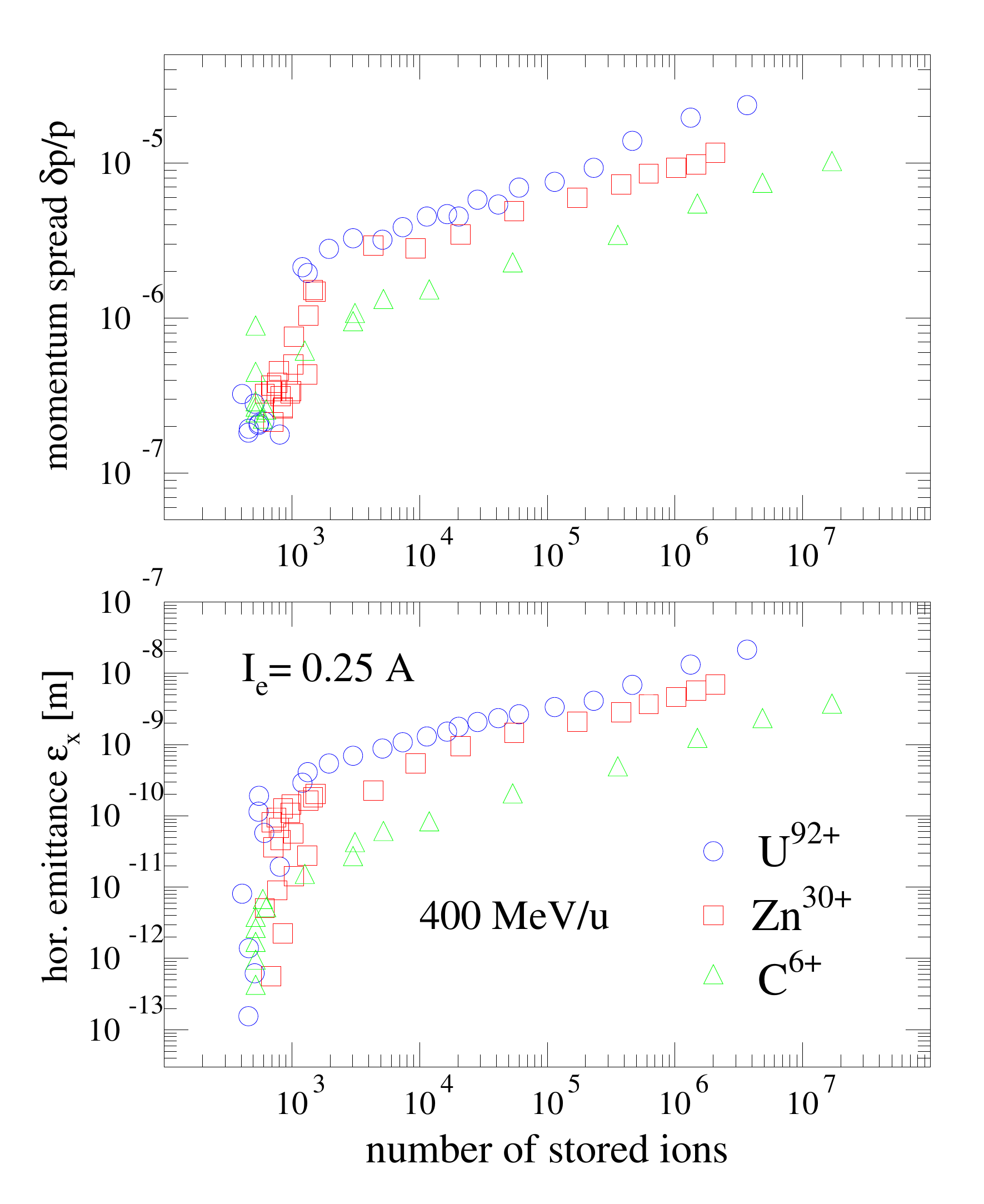}
\caption{(Colour online) A discontinuous reduction of momentum spread and transverse emittance of bare heavy ion beams cooled by electron cooling indicates a transition from an IBS dominated beam to an ordered beam. The cooling rate defined by the applied electron current of 250 mA results in the ordering effect for stored particle numbers below about 1000. The reduction factor is larger for ions with larger charge. Taken from \cite{Steck-2004}.}

\label{step}
\end{figure*}

	  A unique situation with respect to beam parameters of stored cooled beams was observed in experiment with low intensity heavy ions. For electron cooled highly charged ions which are subject to the highest IBS rates, a strong sudden reduction of the momentum spread was observed when the number of stored ions was decreased below a certain value \cite{prlorder}. This discontinuous behavior can be interpreted as a transition from a state when the IBS rate decreases below the cooling rate provided by the electron beam. At lowest intensities, electron cooling dominates the ion beam properties and an equilibrium is reached between the ion and electron temperature which is dominated by the electron temperature. Direct measurements of the revolution frequency distribution for the low intensity beam revealed that the frequency spread measured by Schottky noise detection is limited by the ripple of the dipole power converters. The current noise results in fluctuations of the revolution frequency which is larger than the spread due to the energy spread of the beam. 
	  The sudden reduction of the momentum spread indicates a transition from a gaseous to an ordered ion beam \cite{Hasse-2003}, the ions are lined up longitudinally and IBS is suppressed by cooling.  In further experiments it could be demonstrated that the transverse emittance also shrinks in a discontinuous way (Figure \ref{step}).

	  The very small momentum spread of the ordered ion beam has been achieved for ion numbers up to $10^4$. The experiments showed that the maximum number of ultra-cold ions depends on the applied cooling rate which is determined by the electron current \cite{prlordercryring}.
	  A significant dependence on the charge of the ion could not be observed, most likely due to the dependence of both cooling rate and IBS rate on the charge and mass of the ion. The ordered beam has a momentum spread below $10^{-6}$ which provides excellent conditions for precision experiments which are sensitive to the momentum spread. A prominent example is Schottky Mass Spectrometry (SMS) \cite{sms}, see Section \ref{s:sms}, the measurement of nuclear masses with cooled rare isotope beams by measurement of the revolution frequency by means of the Schottky noise. The determination of the revolution frequency with an accuracy in the range of $10^{-7}$ results in accurate relative measurement of masses which depends on the ion optical $\alpha _p$ value and consequently results in a relative mass accuracy down to a few times $10^{-7}$. The existence of the ultra-cold beam at small particle numbers with the very small frequency spread of the longitudinal Schottky noise signal is key to the achieved resolution in mass measurements and separation of isomeric states \cite{Franzke-2008}. 

	  Experiments also proved that the ultra-cold beam has a transverse emittance in the range $10^{-6}$ to $10^{-7}$~mm~mrad. This feature of the ultra-cold beam, until now, has not been exploited in experiments. It would allow for ultimate precision for the determination of the interaction point in in-ring reaction studies.
	  
	  The operation of the storage rings with high intensity beams conflicts with the achievement of ultimate resolution. The increase of the phase space volume of the beam due to IBS results in larger momentum spread and larger transverse emittance. High intensity operation is a trade off between luminosity and resolution.
	  High intensity stored beams also suffer from additional processes like interaction with impedances of the accelerator components and space charge effects which can result in beam losses. The achievement of a small impedance of the machine is in conflict with the necessity to install experimental devices in the ring which introduce additional impedances, thus increasing the total impedance value of the storage ring. Higher beam intensities can be stabilized by feedback systems which damp coherent motion of the beam particles due to impedances. For the typical cooled beam parameters coherent oscillations start for ion currents of a few milliAmperes of coasting beam, for bunched beam with larger local ion currents the threshold for collective effects is even lower and is proportional to the increased local current of the bunched beam.

	  	\subsection{Beam Lifetime}
	  	
	  The lifetime of the stored beam is an important property for experiments in storage rings. High intensity beams and high brightness beams can have limited lifetime due to interaction with machine impedances and due to space charge, particularly at low energies space charge forces limit the intensity for cold beams. At higher energies the main limitation comes from interaction of the ions with the residual gas or an internal target with which the ion beam interacts. The process of interaction with residual gas and an internal target is basically the same and depends on the stored particle species and the beam energy. Highly charged ions tend to capture an electron in the interaction with matter. The capture rate depends on the ion charge and the energy of the ion. Partially stripped ions can be ionized by the target. After changing the charge in both cases, capture and ionization, the ion will be lost at the limited momentum acceptance of the storage ring. Even in the case of large momentum acceptance and the ability to store ions with different charge states the orbits of the charge states will differ significantly and the situation has to be studied by ion optical calculations in order to have clear experimental conditions and to benefit from the large momentum acceptance. The rate of charge changing is proportional to the target density. The residual gas affects all stored particles and consequently storage rings which are operated with heavy ions aim for a gas pressure in the low $ 10^{-11} $ mbar range or better, particularly if particle storage at lowest energies is considered. Therefore the vacuum systems are designed as ultra-high vacuum systems with special choice of all materials installed and the necessity to bake out the vacuum system in situ is an indispensable procedure to reach ultra-high vacuum conditions in workable time periods. The rules for ultra-high vacuum components apply also to all experimental set-ups which are installed in the vacuum system of the ring and consequently contribute to the total vacuum system outgassing rate.
	  
	  For highly charged ion beams which are cooled by electron cooling the main loss mechanism is electron capture of free electrons during the passage through the cooling section. The dominant capture process is by radiative electron capture~\cite{rec, Eichler-2007}, especially for bare ions, but other capture process, e.g. dieletronic recombination for incompletely stripped ions, see Section \ref{s:atomic}, can contribute significantly to the recombination rate or even dominate the loss rate \cite{dr}. The recombination with electrons of the electron beam is proportional to the electron density. Thus stronger cooling results in faster particle losses, but the ratio of the cooling rate and recombination rate is not affected for a constant normalized emittance of the ion beam. Even for the ions of highest charge the lifetime is two to three orders longer than the cooling time and most of the experiments are not seriously hampered. In experiments which are aiming at long storage time, e.g. when the ion beam is used to breed another beam component in the ring, the cooling rate has to be compromised for longer beam lifetime.
	  
	  Another regime is the operation of storage rings with high energy ions which still have a significant number of bound electrons. A typical case is experiments using laser systems either for precision spectroscopy, fundamental issues or laser cooling. The lifetime of these ions is limited by the high cross section for stripping in the residual gas. The requirement for this type of experiment is in accordance with that for operation of highly charged ions at lowest energy and requires a vacuum pressure of $10^{-11}$ mbar or better. Finally, storage rings were employed in experiments with low energy molecular beams and for reasonable beam lifetime the same stringent vacuum requirements are valid.
	  
	  \subsection{Beam Accumulation}
	  \label{s:accu}
	  
	  The intensity required by experiments in storage rings is not always available from the injector feeding the ring. The injector has to deliver the required beam current in a time period which is matched to the injection method of the storage ring. The lack of intensity to fill the storage ring is usually more pronounced for experiments using radioactive ion beams. Methods to accumulate secondary beams were first developed for the preparation of antiproton beams which were required for high luminosity operation in proton-antiproton colliders. More recently various additional methods of beam accumulation have been demonstrated which were conceived for experiments with heavy ion and rare isotope beams. The accumulation methods are based on beam cooling which allows one to overcome the limit of the intensity due to the value defined by the acceptance of the storage ring and the emittance of the injected beam. The general scheme is based on the ability provided by cooling to force particles within the acceptance to a certain phase space volume and thus emptying a major part of the acceptance for repeated injection. This makes beam accumulation an indispensable method to provide useful beam intensities, particularly for radioactive ion beams. 
	  
	  In low-energy storage rings a routine method to inject the beam into the ring is transverse multiturn injection. This method allows filling of the transverse ring acceptance during a time corresponding to several turns, typically some ten turns, with a beam of smaller emittance. The well-established standard method uses horizontal multiturn injection, but the combination with vertical phase space painting is an option, which requires the use of a tilted injection septum and a sufficiently large acceptance of the storage ring in both transverse planes. Due to cooling it is possible to repeat the injection after compressing the stored beam in all degrees of freedom to a much smaller phase space volume~\cite{mmti}. Consequently, the compression to the smaller phase space volume is repeated for even more injections. The repeated multiturn injection has been used with beams from electrostatic accelerators, linear accelerators and cyclotrons. The particles accumulated with this method range from protons, unpolarized and polarized, to highly charged heavy ions.

	The efficiency of the accumulation depends on the emittance of the incoming beam and the acceptance of the storage ring, but also the lifetime of the beam which needs to  be significantly longer than the cooling time affects the performance of the accumulation procedure. The gain factor depends on the beam lifetime. Long-lived secondary particles at high energy are more suitable for accumulation than short-lived isotopes. Low energy beams should have a long lifetime with respect to the vacuum to allow a large gain factor.  
	 
	 Another non-Liouvillean method is charge changing injection. This is used routinely for protons by injecting H$^-$ ions from the injector and passing them through a stripping foil at the injection point of the ring. This method has also been demonstrated for light ions by ionizing the ions in a stripper foil to the bare charge state and thus matching the injected particle after stripping to the magnetic rigidity of the storage ring \cite{stripinject}. The particle can pass several turns through the stripping foil until the closed orbit in the ring is shifted to a position outside the foil, similar to the orbit shift during multiturn injection. This method, however is limited for heavier ions by the energy deposition in the stripper foil during successive passages. The energy loss in the foil results in deceleration of the ions during their first turns in the ring.
	 
	 The accumulation can also be performed in longitudinal phase space. Already in the early concepts for antiproton accumulation a ring with large momentum acceptance was filled with coasting beam of smaller momentum spread which was compressed in the phase space by stochastic cooling. This has later also been applied in ion storage rings with appropriate momentum acceptance. The method is based on harmonic rf systems and the application of cooling to decrease the huge momentum spread (up to a few percent) of the stacked beam. By application of modern broadband rf systems for the control of the longitudinal particle position in the ring, the method can be extended to filling injected beams into a fraction of the ring circumference. Barrier bucket systems provide an rf potential which can vary from sinusoidal in time to nearly rectangular defining a gap in the circulating beam for the injection of additional particles. The ring fraction which is not filled by the injection process is used to store and accumulate the beam after cooling. This has been demonstrated both with electron and stochastic cooling \cite{bbaccuec, bbaccusc}. Stochastic cooling is limited by the fact that a large number of particles in a short longitudinal distance results in a reduction of the stochastic cooling rate, whereas electron cooling is independent of the ion density. Both cooling methods reduce the momentum spread of the beam and force the particle into the stable longitudinal phase space region defined by the rf waveform.
	 An example of a reaction experiment at the ESR, where the accumulation of secondary particles was decisive is discussed in Section \ref{S:reactions}.
	 
	 A constraint to the accumulation of high intensity beams is given by the momentum spread of the cooled accumulated beam which has to be small enough to keep the particles with a given rf amplitude in the available longitudinal phase space area. For stochastic cooling, the cooling rate decreases with the number of stored ions and will not be able to cope with a constant injection rate during the increase of the accumulated intensity. For electron cooling limitations can come from other processes, high phase space density approaching the space charge limit or the saturation of the accumulation process due to the lifetime limited by the recombination rate of the ions with the electrons.
	 
	 \subsection{Beam Deceleration }
	 \label{s:deceleration}
	 
	 Storage rings are usually equipped with similar power converter and rf systems as synchrotrons which are preferentially used to increase the beam energy. The storage rings with low energy injectors accelerate the beam in a synchrotron mode to the energy required by the experiment. A special feature of the storage rings, which allow injection at high energy due to the availability of a high beam energy from the injector, is the operation in deceleration mode. This employs the usual ramped components of synchrotrons operated synchronously from high to low values corresponding to the required reduction of beam energy. The main motivation to decelerate the beam comes from experiments which prefer highly charged ions or secondary beams at lowest energies. The high energy of the primary beam is necessary to produce high charge states or secondary particles with large rates, the low energy after deceleration allows studies of certain processes or improves the resolution in the observation of the interaction of beam particles with a target.

\begin{figure*}[t!]
\centering\includegraphics[angle=-0,width=0.8\textwidth]{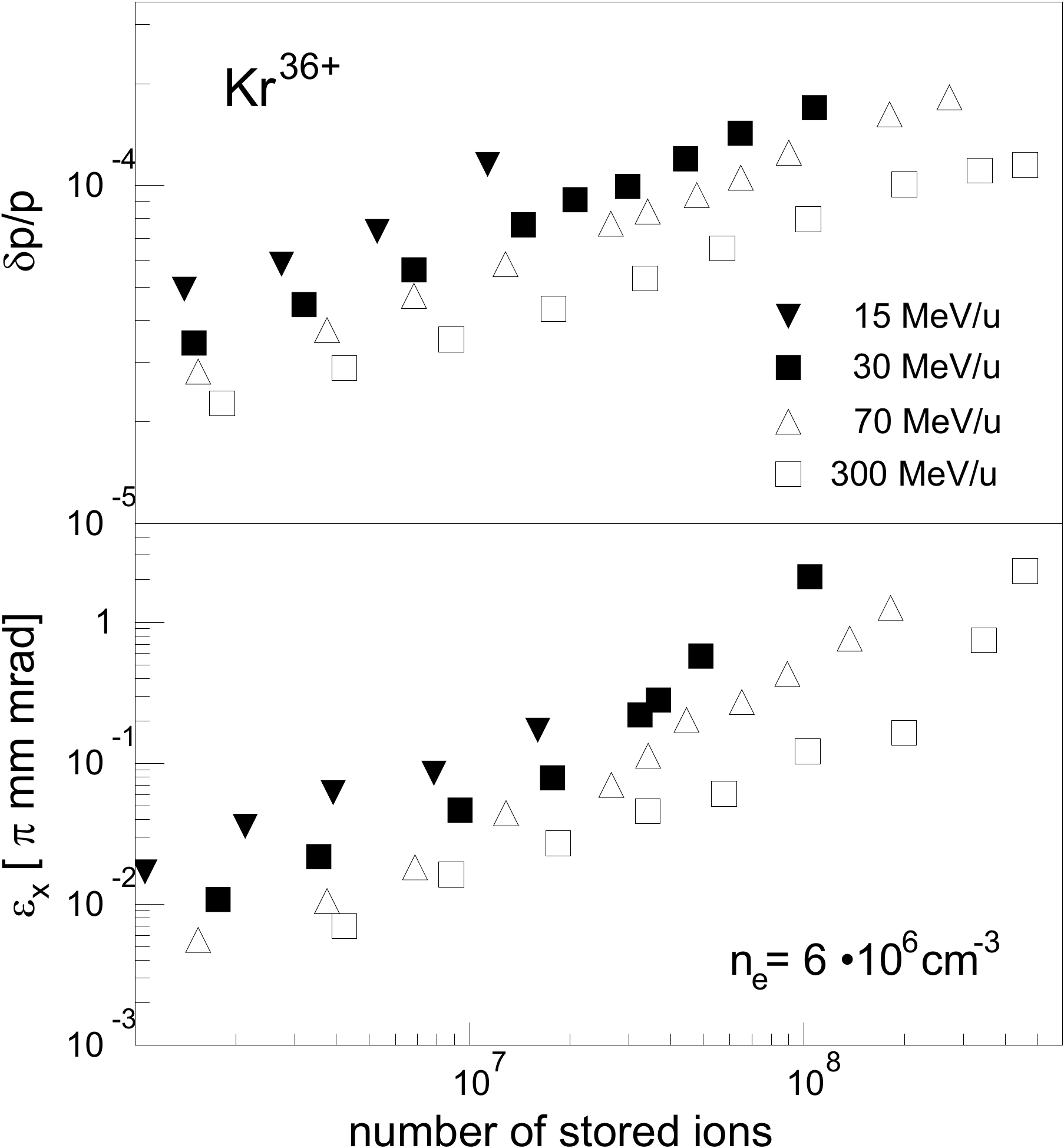}
\caption{ Beam parameters of cooled bare krypton ions stored in the ESR and cooled by electron cooling at different energies with a fixed cooling rate. The growth with increasing number of stored ions and for lower energy of the beam is caused by intrabeam scattering. Adopted from \cite{esrdecel}.}

\label{equi-abb}
\end{figure*}

	An important benefit for the operation in a deceleration mode comes from the availability of beam cooling. Beam cooling provides beams of small momentum spread and emittance. As the beam emittance unavoidably grows adiabatically during deceleration, unlike the adiabatic shrinking of beam emittance during acceleration, the deceleration without beam cooling results in losses in the course of the deceleration process. Such losses can be avoided for beams which are cooled before deceleration and cooling increases the efficiency of the deceleration process.  This is illustrated in Figure~\ref{equi-abb}, which shows that even with cooling the emittance at lower energies is increased. It shows the dependence of the beam quality on beam energy. In these measurements the electron density was kept constant by varying the electron current according to the change of electron density with beam velocity. For the consequently constant cooling rate, the equilibrium values improve at higher energies as expected because of the dependence of the IBS rate on energy. In addition, cooling is substantial as cooling compensates any additional emittance growth during deceleration, either the natural adiabatic growth or due to additional heating by machine imperfections.
	
	The first application of a storage ring for deceleration was the LEAR storage ring at CERN for the deceleration of antiprotons~\cite{leardecel}. Similar to its predecessor LEAR the Antiproton Decelerator (AD)~\cite{ad} is the first storage ring in a staged arrangement of decelerators for bringing antiprotons to rest, e.g. capturing them in a trap. In a similar fashion the ESR ring was designed and commissioned for the deceleration of highly charged ions or rare isotopes which are used for in-ring experiments or after extraction and further deceleration in the HITRAP linear decelerator for transfer into the trap \cite{Kluge-2008, Herfurth-2015}. Another branch of the deceleration concept at the ESR is the transfer of decelerated highly charged ions up to bare uranium from the ESR to the CRYRING@ESR where the stored beam, optionally after further deceleration in CRYRING@ESR, is used in physics experiment in the energy regime of a few MeV/u or even lower. In the ESR the feasibility of storage of heavy ion beams after deceleration has been demonstrated with storage over time periods which allow experiments using the stored beam in combination with the internal target, see Section \ref{s:pg}. 

	A typical deceleration cycle in the ESR consists of the injection of ions at an energy of 400 MeV/u, which is matched to the operation of the stochastic cooling system right after injection, followed by deceleration to an intermediate energy of 30 MeV/u~\cite{esrdecel}. At 30 MeV/u electron cooling is applied and the rf harmonic number is changed thus allowing further deceleration with the same rf system to a minimum energy of 3 MeV/u. The typical cycle time for the deceleration from 400 to 3 MeV/u is about 30 s, determined by the required cooling time and fastest ramp rate in deceleration mode which was achieved so far. The efficiency of the deceleration varies with beam species, beam intensity and the final energy after deceleration and is in best cases up to 50~\% including all kinds of losses. 
 The operation of highly charged ions at low energy is clearly limited by recombination losses in the residual gas and the decrease of the beam lifetime with the energy of the beam. For the typical ESR vacuum in the $10^{-11}$~mbar range lifetimes of a few seconds for bare heavy ions at the lowest energy were demonstrated which are increasing with the square of the energy and inversely to the charge of the stored ion. The beam lifetime at lowest energies is dominated by the residual gas, recombination losses with the electrons of the cooling system are of minor importance.
	
	For experiments which are performed with the decelerated beam stored in the ESR electron cooling is applied down to the lowest energies. The beam parameters of the stored ion beam are dominated by the equilibrium between intrabeam scattering and cooling. For a constant cooling rate Figure \ref{equi-abb} shows the beam parameters depending on the particle number for different energies. The internal target is preferentially operated at reduced target density which is matched to the shorter beam lifetime at low energies. Particularly light target particles are well suited for experiments with low energy beams. The low energy reaction products require detection systems which are installed in the vacuum system of the storage ring. The use of thin windows to separate the volume for the installation of the detectors for the low energy ions from the ring vacuum is ruled out as the reaction products would be stopped in the window, see Section \ref{s:pg}. The necessity to install detectors in the ring vacuum containers results in very stringent requirements on the vacuum compatibility or special solutions with detectors which will be moved into the vacuum system through ports.	  
\section{Production of Secondary Beams}

As mentioned in Section \ref{s:intro}, we refer to `secondary particles' as those which do not readily exist on Earth.
They have to be specially produced in nuclear or atomic reactions at dedicated accelerator facilities.
Such secondary particles can either be stable elements in specific high atomic charge state, short-lived radioactive nuclides, antiprotons, etc.
The production of beams of secondary ions is an essential prerequisite for the experiment.
This work is restricted to the discussion of experiments with highly charged (radioactive) ions which is relevant to the physics programs at the existing heavy-ion storage rings.
Special issues related to the production and separation of beams at the future storage rings will be addressed in the corresponding sections.

\subsection{Production of highly-charged ions}
Whereas intense beams of highly charged light ions can directly be produced in modern ion sources or electron-beam ion traps/sources (EBIT/S),
the efficient production of beams of highly charged heavy ions is presently done by stripping bound electrons off the swift projectiles when passing through a thin stripper target~\cite{Sigmund-2004}.
For this purpose, low charged ions from an ion source are accelerated to high energies by employing linear accelerators, cyclotrons and/or synchrotrons. 
The accelerated high-energy beam of primary particles is then focused onto a dedicated stripper target.
In first order, the so-called Bohr criterion \cite{Sigmund-2004} can be used to estimate the velocity needed to achieve a required charge state.
The criterion predicts that bound electrons are stripped off the projectile most efficiently during the penetration through matter if their ``classical'' orbital velocities $\beta=v_0/c=Z\cdot\alpha/n$, 
are in the order of the velocity of the projectile,
where $Z$ is the proton number, $\alpha$ the fine structure constant $\alpha=1/137$, and $n$ is the principal quantum number. 
For example, to produce the heaviest available stable beam, uranium, as a bare (fully-ionized) ion the velocity $\beta=v_0/c\approx0.67$ is needed, which corresponds to primary uranium beam energy of at least 325 MeV/u. 
If the stripper target is thick enough, the charge state distribution at the exit of the target becomes independent of the initial charge state of the projectile, 
that is the probabilities for stripping and pick-up of electrons are equal leading to the equilibrium charge state distribution.
The equilibrium charge state distribution depends on the proton number of the projectile and of the stripper material and on the exit energy from the target \cite{Shevelko-2020}.

For many experiments, the production of ions with one- (hydrogen-like, H-like), two- (helium-like, He-like), three- (lithium-like, Li-like) and so on bound electrons is required.
In such cases it may be efficient to select stripper foils to be much thinner than the equilibrium thickness.
In first order, the thickness of the foil has to be equal to $n\cdot\langle x\rangle$, where $n$ is the number of electrons to be removed from the projectile and 
$\langle x\rangle$ the mean free path between atomic collisions in the target.

\subsection{Production of exotic nuclei}
Dependent on the nuclei of interest, various nuclear reactions can be employed for their production.
Focusing on the experiments at the existing storage rings, two main reactions, projectile fragmentation and in-flight fission, are employed.

In the projectile fragmentation reaction, by interaction with target nuclei several nucleons are removed from the projectile.
In a simplified picture, the part of the projectile corresponding to the geometrical overlap with the target nucleus is suddenly cut off.
A highly excited fragment travels nearly with the same velocity and nearly in the same direction as the projectile.
The fragment de-excites by emission of neutrons, charged particles and $\gamma$-quanta.
Many different nuclei are produced in the reaction.
The production cross-section decreases with the number of removed nucleons.
Furthermore, neutron-deficient nuclei are produced with higher probabilities than neutron-deficient ones.
Therefore, the primary beam is typically selected to be as close to the nuclei of interest as possible.
The projectile fragments are kinematically focused in the forward direction.
The energy spread is different for each fragment and, due to Fermi momenta of the nucleons, depends on the number of removed nucleons.

In-flight fission reaction is applicable to uranium beams, in which the projectile disintegrates into two fragments. 
Stable $^{238}_{~92}$U$_{146}$ has more neutrons ($N=146$) than protons ($Z=92$).
The neutron-to-proton ratio of the fragments remains nearly the same as the one of the parent $^{238}$U. 
Therefore, fission reaction is typically used to produce neutron-rich medium mass nuclei \cite{Bernas-1994,Enqvist-1999}. 
Different from the projectile fragmentation reaction, the emittance of the beam after fission is dramatically larger.
In the centre of mass system the kinematics of fission fragments is determined by Coulomb repulsion.
In the case of uranium fission the total kinetic energy of fragments is about 170 MeV.

The thicknesses of the targets are typically in the order of one to a few g/cm$^2$. 
There is a compromise to be found in selecting the target thickness between the production yield of the rare isotopes and emittance growth.
Carbon and beryllium targets are often used.
The rare fragments emerge from the target at relativistic energies and thus as highly-charged ions. 

The production rates of rare isotopes can be well estimated with modern codes.
For projectile fragmentation reaction the EPAX3 parametrisation \cite{Summerer-2000} and for fission reaction the ABRABLA code \cite{Gaimard-1991, Schmidt-2002} are frequently used.

\subsection{Separation of secondary beams}
In principle, in all atomic and nuclear reactions discussed above the ions of interest are accompanied by unwanted contaminants.
In the case of highly-charged stable ions, these are the ions of the same element in other charge states.
To isolate the beam of interest is relatively easy and is achieved after a dipole magnet.
The requirements to the resolving power of the separation magnetic system are moderate since after the thin stripping foil the ions travel with nearly the same velocities and the energy as well as angular straggling are small \cite{Geissel-1995d, Morrissey-1989}.

In the case of the rare nuclides, the separation of nuclei of interest from contaminants requires dedicated in-flight fragments separator facilities.
The in-flight separation method is based on the fact that, due to the kinematics of the reaction, the products are mainly emitted in the forward direction. 
Depending on the energy range, the reaction products will have characteristic charge state distributions.
To boost a specific charge state of interest, foils of special materials and thicknesses may be used after the production target.
The cocktail beam of various fragments is analyzed by a magnetic system dispersing them by their momenta.
However, pure electromagnetic separation does not allow the separation of isobaric contaminants present in the same atomic charge state.
Therefore, for the spatial separation of mono-isotopic beams, dedicated energy degraders are used in addition to the electromagnetic analysis.
According to Eq.~(\ref{BBloch}), the atomic energy loss ($\Delta E$) of the ions penetrating through matter depends on their $Z^2$ \cite{Geissel-2002d}. 
In this way the magnetic rigidity ($B\rho$) analysis is performed before and after the degrader. 
This is the so-called $B\rho-\Delta E-B\rho$ separation method \cite{Geissel-1992d}. 
A clear advantage of the in-flight separation is the short flight-time between the 
production target and the exit of the separator. This time is in the microsecond 
region or less, and thus allows study of the most exotic nuclei with shortest lifetimes. 
The disadvantage is that the nuclear reactions lead to inevitable phase-space enlargement of the separated beams \cite{Geissel-1995d}.
Models to calculate momentum distributions of secondary beams can be found in \cite{Goldhaber-1974,Morrissey-1989}.
All presently operating storage rings are connected to in-flight separators.
The only project to construct a low-energy storage ring at ISOLDE/CERN is based on the complementary, ISOL separation method \cite{Grieser-2012}.
	 
\section{Status of Storage Rings}
\label{s:rings}
	
\subsection{Existing Storage Rings for Atomic and Nuclear Physics Experiments}	

\subsubsection{The medium- and low-energy storage ring complex at GSI in Darmstadt}
	
	After a very active decade at the end of the last century several of the storage rings of the first generation have been shut down. At present a few rings continue operation and pursue a broad physics program. The ESR storage ring at GSI will continue to serve for a variety of atomic and nuclear physics experiments providing highly charged ions and rare isotopes. The ESR allows the use of beams in a wide energy range. Already the injection energy from the synchrotron SIS is variable and allows a large flexibility in the choice of the charge state after stripping the ions in the beam line. The primary beam from SIS can also be used to produce and separate rare isotopes in the in-flight separator FRS \cite{Geissel-1992}. After injection into the ESR the beams can be accelerated or decelerated to any energy which corresponds to a magnetic rigidity in the range 0.5 to 10 Tm.

\begin{figure*}[h!]
\centering\includegraphics[angle=-0,width=0.9\textwidth]{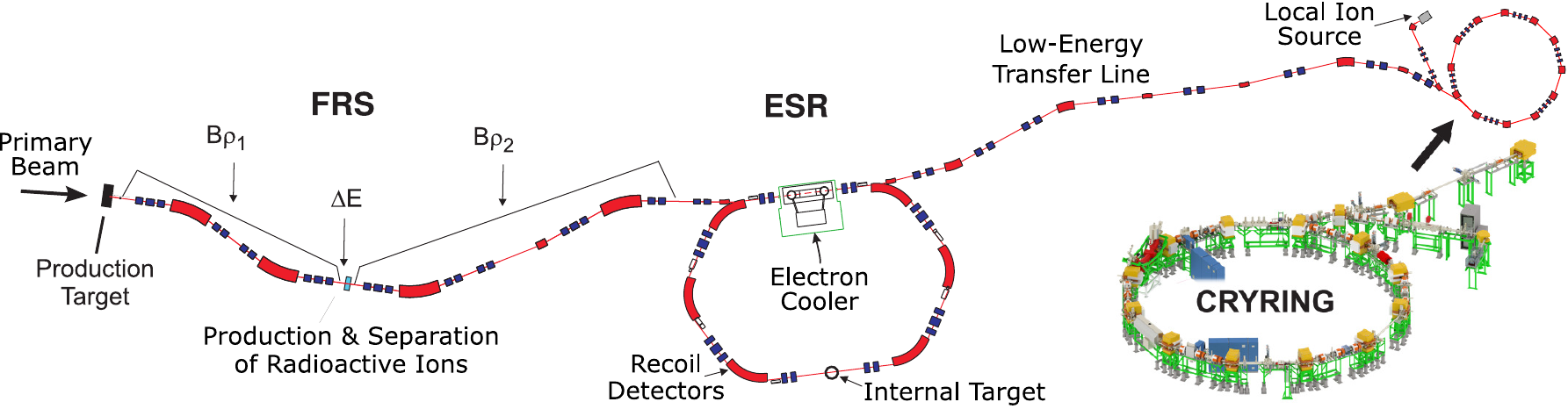}
\caption{(Colour online) The extended storage ring complex with the combination of ESR and CRYRING@ESR. Stable highly charged ions or radioactive beams from the fragment separator FRS are injected into the ESR. The experiments are conducted in the ESR at variable energies ranging from about 400 MeV/u down to below 10 MeV/u. At an energy of about 15 MeV/u the ions can be transferred to CRYRING@ESR, where they can be further decelerated and stored at variable energies down to 100 keV/u.}

\label{esr-cry}
\end{figure*}

The addition of the former CRYRING (Figure \ref{esr-cry}) as another stage for deceleration of heavy ions and rare isotopes will open an extended scope for experiments with low energy stored and cooled heavy ion beams in the new CRYRING@ESR facility which is part of the FAIR Phase-0 experimental program~\cite{cryringatesr}. The CRYRING was formerly operated with beams from a linac which were injected at rather low energy. Consequently an excellent vacuum was required and routinely achieved. Fast acceleration of the injected low energy beam was another feature of CRYRING. Both aspects make it the most adequate storage ring for the deceleration and storage of highly charged ions with final energies in the order 100~keV/u.

The two-stage deceleration chain with ESR and CRYRING@ESR is presently under commissioning and is expected to be available in the coming years for physics experiments. Both rings are equipped with electron cooling and allow for experiments at internal targets. Both rings also have components for slow extraction of the decelerated ion beams to support fixed target experiments with the small emittance of cooled stable or radioactive ions.

	A particular operation mode only available with highly charged ions at the ESR is the ultra-slow extraction of the cooled beam by use of the charge changing interaction of the stored particles with the internal target or the electrons of the cooling system. This enables a truly continuous beam for fixed target experiments with the excellent quality provided by cooling. The energy can be chosen in the range which is available for beam storage and the extracted current can be controlled by the density of the target or the electron beam which are used to produce down charged ions after electron capture.
	
	\subsubsection{The medium and low-energy storage ring complex at RIBFL in Lanzhou}
\begin{figure*}[t!]
\centering\includegraphics[angle=-0,width=0.8\textwidth]{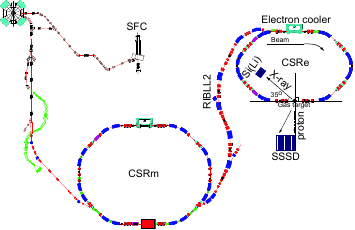}
\caption{(Colour online) 
Schematic layout of the RIBLL-CSR facility at the Institute of Modern Physics in Lanzhou.
Primary beams are accelerated to relativistic energies by two cyclotrons and a heavy-ion synchrotron CSRm.
The reaction target is placed at the entrance of the in-flight fragment separator RIBLL2 and transported to the cooler-storage ring CSRe.
Components of the setup for the reaction studies are indicated, see Section \ref{S:reactions}.
}
\label{F:Lanzhou}
\end{figure*}

	The CSR facility at IMP Lanzhou serves a similar class of experiments with cooled heavy ion beams. The layout of the facility is given in Figure \ref{F:Lanzhou}. It also allows the production of  rare isotopes by in flight separation and injection into the storage ring CSRe which is also equipped with an internal target and electron cooling. Similar to the ESR at GSI, the CSRe can be operated in a dedicated ion optical mode with the transition energy $\gamma_t$ equal to the relativistic beam energy $\gamma$. This special ion optical mode results in the isochronous revolution of particles in a momentum acceptance of typically $ \pm 0.25 $ \% which is suited for the measurement of short-lived rare isotopes either with a destructive time of flight detector or using highly sensitive resonant cavities for Schottky noise detection in time intervals shorter than the particle lifetime, see Section \ref{s:srms}. 
	 
	\subsubsection{The medium-energy storage ring at RIBF in Wako}

	 A dedicated ring for operation in isochronous modes is the RareRI ring at RIKEN which has been commissioned recently \cite{rareriring}. It is designed using components of the former TARN-II ring for the operation in the isochronous mode for mass measurements of rare isotopes produced with primary heavy ion beams accelerated in the superconducting cyclotron SRC and produced in the separator BigRIPS. Much care was spent on the tuning of the magnetic system for the isochronous revolution in a momentum range of $ \delta p/p = \pm 0.5 ~\%$  A special feature of this ring is a dedicated injection kicker system which is selectively triggered when a certain isotope intended to be stored in the ring  passes a detector in the transfer line.

	 \subsection{New Storage Rings}
	 
	 There are presently two funded accelerator projects which employ storage rings with the typical concepts and technology of the existing storage rings.
	 In the present scope of the Facility for Antiproton and Ion Research (FAIR)~\cite{fair}, the so-called Modularized Start Version \cite{GreenPaper}, two new storage rings will be constructed, the Collector Ring (CR)~\cite{cr} and the High Energy Storage Ring (HESR)~\cite{hesr}. Former plans to build even more storage rings are postponed to a later stage of the project due to funding issues \cite{GreenPaper}. The two rings will preferably be used with antiprotons and highly charged stable and/or radioactive ions. The FAIR facility will provide 3-4 orders higher intensities for rare isotopes produced from a primary heavy ion beam by bombardment of a solid target and in flight separation. Two large acceptance separators, one for rare isotopes and one for the antiproton beam are foreseen. With a primary high intensity short proton bunch at 29 GeV antiprotons are produced by bombardment of a thick copper target and by  subsequent separation of antiprotons of a kinetic energy of 3 GeV. The rare isotopes are produced in a thick carbon target. They will be separated in flight with the new superconducting fragment separator Super-FRS \cite{Geissel-2003s} and at an energy of 740 MeV/u will be transported to the CR. Both secondary beams, rare isotopes and antiprotons, arrive with large emittance and momentum spread in a very short bunch ($\le$ 100 ns) at the CR. 
	 
	 By design, the CR is a ring with large acceptance, both transversely and longitudinally, for the storage of secondary beams at a maximum magnetic rigidity of 13 Tm, aiming at the injection of antiprotons with an energy of 3 GeV and rare isotopes with an energy of 740 MeV/u. It is equipped with a dedicated rf system for fast bunch rotation to reduce the momentum spread of the incoming short bunch and stochastic cooling in order to be able to prepare the secondary beam in a short time interval (about 10 s for antiprotons and 1 s for rare isotopes) for transfer to the second new storage ring HESR. As an additional feature, the flexible ion optical layout allows operation of the CR in isochronous mode for mass measurements of short-lived rare isotopes benefiting from higher intensities and consequently availability of short-lived rare isotopes far off stability. 
	 
	 The cooled beam from the CR is transferred and injected into the HESR with the maximum rigidity of 13 Tm. Antiprotons can be accumulated in the HESR with a combination of a barrier bucket rf system and stochastic cooling. The goal is to achieve up to $10^{10}$ stored antiprotons in the HESR. Stored particles can be accelerated or decelerated in the HESR with a ramp rate of 0.025 T/s to any energy which corresponds to the rigidity range 5 to 50 Tm.  More recently the option to operate the HESR with ion beams was investigated \cite{Stohlker-2014, Stohlker-2015}. In a similar way as with antiprotons the HESR can also be used for experiments with stored highly charged stable and/or radioactive isotope beams. The ion beams will be injected at the CR energy of 740 MeV/u and can be accelerated or decelerated to energies in the range 200 MeV/u to 5 GeV/u. The HESR will be equipped with stochastic and electron cooling and internal targets and therefore experiments with cooled stored ion and antiproton beams will be possible.  
	 
	 The second major storage ring facility which is proposed and funded is the High Intensity heavy ion Accelerator Facility (HIAF) \cite{hiaf} proposed by IMP Lanzhou which will be constructed in the Huizhou area in the south of China. The main interest is in the fields of atomic, nuclear and plasma physics. The primary beam section consists of a new superconducting heavy ion linac which injects into the booster synchrotron BRing. The BRing uses two plane multiturn injection and has the option of electron cooling of the injected beam in order to reach highest intensities of heavy ions. After acceleration in BRing the beam can either be transferred directly or via a fragment separator of rare isotope beams to the SRing, the second ring in the HIAF concept. SRing is foreseen to be equipped with cooling and internal target. As an option the SRing can be extended in a later stage of the project by a third ring (MRing) which would allow experiments with merged co-propagating ion beams. At the time of writing this review, the full scope of the HIAF facility is still under discussion.
	 
	 In addition, the Nuclotron-based Ion Collider fAcility (NICA) which is presently under construction at JINR Dubna resembles various aspect of storage rings~\cite{nica}. It uses a new booster synchrotron which is equipped with electron cooling for accumulation at the injection energy with heavy ion beams coming from a new heavy ion linac. From the booster the heavy ions are transferred to the existing Nuclotron ring which allows the acceleration of heavy ions to the energy at which the ion beams in the NICA collider are stored and collided. In the first phase it is planned to collide counter-propagating gold beams with an energy up to 4.5 GeV/u per beam, in a later stage also the collision of other ions or polarized proton beams is envisaged.
	 
	 Another very recent proposal by JINR Dubna, the DERICA project, considers the construction of a facility for the acceleration of rare isotopes in a chain of linear accelerator and booster synchrotron and the storage of the rare isotopes in a storage ring~\cite{derika}. The main goal is the collision of the stored rare isotopes with electrons which are circulating in an electron storage ring which has a common interaction point with the ion ring. The rare isotope-electron collider concept was proposed in previous storage ring projects, but was never realized  because of limited funding \cite{Antonov-2011}. 
	  
	  At CERN/ISOLDE, there is a standing proposal to construct a dedicated low-energy storage ring \cite{Grieser-2012}. The major goal is to perform in-ring experiments with stored and cooled short-lived nuclides \cite{Butler-2016}. The peculiarity of this project is that ISOLDE is based on the Isotope-Separation On-Line method, where the radioactive nuclides are produced at rest in a spallation reaction of 1-1.5 GeV protons on a thick target, extracted and transported as singly-charged ions, charge-bred in a dedicated EBIS, accelerated to a few MeV/u energies by a dedicated HIE-ISOLDE linac and then injected into the storage ring as highly charged ions.

\section{Experiments at Heavy-Ion Storage Rings}
\label{s:masses}

The unparalleled experimental capabilities of heavy-ion storage rings enable a broad range of precision studies with stable and radioactive secondary beams \cite{Litvinov-2013}.
Whereas in the past the majority of conducted experiments could easily be placed into a specific research field like atomic physics, nuclear structure or tests of fundamental symmetries, investigations today are more often at the intersection of different fields.
	 
\subsection{Storage ring mass spectrometry}
\label{s:srms}
Atomic nuclei are quantum-mechanical objects, in which two types of fermions, protons and neutrons, are bound in a complicated way through the interplay of
strong-, weak-, and electromagnetic- fundamental interactions \cite{Bohr-1998}.
The result of this interplay is reflected in the nuclear binding energy ($BE$), which is one of the basic properties of the nucleus. 
It is connected to its mass, $M$, in a straightforward way through 
\be M(Z,N)=Z\cdot m_p+N\cdot m_n - BE(Z,N),\label{mass}\ee
where $Z$ and $N$ are the numbers of protons and neutrons in the nucleus, respectively, and $m_p$ and $m_n$ are respectively the proton and neutron masses. 
Masses of the neighbouring nuclei form the mass surface, which in general exhibits a smooth behaviour \cite{Blaum-2006}.

Since the very beginning of nuclear physics, the knowledge of binding energies was decisive for understanding the nuclear structure \cite{Bohr-1998}.
The nuclear shell closures, pairing correlations, or onsets of deformation are seen as irregularities on the smooth mass surface.
Various mass filters have been proposed for systematic studies of nuclear properties.
For instance, one and two neutron (proton) separations energies are routinely employed to determine 
the borders of nuclear existence (driplines), see, e.g. \cite{Novikov-2002}, 
for studying the evolution of the known neutron (proton) shell gaps and the search for new shell closures, see, e.g. \cite{Wienholtz-2013,XingX-2019}.
Other filters can specifically be built to investigate odd-even staggering, which is related to pairing correlations \cite{Madland-1988,Litvinov-2005a}, 
proton-neutron interaction strengths, see, e.g. \cite{Zhang-1989, Cakirli-2006, Cakirli-2009}, etc.
In addition to nuclear structure, masses are an essential input to modelling of stellar nucleosynthesis processes, see, e.g. \cite{Schatz-2013, Langanke-2013, Atanasov-2015b, Vilen-2018}.
It is therefore no surprise that precision mass measurements of exotic nuclei are pursued at basically all existing as well as future radioactive-ion beam facilities.

Various techniques have been developed and are successfully applied dependent on the facility type and nuclei of interest.
To address various physics applications one requires different levels of mass accuracy.
For instance, for investigations of nuclear structure like pairing effects or shell evolution a typical relative mass accuracy of $\delta M/M\approx10^{-7}$ is sufficient while for tests of fundamental symmetries or neutrino physics one often requires $\delta M/M<10^{-10}$ or even smaller \cite{Blaum-2006}.
In general, the Penning trap mass spectrometry (PTMS) \cite{Blaum-2006}, applied to the trapped ions nearly at rest, is presently superior in terms of achievable mass precision and accuracy. 
The conventional PTMS detection scheme is the time-of-flight ion-cyclotron resonance technique (TOF-ICR).
In TOF-ICR, a relatively large number of nuclei of interest is needed and it can therefore be applied to nuclides with relatively high production yields \cite{Atanasov-2015b}. 
Since very recently, the phase-imaging ion-cyclotron-resonance technique (PI-ICR) \cite{Eliseev-2013} is employed in PTMS.
With the introduction of this new technique the mass resolving power could significantly be increased \cite{Eliseev-2014, Nesterenko-2018,Manea-2020} and, furthermore, 
the number of needed ions of interest can in principle be reduced to just a few \cite{Eliseev-2015, Kaleja-2020}.
In addition to PTMS, the multi-reflection time-of-flight (MR-TOF) technique \cite{Wollnik-2013, Wolf-2013, Schury-2013} 
is being intensively developed worldwide and is more often used for mass measurements of short-lived nuclei.
First mass measurements with MR-TOF \cite{Wienholtz-2013, Atanasov-2015, Hornung-2020} reach relative mass accuracies in the order of $\delta M/M\approx10^{-7}$ and indicate the huge potential of this method.
For more details the reader is referred to \cite{100YMS} and references cited therein.
On the one hand, to be studied by PTMS or MR-TOF,  the half-lives of the nuclides of interest need to be at least a few milliseconds. 
Furthermore, only a few different nuclides of interest or even a single nuclear species can simultaneously be addressed.
On the other hand, in the storage ring mass spectrometry (SRMS) several tens of different nuclei in various charge states can be measured at the same time.
Since the SRMS is used on fast in-flight separated ions, the nuclides with half-lives down to a few ten microseconds can be studied.
Mass resolving powers of $\delta M/M\approx10^{-6}-10^{-7}$ are routinely achieved dependent mainly on the available reference masses. 
The biggest advantage of SRMS in comparison to all presently employed mass measurement techniques is the small number of short-lived ions 
needed to achieve a successful high-precision determination of their mass.
Often, just a single stored ion is sufficient \cite{Chen-2009a}.  

Storage ring mass spectrometry is applied at the FRS-ESR, RIBLL2-CSRe and Big-RIPS-R3 facilities \cite{Franzke-2008, Bosch-2013, Zhang-2016}.
The storage-ring mass spectrometry is based on the measurement of the revolution frequencies of the ions stored in the ring.
By inspecting Eq.~(\ref{eqx}), it is clear that the mass-to-charge ratios of the ions are directly related to their revolution frequencies.
However, there is a second term on the right hand side of this equation which contains the velocity spread of the ions.
The magnitude of this term determines the mass resolving power and has to be made as small as possible. 
There are two ways to minimize this term.

\subsubsection{Schottky mass spectrometry}
\label{s:sms}
The first method is based on the cooling the beam by electrons, see Section \ref{s:cooling}.
On the one hand, since the secondary beams are produced with relatively large emittance, 
typically larger than the injection momentum acceptances of the corresponding storage rings,
the electron cooling takes at least several seconds.
On the other hand, the large velocity spread enables the simultaneous injection and storage of multiple ionic species with different $M/Q$ values but matching velocities, such that their magnetic rigidities $B\rho=Mv\gamma/Q$ are equal to the selected injection $B\rho$ of the machine \cite{Radon-2000}.
Thus, the entire injection acceptance, frequency bandwidth, of the storage ring is filled with various exotic nuclei.
By analyzing the revolution frequencies of nuclei with well-known and unknown masses, the latter are determined through Eq.~(\ref{eqx}).
The ESR acceptance corresponds to $\Delta(M/Q)/M/Q=\pm1.5\%$ \cite{Litvinov-2005b}.
This acceptance is large enough to enable the same nuclide to be stored in several (up to three) atomic charge states. 
Hence, its mass can be determined by comparing to different reference nuclides.
To profit from this redundant information, a special maximum likelihood procedure has been developed to simultaneously analyze all measured data from one or several experiments \cite{Radon-2000}. 

\begin{figure*}[t!]
\centering\includegraphics[angle=-0,width=0.8\textwidth]{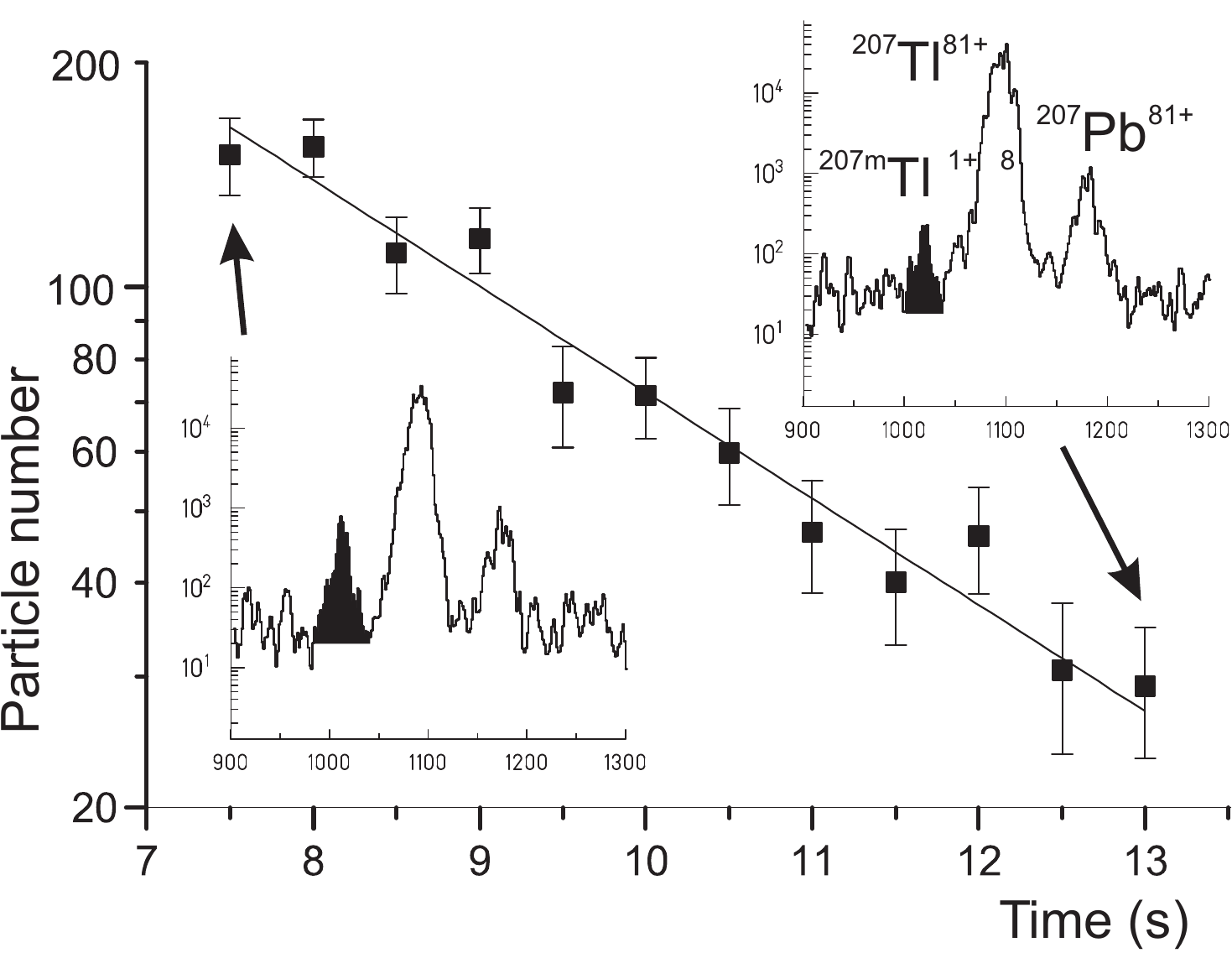}
\caption{(Colour online) 
The decay curve of the short lived isomeric state of $^{207m}$Tl$^{81+}$ as measured by time-resolved Schottky mass spectrometry, see Section \ref{s:t12}.
Schottky frequency spectra of the $A=207$ isobars, $^{207m}$Tl$^{81+}$, $^{207g}$Tl$^{81+}$ and $^{207}$Pb$^{81+}$, are shown in the inserts for the beginning of the measurement and at the end of the measurement, when the isomeric state nearly completely decayed.
This measurement is the result of the combined stochastic pre-cooling and electron cooling to address short-lived nuclides.
The figure is from \cite{Geissel-2004}.
}
\label{F:206Tl}
\end{figure*}

To access shorter-lived nuclei, the combined application of the fast stochastic cooling 
followed by the electron cooling can be used to reduce the cooling time for highly charged ions to a very few seconds.
This has been demonstrated in \cite{Geissel-2004} on the example of the isomeric state in $^{207}$Tl which has a half-life $T_{1/2}=1.33(11)$~s \cite{NNDC}, see Figure \ref{F:206Tl}.
The disadvantage of this operation is that the stochastic cooling cools the ions within a limited frequency bandwidth and heats those lying outside of it \cite{Nolden-2004}.

Conventionally, the mass spectrometry applied to electron cooled beams is called Schottky mass spectrometry (SMS), 
getting part of the name from the Schottky-noise diagnostics employed to measure the revolution frequencies of the stored ions, see Section \ref{S:BeDi}.
This, however, is no longer strictly true since the cavity-based Schottky pick-ups were shown to be fast enough to operate also in the 
isochronous mass spectrometry described below \cite{Walker-2013, Tu-2018}.

Exotic nuclei are stored as low-intensity beams (often as just single particles) dependent on their production cross-sections and transmission efficiency.
Electron cooling of such beams leads to one-dimensional crystalline beams with extremely small velocity spreads of the order $\Delta v/v = 10^{-7}$, see Section \ref{s:bpar}.
Typical mass resolving powers achieved in the SMS experiments were about 750000 and the typical relative mass accuracies were about $\Delta M/M=(1-5)\cdot10^{-7}$.

\begin{figure*}[t!]
\centering\includegraphics[angle=-0,width=0.8\textwidth]{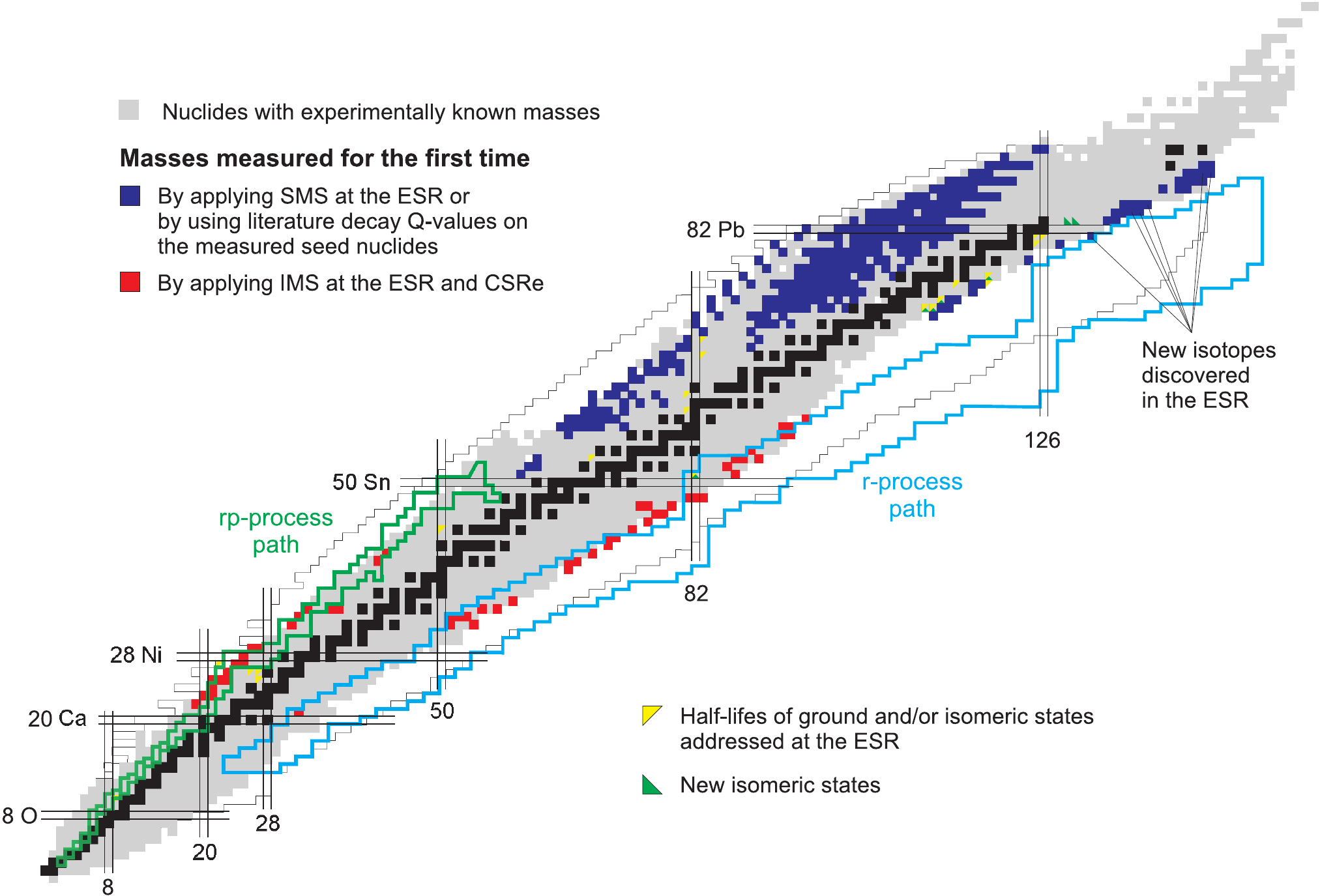}
\caption{(Colour online) 
Nuclidic chart illustrating the contribution of experiments at storage rings to the present knowledge of nuclear masses and isomeric states as well as to radioactive decays of highly charged ions.
The experimental data on masses measured for the first time at the ESR and CSRe are taken from 
\cite{Radon-2000, Novikov-2002, Litvinov-2005b, Tu-2011a, Chen-2012, Zhang-2012, Yan-2013, Shubina-2013, Xu-2016, Knoebel-2016, Knoebel-2016b, ZhangP-2017, Fu-2018, Xing-2018, Zhang-2018a}. 
The first measurements were performed at the R3, but the data analysis is still ongoing.
We note that numerous improved values were also obtained which are not shown here. 
However, especially on the neutron-rich side of the chart, the newly determined results show dramatic differences to previously measured masses \cite{XingX-2019}.
The experimentally known masses are taken from the latest Atomic Mass Evaluation 2016 \cite{AME-2016}.
This figure is updated from the original chart in \cite{Bosch-2013} to include all data available to date.
}
\label{F:nucl_chart}
\end{figure*}

The clear advantage of the SMS measurements is the ultimate sensitivity to single stored ions.
One of the highlight experimental results is the measurement of the mass of $^{208}$Hg nuclide, 
which was produced only once as a hydrogen-like $^{208}$Hg$^{80+}$ ion within a two-weeks long experiment, 
and whose mass was determined from this single event with merely 30~keV uncertainty \cite{Chen-2009a}.
Another advantage is the large amount of different ion species that can be studied in a single experiment.
About 300 masses were obtained for the first time with the SMS in merely 5 experiments \cite{Radon-2000, Novikov-2002, Litvinov-2005b, Chen-2012, Shubina-2013}, see Figure~\ref{F:nucl_chart}.

In the latest experiment on neutron-rich $^{238}$U projectile fragments, the region between Pb and U was addressed \cite{Chen-2012}.
The existence of several new isotopes was unambiguously shown through their precision mass measurement \cite{Chen-2010}.
In addition to the ground states the masses, and thus the excitation energies, of isomeric, metastable states of nuclei can be addressed.
Several isomeric states were discovered at the ESR \cite{Chen-2010, Chen-2013}.

The conventional SMS mass measurements are performed only at the FRS-ESR.
The disadvantage of the technique with cooled beams is that the cooling process takes at least one second.
On the one hand, since most of the nuclides of interest today are shorter-lived, the application of the SMS for mass measurements is limited.
The area of neutron-rich nuclides in the lead region is one of the few remaining regions on the chart of nuclides where it can still successfully be applied.
On the other hand, the search for long-lived isomeric states is probably the most promising future use of the SMS. 
Here, the sensitivity of the SMS and the non-destructive nature of Schottky detection bring in unrivalled capabilities.
For instance, long-lived exotic isomeric states are expected in the $A=180$ neutron-rich nuclides \cite{Walker-1999, Walker-2001}.
Production of the isomeric states of interest is one particle per hour/day or even week \cite{Walker-2020}.
Taking into account expected long lifetimes of the order of many minutes to days or even longer, 
the detection of such exotic species is hardly possible with any other technique.
The first experiments at the ESR addressing $^{197}$Au projectile fragments led to the discovery of several such isomeric states, 
including the exotic four-quasiparticle state in $^{186}$Hf \cite{Reed-2010, Reed-2012}.
In the future these measurements will be extended to $^{188}$Hf where an extremely long-lived isomer is predicted to exist. 

Details on SMS measurements conducted at the ESR to date can be found in reviews \cite{Bosch-2006, Franzke-2008, Bosch-2013, Zhang-2016}.
Nuclear masses obtained with SMS directly as well as with help of known energies of proton- or $\alpha$-decays are indicated in Figure~\ref{F:nucl_chart} with dark blue color.
From the performed experiments, only the data on neutron-deficient $^{152}$Sm projectile fragments remain unpublished \cite{Litvinov-2006c, Yan-2014}.

\subsubsection{Isochronous mass spectrometry}
\begin{figure*}[t!]
\centering\includegraphics[angle=-0,width=0.45\textwidth]{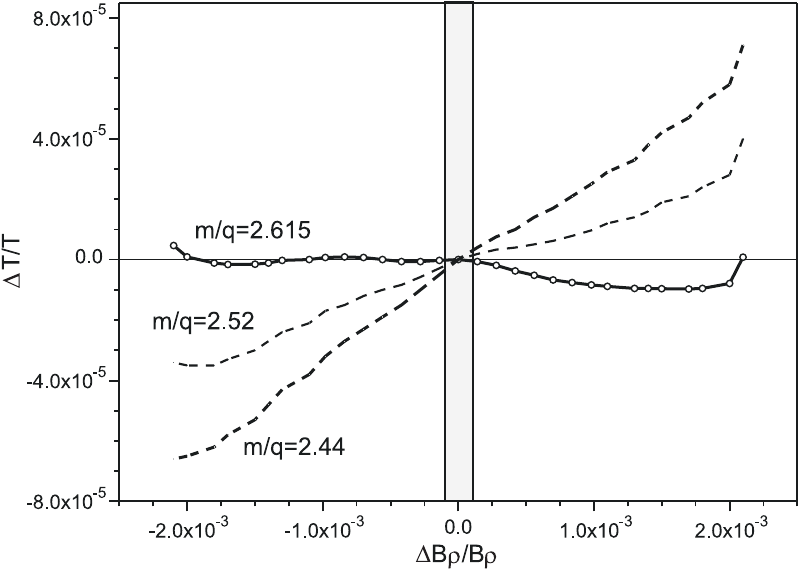}
\centering\includegraphics[angle=-0,width=0.45\textwidth]{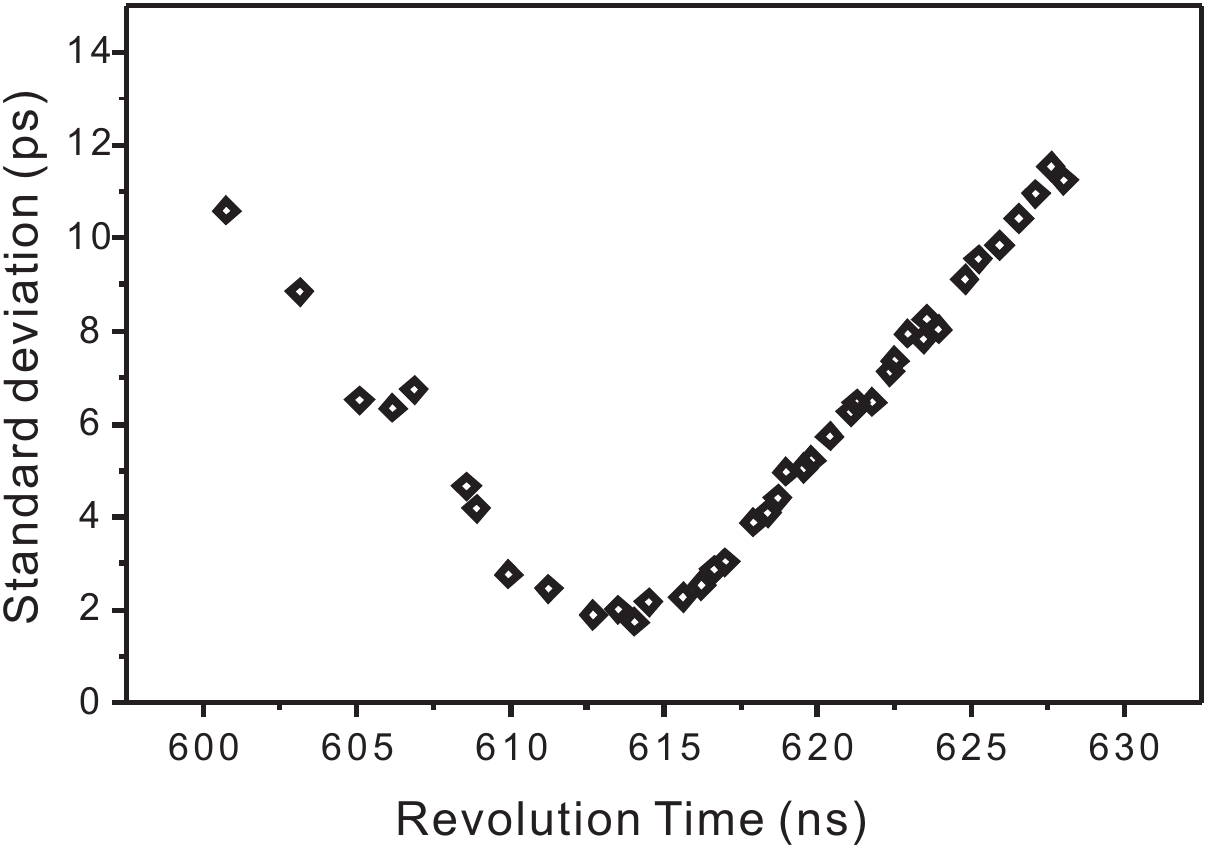}
\caption{(Colour online) (left) Isochronicity curve of the ESR. The ESR was tuned into isochronous mode for primary $^{238}$U$^{91+}$ ions which have $M/Q=2.615$. The velocity of the beam (magnetic rigidity) was changed by changing the voltage applied to accelerate electrons in the electron cooler. The frequency of the beam (revolution time) was measured with the SMS. The curves for other $M/Q$ ratios were transformed from the measured one by assuming that the particles with the same magnetic rigidity must have the same orbit length in the ESR. The grey band indicates a selected $B\rho$ window of $\Delta (B\rho)/(B\rho) = 1.5\cdot10^{-4}$, see text. Adopted from \cite{Bosch-2013b}. (right) Standard deviations of the revolution time distributions for different neutron-deficient $^{58}$Ni projectile fragments measured in the CSRe. The widths depend strongly on the mean revolution time, which is directly related to the magnetic rigidity. Taken from \cite{Xu-2013}.
}
\label{F:Isochronicity}
\end{figure*}

The Isochronous Mass Spectrometry (IMS) is based on the tuning of the storage ring into the isochronous ion-optical mode.
The ions of interest are injected with energies corresponding to $\gamma=\gamma_t$, where $\gamma_t$ is the transition energy.
The second term on the right hand side of Eq.~(\ref{eqx}) disappears (in first order).
A simplified understanding of the method is that the faster ion of a given ion species travels on an orbit longer than the slower ion of the same species.
As a result the velocity spread of the ions is compensated by the lengths of their closed orbits.
Hence, the revolution frequencies of the ions of same ion species are the same and depend only on the $M/Q$ ratio.
This mode is characterized by large ring dispersion functions.
Since no cooling is required, the method can be applied to very short-lived nuclei.
In general, several ten revolutions (typically one revolution is $\sim0.5~\mu$s) should be sufficient to determine the revolution frequency (revolution time).
Since the masses of nuclides decaying via strong interaction ($\alpha$-, proton-, etc. -decays, fission) are typically addressed by dedicated spectroscopic methods, 
the main goal is to measure $\beta$-decaying exotic nuclei, whose characteristic lifetimes are in the millisecond range.
This is the main strength of the IMS, that it can be applied to all nuclides in between the proton and neutron driplines.
As compared to the SMS, the ESR acceptance in the isochronous mode corresponds to $\Delta(M/Q)/M/Q=\pm7\%$.
The isochronous mode is characterized by the so-called isochronicity curve, 
which illustrates the behaviour of $\gamma_t$ versus the magnetic rigidity of the ring.
Ideally, the isochronicity curve should be flat with $\gamma_t=const$ across the ring aperture.
In reality, however, it has a complicated shape which reflects the inhomogeneities and higher order magnetic fields.
The isochronicity curve for primary $^{238}$U$^{91+}$ ions measured in the ESR is illustrated in the left panel of Figure \ref{F:Isochronicity}.

The IMS was proposed \cite{Wollnik-1987} and experimentally verified at the ESR \cite{Hausmann-2000, Hausmann-2001} and is now routinely applied also at the CSRe and the R3.
The masses measured so far with this technique are indicated in Figure~\ref{F:nucl_chart} with red color.

\begin{figure*}[t!]
\centering\includegraphics[angle=-0,width=0.8\textwidth]{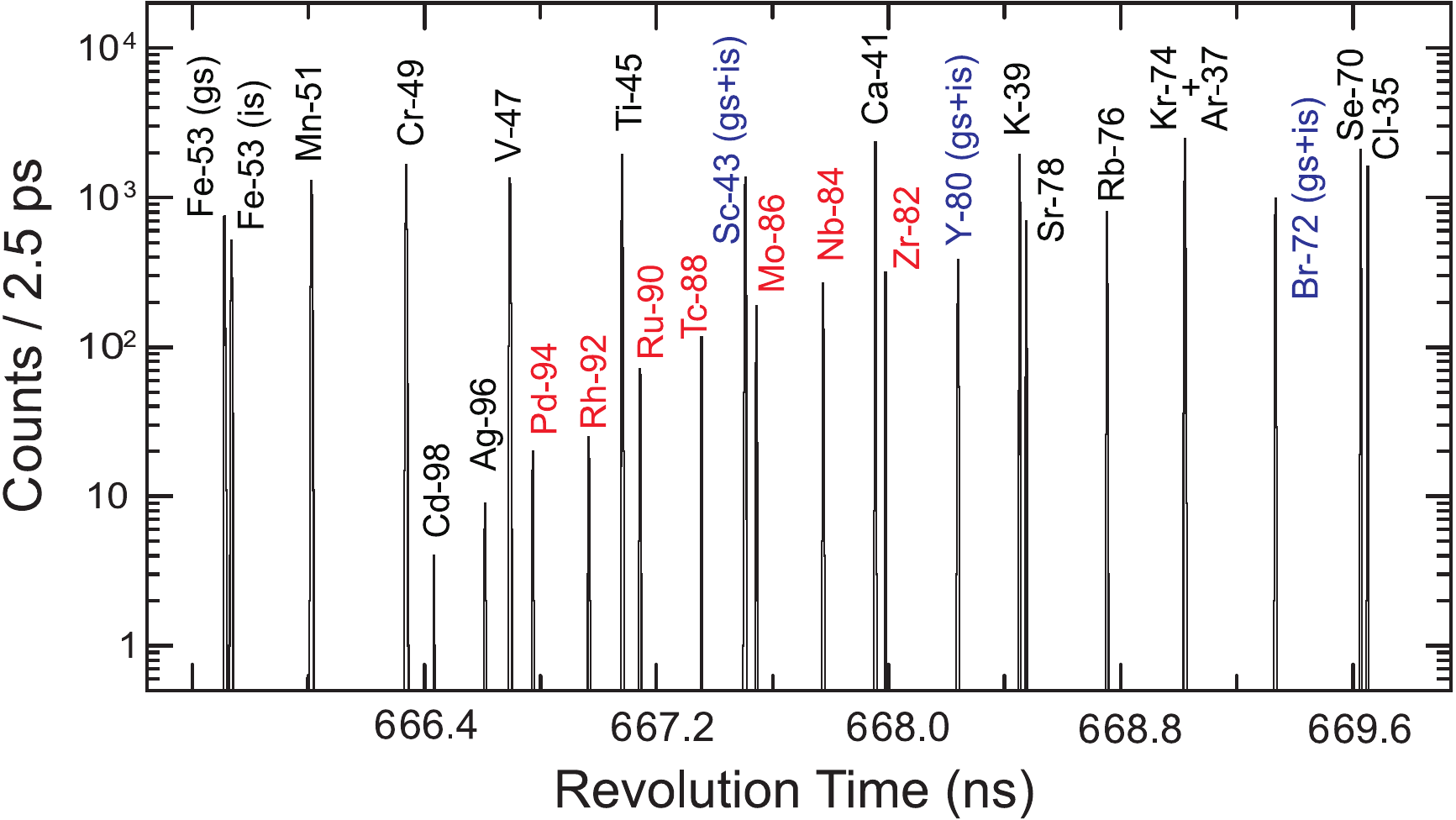}
\caption{(Colour online) Revolution time spectrum of $^{112}$Sn projectiles measured in the CSRe.
Figure illustrates a part of the spectrum published in \cite{Xing-2018}.
}
\label{F:spectrum_ims}
\end{figure*}

\begin{figure*}[t!]
\centering\includegraphics[angle=-0,width=0.5\textwidth]{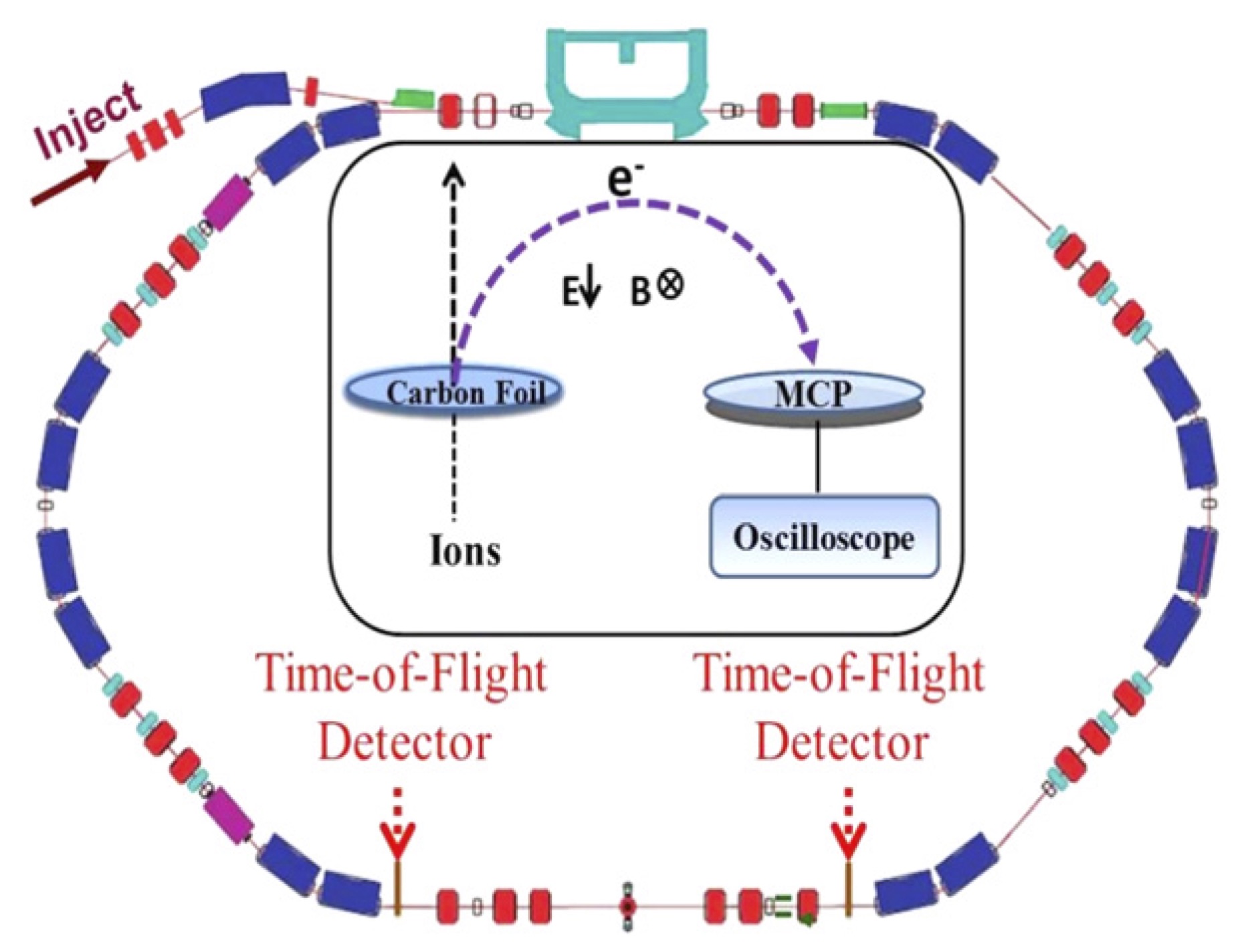}
\caption{(Colour online) Schematic illustration of the arrangement with two ToF detectors in the CSRe \cite{Xing-2015}.
}
\label{F:2tof}
\end{figure*}

The revolution frequencies are conventionally measured by dedicated Time-of-Flight (ToF) detectors \cite{Trotscher-1992, Mei-2010, ZhangW-2014}.
The ToF detector is based on the detection of secondary electrons emitted from a thin foil by passing through it swift ions.
Typical thicknesses of the employed foils are in the order of a few $\mu$g/cm$^{2}$.
Such very thin foils enable several hundreds of revolutions of the ions.
At each turn the ions loose some energy and finally leave the momentum acceptance of the ring.
This means that the ions slowly change their orbits and thus probe a considerable part of the isochronicity curve.
The secondary electrons released from the foil are guided by perpendicularly arranged electric and weak magnetic field to a set of microchannel plates (MCP), see the insert in Figure \ref{F:2tof}.
The electrons from the MCP are collected on the anode which is directly connected to the input of a fast oscilloscope without any intermediate signal manipulations.
The signal from the MCP is continuously sampled for a few hundreds microseconds triggered at each injection of fresh ions into the ring.
Such an arrangement allows the extraction of time stamps with the highest time resolution.
The amplitudes of the signals depend on the $Q$ of the ions and may be used to separate events from nuclear species with very close $M/Q$ values, which otherwise cannot be resolved by their times of flight \cite{Shuai-2014}.
In the offline analysis the obtained time stamps are assigned to individual ions \cite{Tu-2011b}. 
The revolution time of each ion is obtained as a slope of the fit of these time stamps versus revolution number.
Such a fit is typically a polynomial function which takes into account that the isochronicity curve is not strictly constant. 
The extracted revolution times are put together into a histogram forming the revolution time spectrum.
An example of a revolution time spectrum is illustrated in Figure~\ref{F:spectrum_ims}  for the case of mass measurement of $^{112}$Sn projectiles in the CSRe \cite{Xing-2018}.


The small injection acceptance of the ring allows for probing the momentum distribution of the fragment beam in front of the ring.
For instance, by comparing the isomeric ratios of $^{53}$Co and $^{53}$Fe in the ESR and CSRe produced in fragmentation reactions of different projectiles 
conclusions could be drawn on the angular momentum population  \cite{Tu-2017} as well as on the constancy of the isomeric ratio for different longitudinal momenta \cite{Tu-2015}.
The measured odd-even staggering of fragment yields allowed for constraining theoretical reaction models \cite{Mei-2014, Mei-2016}.

As can be seen in the left panel of Figure \ref{F:Isochronicity}, the isochronicity curve changes with changing $M/Q$ ratio of the ions.
This is due to the fact that the magnetic rigidity $B\rho$ at the injection into the ring is fixed and the injection acceptance is much smaller than the beam emittance.
In this case the ESR is tuned to be isochronous for the primary $^{238}$U$^{91+}$ ions with $M/Q_{i}=2.615$.
With increasing difference $M/Q-M/Q_i$ also the difference $|\gamma-\gamma_t|$ increases and, hence, the isochronous condition, see Eq.~(\ref{eqx}), is no longer fulfilled.
This is reflected by increased widths of the measured revolution time distributions as, for example, shown in the right panel of Figure \ref{F:Isochronicity} for $^{58}$Ni projectile fragments measured in the CSRe \cite{Xu-2013}.
The minimal width translates into the maximal mass resolving power.
The measured systematics of the widths is well understood and was reproduced in simulations \cite{ChenRJ-2015, Xing-2019}.
Typically, in the analyses of ESR and CSRe data only a part of spectrum was analyzed where the widths were below about 2 ps.
It is clear that a big portion of measured data cannot be used.
It has been realized that an additional measurement of the velocity or magnetic rigidity of each ion would allow for correcting the effect of the non-isochronicity \cite{Geissel-2005}.
At the ESR it has been realized by collimating the beam in the FRS \cite{Geissel-2006}.
The beam emittance was reduced by a set of slits to $\Delta (B\rho)/(B\rho) = 1.5\cdot10^{-4}$, see the gray band in the left panel of Figure \ref{F:Isochronicity}.
Indeed, the resolving power was significantly increased, though at a cost of dramatically reduced transmission.
Another proposed solution is to install two ToF detectors in the straight section of the ring.
Since the distance between the detectors can be measured with high precision \cite{Yan-2019}, the difference of time signals for the same ion provides directly its velocity.
Due to space restrictions, it is not feasible to install two ToF detectors in the ESR. 
Such a setup is foreseen in the CR where it has been considered from the very beginning of the ring design.
The development of this setup is pursued by the ILIMA Collaboration \cite{Walker-2013} within the NuSTAR pillar \cite{NuSTAR} of the future FAIR project \cite{Bosch-2003f, Durante-2019}. 
Two ToF detectors were recently installed at the CSRe \cite{Shuai-2016}, which boosted the mass resolving power and thus the precision of mass determination.
The first experiments addressing neutron-deficient $^{78}$Kr \cite{Xing-2015} and neutron-rich $^{86}$Kr projectile fragments were performed and data analysis is in progress.
In addition to nuclear mass measurements, the information from two ToF detectors can be used for accelerator-based investigations of the CSRe, see, for example, \cite{ChenRJ-2018, Ge-2018}.

Another way to obtain information on magnetic rigidity of each ion is based on transversely-sensitive Schottky-noise diagnostics \cite{ChenX-2014, ChenX-2015}.
The design of such a cavity has been completed \cite{ChenX-2016}.
However, no such device was installed in any operational ring to date.

\subsection{Nuclear lifetime spectroscopy}
\label{s:t12}
Investigations of radioactive decays of highly charged ions was among the motivations for the construction of the ESR \cite{Kienle-1993}.
These studies are based on the capability of a ring to store, cool and keep radioactive particles 
in a defined high atomic charge state for a long period of time.
Measurements of decay rates of highly charged ions were planned to be performed in Super-EBITs \cite{Elliott-1993}.
However, ions in an EBIT have a distribution of charge states, which prevents precision experiments 
where the control of the charge state is essential.
Since recently there are proposals for lifetime studies of highly charged ions in Penning traps \cite{Lennarz-2014,Leach-2014,Leach-2015}.
As of today, apart from in-flight measurements on very short-lived nuclides, see, for example, \cite{Phillips-1989, Phillips-1993,Attallah-1997}, 
extensive studies of radioactive decays, especially the weak decays, of highly charged ions is done in heavy-ion storage rings.
Interest in such investigations is manyfold.
Highly charged ions enable studies of the influence of bound electrons on the radioactive decays, 
which are complicated in neutral atoms due to complex interactions of many electrons in the atomic shell.
Especially interesting are fully-ionized or heavy hydrogen-like ions, 
which represent clean quantum-mechanical systems in the initial and final states.
On the one hand, it is obvious that decay channels involving bound electrons are disabled in fully-ionised atoms~\cite{Litvinov-2003, Irnich-1995}.
On the other hand, new decay modes--strongly suppressed or disabled in neutral atoms--can open up if bound electrons are removed.
Another motivation for such studies are the nucleosynthesis processes in stars, 
where the involved nuclides are usually highly-charged due to high temperatures and high densities of the corresponding environments. 
For instance, in the $s$-process along the valley of $\beta$ stability the mean ``temperature'' ($kT$) 
amounts to about 30~keV and in the explosive $r$-process it approaches 100~keV \cite{B2FH}.  

Decay studies of highly charged ions are routinely performed at the ESR. 
First measurements were conducted in the CSRe \cite{Tu-2018b}.
Storage rings were intensively used to investigate electromagnetic transitions and first studies of $\alpha$-decays were performed \cite{Nociforo-2012}.
Extensive reviews on the ESR measurements exist \cite{Litvinov-2011, Bosch-2013, Atanasov-2013, Atanasov-2015}.
Here, only a brief introduction of the main concepts is given. 
The discussion is restricted to the newest developments in the studies of electroweak decays.
As discussed in Section~\ref{s:masses} and reflected by Eq.~(\ref{eqx}), 
the revolution frequencies depend on the $M/Q$ ratio of the ions.
In the decay the mass-over-charge ratio changes, 
which inevitably leads to a change of the revolution frequency in the ring between parent and daughter ions.
In the case the frequencies of both, the parent and the daughter, ions lie within the storage acceptance of the ring, 
they can be addressed by the time-resolved non-destructive Schottky spectrometry~\cite{Litvinov-2004a}.
If the trajectories of the daughter ions lie outside of the ring acceptance, 
they can be blocked by particle detectors, typically after the first dipole magnets downstream long straight sections, see Section \ref{S:BeDi}.
The beam of parent ions continues to circulate undisturbed on the central orbit and its intensity is measured with the Schottky diagnosis or a current transformer,
while the daughter ions are intercepted and counted by the detectors.
In both cases a redundant measurement is achieved in which the decay curve of the parent ions 
and the growth curve of the daughter ions are simultaneously measured.

\subsubsection{Two-body weak decays of highly charged ions}
So far all experimental studies of weak decays of highly charged ions were conducted in the ESR.
Various $\beta$-decay channels can be summarized by 
using $p$, $n$, $e^-$, $e^+$ and $\nu_e$, which indicate the proton, neutron, electron, positron and electron neutrino, respectively
\begin{eqnarray}
p  +  e^-_b   \to    n  +  \nu_e    &   ~~~~\textrm{orbital~electron~capture~(EC), two-body decay;}\label{eqec}\\    
p \to n + e^+ + \nu_e & ~~\textrm{continuum~$\beta^+$~decay~($\beta^+_c$), three-body decay;}\label{eqbp}\\
n \to p + e^- + \bar{\nu}_e & ~~\textrm{continuum~$\beta^-$~decay~($\beta^-_c$), three-body decay;}\label{eqbm}\\
n  +  \nu_e   \to    p +  e^-_b  &      ~~~~\textrm{bound-state~$\beta$-decay~($\beta^-_b$), two-body decay,}\label{eqbb}
\label{betadecays}
\end{eqnarray}
where two- and three-body decays indicate the number of particles in the final state.

Beta-decays proceed along isobaric chains and the mass number $A$ remains unchanged.
From the experimental point of view, the decays above can be subdivided in two categories.
In the three-body decays (\ref{eqbp}) and (\ref{eqbm}) the charge changes by one, whereas in the two-beta decays 
 (\ref{eqec}) and (\ref{eqbb}), the charge state is not altered.
 This means that in the former case the orbits of the daughter ions in the ring 
 and correspondingly their revolution frequencies change significantly as compared to the ones of the parent ions.
 In the latter case, the orbits and the revolution frequencies of both, parent and daughter, ions remain nearly unchanged and the frequency change directly reflects the decay $Q$-value.

\begin{figure*}[t!]
\centering\includegraphics[angle=-0,width=0.8\textwidth]{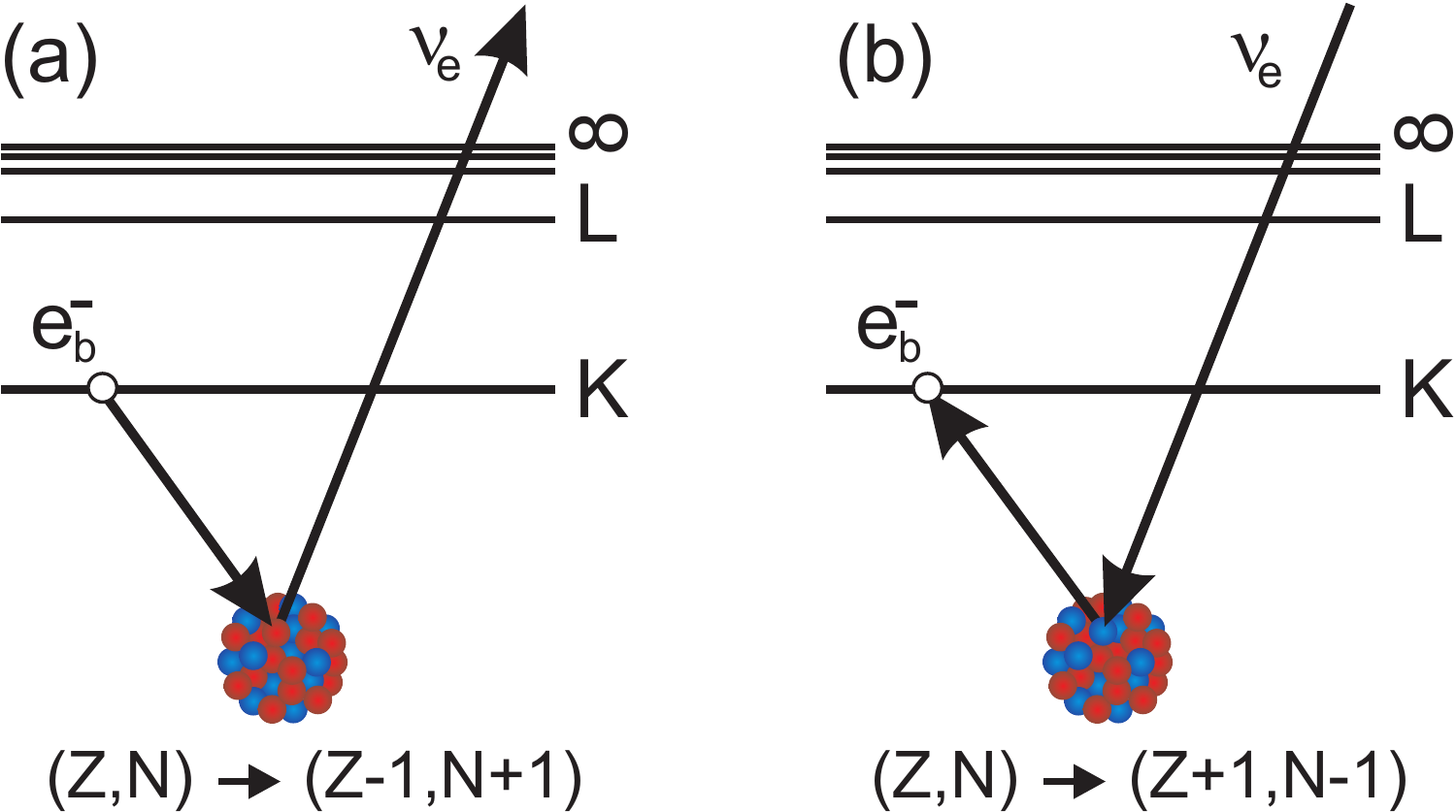}
\caption{(Colour online) 
Schematic view of two-body beta decays. (Left) Orbital electron capture, see Eq.~(\ref{eqec}), in which the bound electron is captured by the nucleus transforming one of its protons into a neutron. The monochromatic electron neutrino is emitted. (Right) the bound-state beta decay, see Eq.~(\ref{eqbb}), in which one neutron in a nucleus decays into proton and a bound electron. The (quasi) monochromatic electron antineutrino is emitted. An electron can be created on K, L, M, etc. atomic shells which affects the final antineutrino energy. The probabilities for different orbitals scale as $1/n^3$, where $n$ is the main quantum number \cite{Bambynek-1977}. Taken from \cite{Litvinov-2011}.
}
\label{F:2body}
\end{figure*}

In the last years, the main focus was given to studies of two-body beta decays.
The two-body beta decays are connected by time-reversal symmetry, see Figure~\ref{F:2body}. 

Being proposed in 1947~\cite{Daudel-1947} the bound-state beta decay, $\beta^-_b$, remained undiscovered until the corresponding experiments in the ESR were conducted.
In particular in nuclear astrophysics this decay mode turns out to be important \cite{Bahcall-1961, Takahashi-1983a, Takahashi-1987a, Takahashi-1987b}.
The first experimental verification of the existence of $\beta^-_b$ decay was shown on the case of stored fully-ionzed $^{163}$Dy$^{66+}$ ions \cite{Jung-1992}.
Neutral $^{163}$Dy are stable \cite{NNDC}. However, when stripped off all bound electrons, the bare $^{163}$Dy nuclei decay with the half-life of merely $T_{1/2}=47^{+5}_{-4}$ days.
This result allowed for determining the conditions during the astrophysical s-process nucleosynthesis \cite{Bosch-2006b} as well as provided constraints on the electron neutrino mass \cite{Jung-1992}.

Neutral $^{187}$Re atoms have long half-life of $T_{1/2}=43.3(7)$ Gy \cite{NNDC}, 
which was the reason to consider the pair of isobars $^{187}$Re/$^{187}$Os 
as a possible system for the model-independent determination of the mean age of the Universe \cite{Symbalisty-1981}.
However, the measured in the ESR half-life of fully-ionized $^{187}$Re$^{75+}$ nuclei $T_{1/2}=32.9(20)$ y \cite{Bosch-1996}
is by 9 orders of magnitude shorter. 
This result made such determination impossible without the knowledge of the charge state of $^{187}$Re throughout the galactic history. 
Still, by employing cosmochemical models a reasonable age of the Galaxy of $T_G = 14(2)$ Gy could be obtained \cite{Takahashi-1998}.

In both above cases only the $\beta^-_b$ decay channel was open, 
whereas in $^{207}$Tl both, $\beta^-_b$ and the three-body $\beta^-_c$ decays were measured in the ESR \cite{Ohtsubo-2005}.
This allowed for the first time to extract the ratio of the two- and three-body $\beta^-$ decay probabilities 
and to compare it with well-understood ratios for $\beta^+$ decay, EC/$\beta^+_c$.
For $^{207}$Tl such comparison showed a very good agreement with a theoretical result obtained by assuming allowed transitions.
However, for the case of $^{205}$Hg the theoretical value is about 25\% smaller than the one measured in the ESR \cite{Kurcewicz-2010}.

From the cases proposed for the ESR studies of the bound-state beta decay only $^{205}$Tl remains \cite{Singh-2018}.
$^{205}$Tl atoms are stable \cite{NNDC} and the bare $^{205}$Tl$^{81+}$ nuclei are expected to decay with $T_{1/2}\approx120$ d \cite{Takahashi-1987b}.
The experiment was proposed in 1980s \cite{Kienle-1993, Henning-1985, Pavicevic-1988} 
but could not be conducted up to now due to insufficient primary beam intensities. 
It is now scheduled at the ESR for Spring 2020.
The half-life of $^{205}$Tl$^{81+}$ is important to obtain the Solar neutrino capture cross-section, 
which in turn may be used for the determination of the integral Solar neutrino flux.
For this purpose, lorandite mineral from the Alchar ore deposit has been suggested as a geological detector \cite{Pavicevic-2010}.
The age of this mineral is 4.31(2) My \cite{Pavicevic-2018}.
Furthermore, the half-life of $^{205}$Tl$^{81+}$ is needed to constrain 
the very end of the s-process nucleosynthesis, namely to determine the destruction rate of $^{205}$Pb \cite{Yokoi-1985}.

Different from the $\beta^-$ decay, where the two-body decay probability increases with increasing the number of electron vacancies, 
the probability of the two-body weak decay on the neutron-deficient side of the nuclidic chart, the orbital electron capture (EC), 
reduces when the number of bound electrons is reduced \cite{Bambynek-1977, Folan-1995}.
In the limiting case of bare nuclei the EC decay is just disabled \cite{Irnich-1995}.
Interesting results were obtained in experiments on H- and He-like ions in the ESR.
It was found that the EC-rate in H-like $^{140}$Pr$^{58+}$ and $^{142}$Pm$^{60+}$ ions is by 50\% larger than in the corresponding He-like $^{140}$Pr$^{57+}$ and $^{142}$Pm$^{59+}$ ions \cite{Litvinov-2007,Winckler-2009},
though the number of bound electrons is reduced from two in He-like ions to just one in H-like ions.
This effect was later confirmed also for the case of the EC decay rate of H- and He-like $^{122}$I ions \cite{Atanasov-2012}.
This counterintuitive result was explained by taking into 
account the defined helicity of the emitted electron neutrinos and the conservation of the total (nucleus plus leptons) angular momentum \cite{Patyk-2008, Ivanov-2008, Siegen-2011}.
Furthermore, the explanation requires that the electron-cooled hydrogen-like ions are present in the ESR in the ground hyperfine state.
According to the theoretical model, the EC decay probability depends strongly on the spin-parities of the initial and final states.
To confirm the theoretical interpretation, several experiments were proposed.
Similar to the cases of $^{140}$Pr and $^{142}$Pm, $^{64}$Cu decays to $^{64}$Ni via an allowed $1^+\to0^+$ Gamow-Teller EC-decay \cite{NNDC}.
However, the magnetic moment of $^{64}$Cu is negative, which leads to the inverted order of hyperfine states.
It is expected that, if all ions are indeed stored only in the ground hyperfine state, there is no allowed EC rate in H-like $^{64}$Cu$^{28+}$ \cite{Litvinov-2008c}.
In $^{111}$Sn, the decay to $^{111}$In is dominated by a single transition $7/2^+\to9/2^+$ \cite{NNDC}.
If stored in the ground hyperfine state, the total angular momentum of the initial state is $F_i=3$.
Since, it is not possible to form $F_f=3$ in the final state, the transition is not allowed 
and the stored in the ring H-like $^{111}$Sn must have a significantly longer half-life
than the one known in neutral atoms \cite{Litvinov-2009a}.
If confirmed in future experiments, the variation of EC decay rates in H-like ions in comparison to the rate in neutral atoms 
can be used, e.g. to determine signs of nuclear magnetic moments, electron screening in $\beta$-decay, etc.
Interesting effects are also expected for the EC decay of Li-like ions \cite{Siegen-2011a}.

\begin{figure*}[t!]
\centering\includegraphics[angle=0,width=0.8\textwidth]{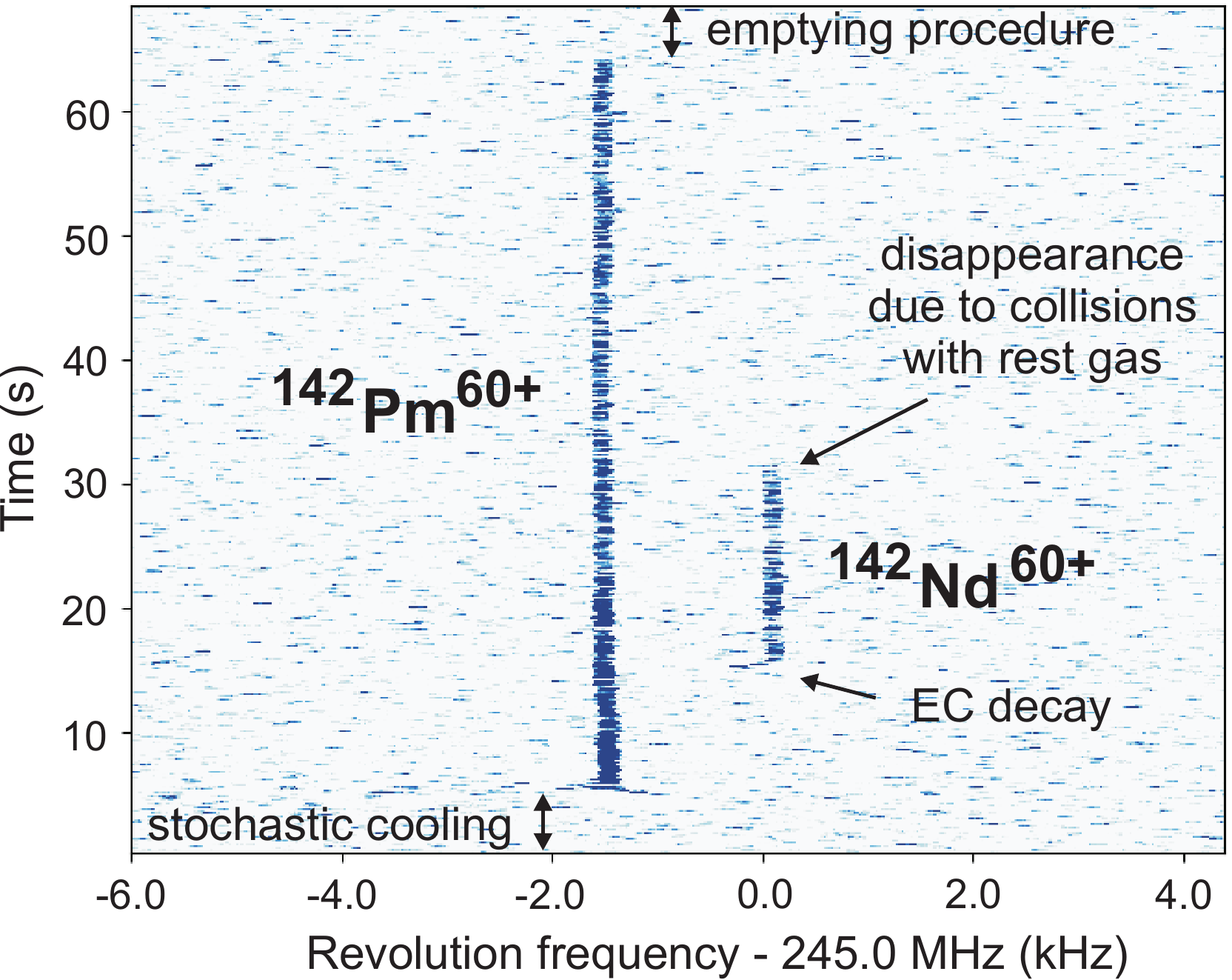}
\caption{(Colour online) 
Measured in the ESR traces of hydrogen-like $^{142}$Pm$^{60+}$ and fully-ionized $^{142}$Nd$^{60+}$ ions.
The time runs from bottom to top.
The ions are injected from the FRS at time $t=0$ s.
The first few seconds are needed for the combined stochastic and electron cooling.
The traces of the individual ions are clearly seen.
At time $t\sim15$ s, one of the $^{142}$Pm$^{60+}$ ions decays via orbital electron capture to the $^{142}$Nd$^{60+}$ ion.
The latter disappears at $t\sim30$ s. Since $^{142}$Nd is stable, the disappearance is due to atomic charge exchange reactions on the residual gas or recombination in the electron cooler of the ESR. Taken from \cite{Ozturk-2019}.
}
\label{F:singles}
\end{figure*}

\begin{figure*}[t!]
\centering\includegraphics[angle=90,width=0.45\textwidth]{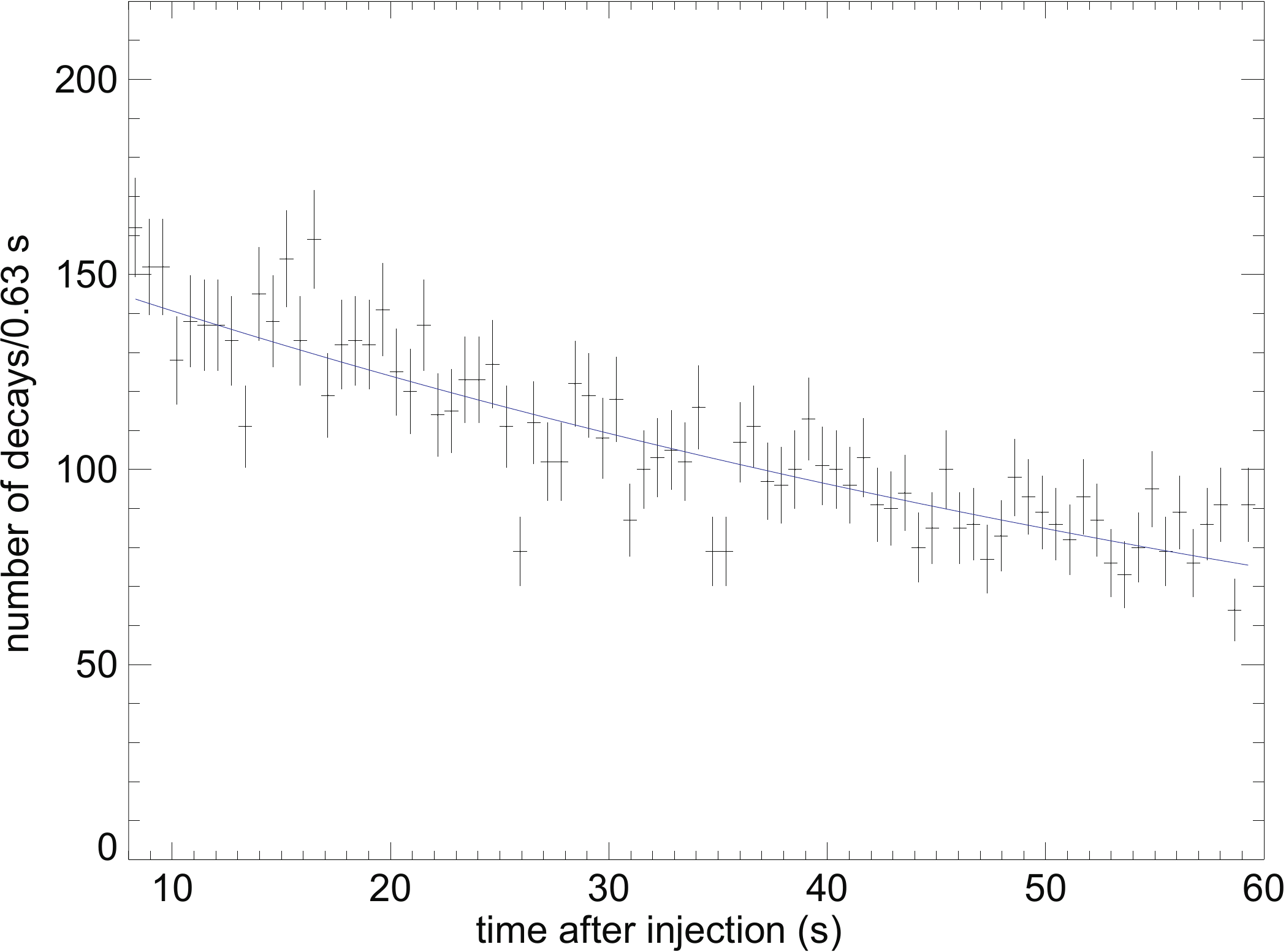}
\centering\includegraphics[angle=90,width=0.45\textwidth]{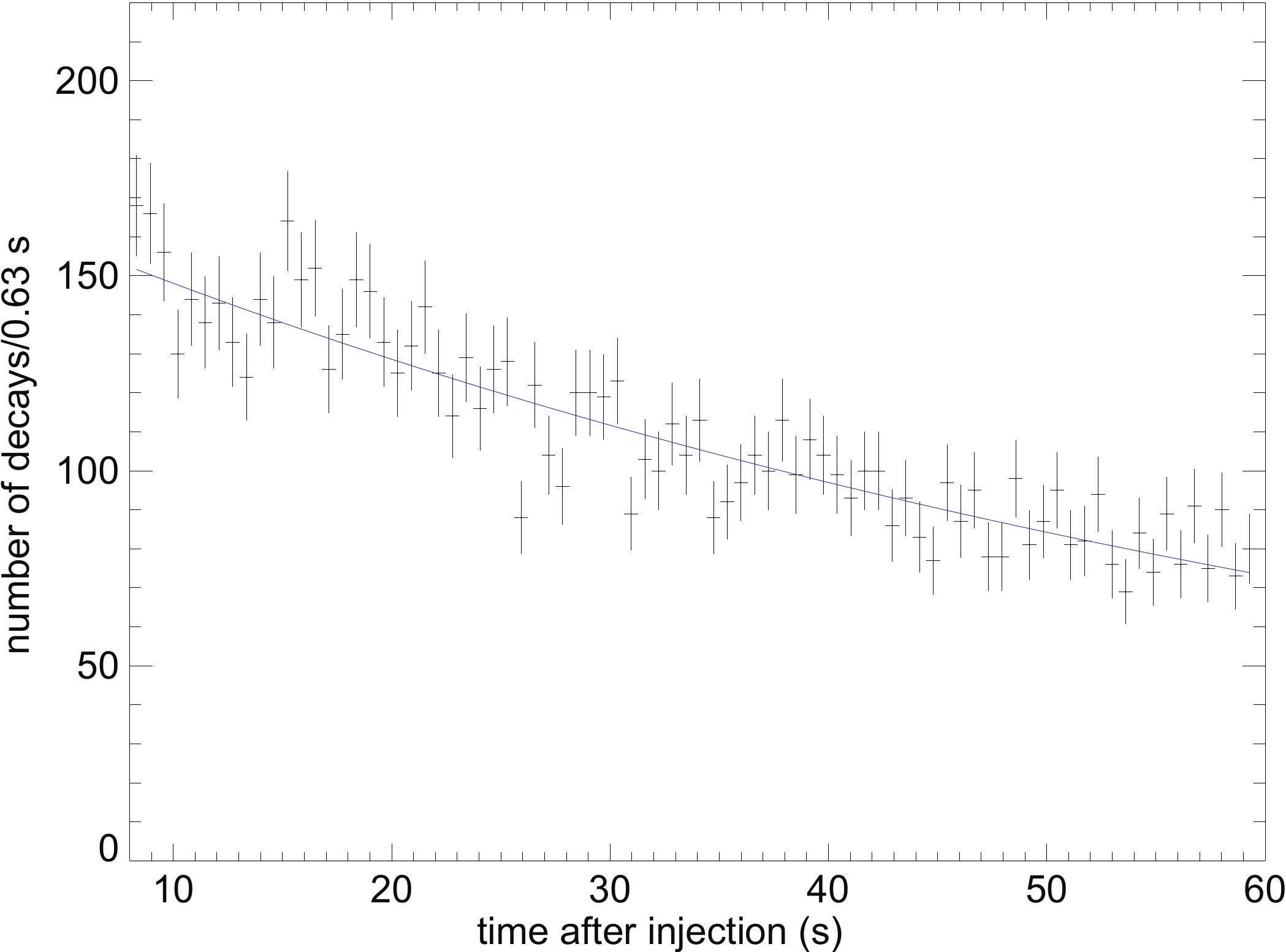}
\caption{(Colour online) 
(Left) Number of EC-decays per 0.63 s as determined in the automatic analysis (8839 EC decays in total). The data points are fitted with a pure exponential function (solid line). (Right) Same as (Left) but for a manual visual analysis (9001 EC decays in total). Taken from \cite{Ozturk-2019}.
}
\label{F:Osc}
\end{figure*}

The development of extremely sensitive non-destructive Schottky-noise diagnostics in the ESR, see Section \ref{S:BeDi}, 
enabled the development of the so-called Single-Particle-Decay Spectroscopy.
In this method, only a few particles are stored in the ring and their individual decays are unambiguously observed.
An example of the decays of single stored ions is shown in Figure \ref{F:singles}.
By employing this method to investigate the two-body EC decay of H-like $^{140}$Pr$^{58+}$ and $^{142}$Pm$^{60+}$ ions,
a surprising observation was made that a 7~s modulation is superimposed on the exponential decay curve \cite{Litvinov-2008a}.
This observation caused an intensive controversial discussion in literature, see, e.g. \cite{Vetter-2008,Ivanov-2008,Faestermann-2009, Merle-2009, Cohen-2009, Giunti-2010, Krainov-2012, Alavi-2015, Gal-2016} and references cited therein.
Several attempts were performed to investigate the observed effect.
Different mass numbers, charge states as well as different decay modes for ions were chosen in different experiments at the ESR.
However, except for the case of hydrogen-like $^{122}$I$^{52+}$ ions \cite{Faestermann-2015,Kienle-2009, Kienle-2009a, Kienle-2010}, 
no statistically significant modulated decays could be observed. 
Also in the case of $^{122}$I$^{52+}$ the signal-to-noise characteristics of the obtained spectra had questioned the overall quality of the measured data.
This resulted that different data analyses did not converge and the final experimental results remained unpublished. 
The repeated experiment on EC decay of hydrogen-like $^{142}$Pm$^{60+}$ ions did not confirm the original modulation parameters \cite{Kienle-2013}.
However, instability of the injection scheme into the ESR might have been present, which could affect the experimental results.
Therefore, the experiment on $^{142}$Pm$^{60+}$ ions was repeated again in 2014 under the conditions that the experimental parameters shall be set as close to the original experiment in \cite{Litvinov-2008a} as possible.
The details on this new experiment can be found in \cite{Ozturk-2019}.
The results of the automatic and visual manual analyses of more than 9000 individual EC decays are illustrated in Figure~\ref{F:Osc}.
If a modulation superimposed on the exponential decay curve is assumed, the best fit gives a modulation amplitude of merely 0.019(15), which is compatible with zero and by 4.9 standard deviations smaller than in the original observation which had an amplitude of 0.23(4) \cite{Litvinov-2008a}. 
With this result, it can be concluded that the existence of the modulation in hydrogen-like $^{142}$Pm is excluded \cite{Ozturk-2019}.
If there is a future research on this topic, it would be interesting to verify the existence of 7-s modulated EC decays in $^{140}$Pr$^{58+}$ ions.
 
\subsection{In-ring nuclear reaction studies} 
\label{S:reactions}

Employing a storage ring for in-ring reaction studies has several distinct advantages as compared to single-path fixed-target experiments \cite{Henning-1997, Muenzenberg-2000, Egelhof-2003}.
First of all it is tempting to recycle the rare ion species which inevitably have small production yields.
On the one hand, very thin internal targets have to be used which are orders of magnitude thinner than fixed solid targets.
Here, the high revolution frequencies of the ions in the order of hundreds of kHz allow for reaching relatively high luminosities, though still smaller than in the fixed-target cases.
On the other hand, thin targets allow low-energy recoils to leave the target undisturbed, thus enabling reaction studies at extremely low momentum transfers.
The cooled beams have small transverse size and high momentum definition, which, combined with small size of the target, 
allows for determining the interaction point with high precision and thus very high angular and energy resolution can be achieved.
Another essential feature of the in-ring reaction studies is very high detection efficiency and extremely low background conditions.
The gaseous or liquid target is of high purity. 
The target is windowless and the background contributions arising from reactions on window material, which are sometimes dominant, are just disabled.

\subsubsection{Nuclear reaction studies at medium/high energies}

Light-ion induced direct reactions, like for example elastic and inelastic scattering, transfer, charge-exchange, or knock-out reactions, 
have been shown in the past, for the case of stable nuclei, to be powerful tools for obtaining nuclear structure information \cite{Bertulani-2004, Bertulani-2007, Bertulani-2009}. 
Since no targets can be produced out of short-lived nuclei, the reactions on the latter are performed in inverse kinematics,
where the beam of exotic nuclei is brought into interaction with a low-$Z$ target.
In the case of proton targets, either polyethylene foils, complicated cryogenically cooled solid \cite{Ishimoto-2002} or a gas filled cell \cite{Vorobyov-1974, Dobrovolsky-2019} are commonly used.
Especially the domain of the low-momentum transfer, if measured with high resolution, brings a wealth of essential nuclear structure information.
It turns out that such measurements at low-momentum transfer, which are highly complicated to address in fixed target experiments, 
are very well suited for being conducted in the storage ring environment, where even very slow target-like recoil particles can be detected.
We emphasize, that such experiments are only possible due to versatile machine capabilities 
and especially the ability to efficiently accumulate and cool rare ion beams \cite{Egelhof-2003, Nolden-2013}, see Sections \ref{s:accu} and \ref{s:bcool}.

\begin{figure*}[t!]
\centering\includegraphics[angle=-0,width=0.8\textwidth]{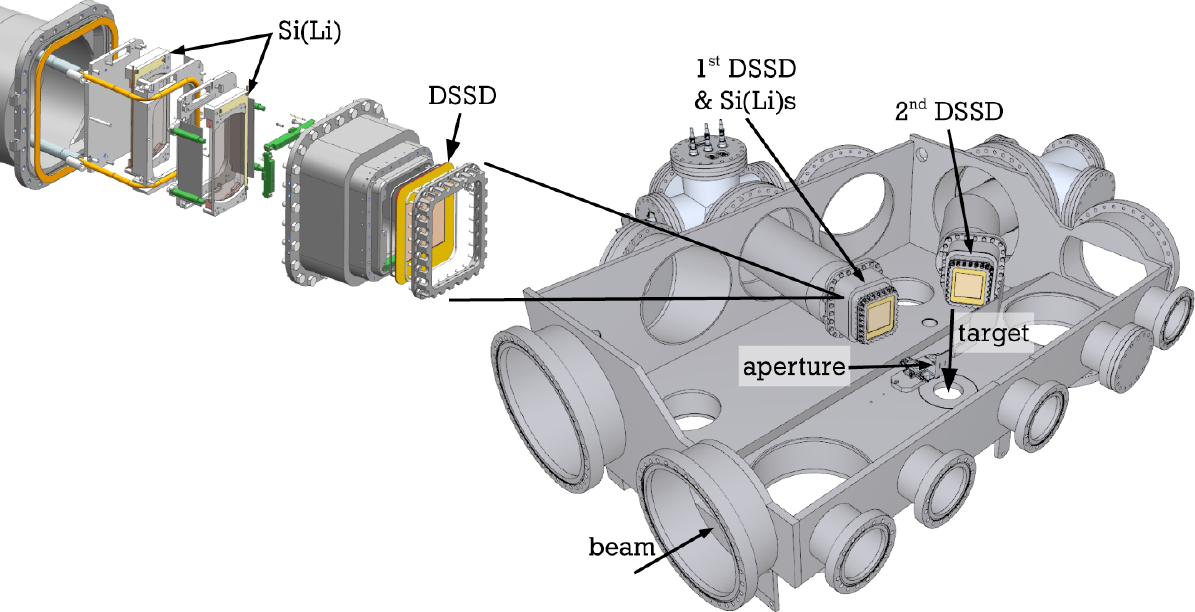}
\caption{(Colour online) Schematic view of the reaction setup implemented at the ESR for detection of low-momentum target-like recoils. 
The beam circulates in the ESR with about MHz frequency and crosses the internal gas-jet target.
To further define the interaction point, a special aperture is introduced.
DSSD stands for double-sided silicon strip detector, which is used as an active vacuum window \cite{Streicher-2011}
to separate ultra-high vacuum of the ring and the auxiliary vacuum in which the Si(Li) detectors are placed.
Two telescope detector arms, see text for more details, cover the laboratory angles of 
$73^\circ < \theta_{lab} < 88^\circ$ (DSSD1), and $27^\circ < \theta_{lab} < 38^\circ$ (DSSD2), respectively \cite{Mutterer-2015}.
The detailed assembly of the telescope is also shown.
The tubes for active water cooling of the detectors during the bake-out of the ring are indicated in orange colour.
The figure is adopted from~\cite{Schmid-2014}.}
\label{EXL_ESR1}
\end{figure*}
First studies of elastic proton scattering on stable as well as unstable stored beams were attempted at the ESR in the early 1990s \cite{Peter-1997}.
However, the low luminosities and technologies available at that time did not allow for a successful measurement.
In the recent years a new state-of-the-art setup for reaction studies has been implemented at the ESR \cite{Moeini-2011,Schmid-2014}.
It is illustrated in Figure~\ref{EXL_ESR1}.
The development of this versatile setup for reaction studies is pursued by the EXL collaboration \cite{EXL} within the NuSTAR pillar \cite{NuSTAR} of the future FAIR project \cite{Bosch-2003f, Durante-2019}.
The ultra-high vacuum compatible scattering chamber was installed at the gas-jet target position of the ESR.
The stored beam circulated with a high frequency in excess of 1~MHz and interacted with the gas target.
At that time, the radial extension of the gas-target was about 5~mm (FWHM).
Therefore, in order to increase the angular resolution a moveable aperture with piezoelectric 
drives was placed directly into the ultra-high vacuum \cite{Mutterer-2015}.

The two telescope arms shown in Figure~\ref{EXL_ESR1} cover the laboratory angles $73^\circ < \theta_{lab} < 88^\circ$ and $27^\circ < \theta_{lab} < 38^\circ$ \cite{Mutterer-2015}.
To be able to detect low-energy recoils it is essential to minimize dead layers and avoid windows.
Therefore, the double-sided silicon strip detectors (DSSD) are used as active windows to separate the ultra-high vacuum of the ESR and the auxiliary vacuum \cite{Streicher-2011}.
All sensitive components which cannot be heated during the bake-out procedure were placed into the auxiliary vacuum.
The whole setup was actively cooled during the bake-out process and operation.
With this technical solution the vacuum pressure of $2\cdot10^{-10}$ mbar on the ring side and $7\cdot10^{-8}$ mbar on the auxiliary side was reached.

For the envisioned reaction studies, the detectors have to be able to cover a large energy range of target-like recoils of several tens of MeV.
If taking proton elastic scattering  $^{56}$Ni$(p,p)^{56}$Ni at 400 MeV/u as an example, the energies of recoiling protons increase from zero to 65~MeV at
laboratory angles of 90$^\circ$ and 75$^\circ$, respectively \cite{Schmid-2015,Egelhof-2015}.
The DSSD detectors are used to measure recoils with low energies of up to a few MeV.
Each detector has an active area of $(6\times6)$ cm$^2$ and has 128 strips (0.5 mm width) on the front side and 64 strips oriented perpendicularly (1 mm width) on the rear side.
To measure the recoils with higher energies, each telescope also contains two segmented 6.5 mm thick Si(Li) detectors. 
More details on the technical realization of this detector setup can be found in \cite{Mutterer-2015}.

In addition, to the setup at the gas-jet, there was a set of six PIN-diodes $(10\times10$ mm$^2$ each),
installed into the ultra-high vacuum of the ESR about 8 m downstream the target \cite{Yue-2011}. 
They were used to detect the beam-like reaction products in coincidence with the events detected by the telescopes.

\begin{figure*}[t!]
\centering\includegraphics[angle=-0,width=0.8\textwidth]{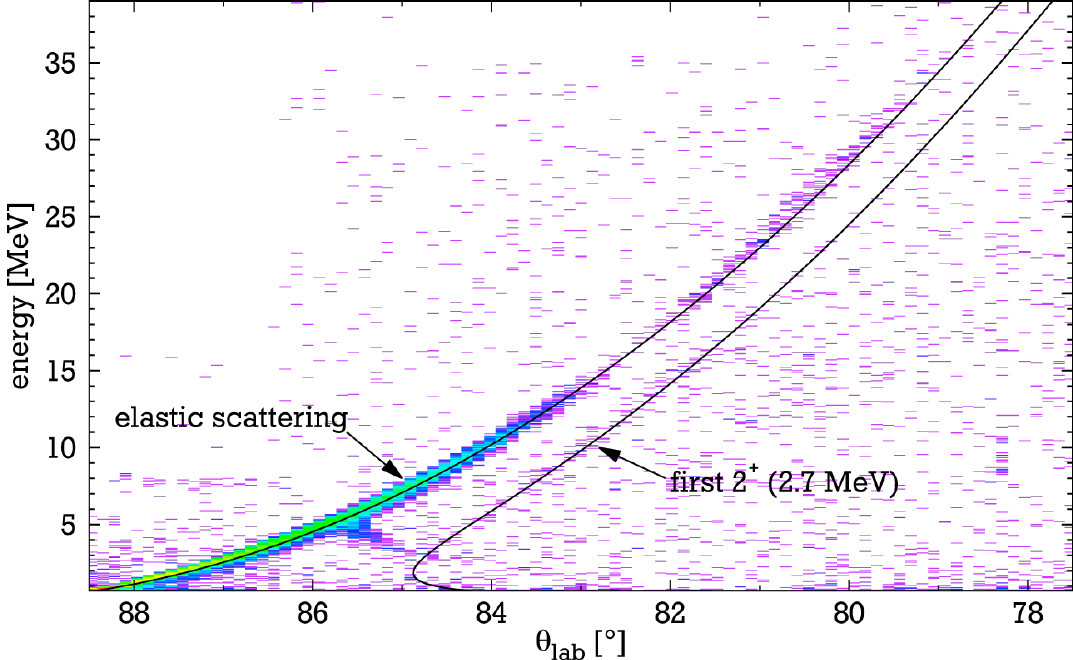}
\caption{(Colour online) Distribution (preliminary) of proton recoils from reaction $^{56}$Ni$+$p measured by the telescope detector (DSSD1) in Figure~\ref{EXL_ESR1}.
The recoils corresponding to the elastic scattering (leaving $^{56}$Ni in the ground state) and the ones corresponding to the excitation of the first $2^+$ state in $^{56}$Ni at 2.7 MeV are indicated.
The obtained angular resolution was only possible by using the aperture, see Figure~\ref{EXL_ESR1}.
The low energies up to about 8 MeV are covered by the DSSD, the energies up to about 35 MeV are covered by the DSSD and the first Si(Li) detector.
To measure higher energies all three detectors were required.
Adopted from \cite{Schmid-2015}.
}
\label{EXL_ESR2}
\end{figure*}
So far only one experimental campaign has been conducted with this reaction setup.
Proton scattering on radioactive $^{56}$Ni beams was investigated \cite{Schmid-2014,Schmid-2015,Egelhof-2015}.
A $^{56}$Ni beam with an intensity of about $7\cdot 10^4$ particles per spill was produced by fragmentation of $^{58}$Ni projectiles in the FRS.
A stacking scheme was applied in the ESR which included the stochastic pre-cooling at 400 MeV/u, bunching, and stacking of several ten $^{56}$Ni pulses \cite{Nolden-2013}.
Finally the intensity of about $5 \cdot 10^6$ $^{56}$Ni ions was accumulated.
Assuming the average density of the H$_2$ target of about $2 \cdot 10^{13}$~atoms/cm$^{2}$, luminosity of about $2 \cdot 10^{26}$~cm$^{-2}\cdot$s$^{-1}$ was reached.
A preliminary measured distribution of recoiling protons on the DSSD1 is illustrated in Figure~\ref{EXL_ESR2} \cite{Schmid-2015}.
The detection of protons with energies up to about 8 MeV is done with DSSD1, see Figure~\ref{EXL_ESR1}.
The measurement of higher energies of up to about 35 MeV required the first Si(Li) detector and 
for even higher energies the summed energy signal from all three detectors has been used. 
The elastic scattering data were obtained for a relatively large angular range, covering the region from small angles up to the second scattering minimum.
The final results, including also the discussion on the nuclear matter distribution, can be found in \cite{Schmid-2019, Egelhof-2020}.
It should be pointed out that this experiment was, even on a world-wide scale, the first of this kind performed with a radioactive beam.

Furthermore, $\alpha$-scattering on much more intense, stable $^{58}$Ni beam stored and cooled at 100 MeV/u has been successfully studied \cite{Egelhof-2015,Zamora-2015,Zamora-2016,Zamora-2017}.
The intention was to identify very low energy (200-300 keV) $^4$He recoils from the excitation of the 
isoscalar giant monopole resonance in $^{58}$Ni with the DSSD2 placed at forward angles.
The results showed that the monopole contribution exhausts 79+12\% of the energy-weighted sum rule (EWSR), 
which agrees with measurements performed in normal kinematics \cite{Zamora-2016}.
Also, the nuclear matter radius of $^{58}$Ni was determined \cite{Zamora-2017}.

\begin{figure*}[t!]
\centering\includegraphics[angle=-0,width=0.55\textwidth]{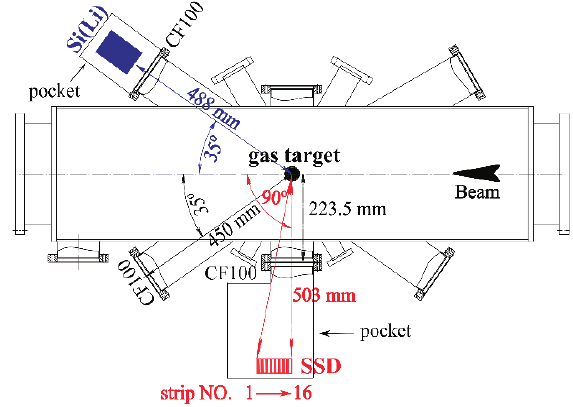}
\centering\includegraphics[angle=-0,width=0.35\textwidth]{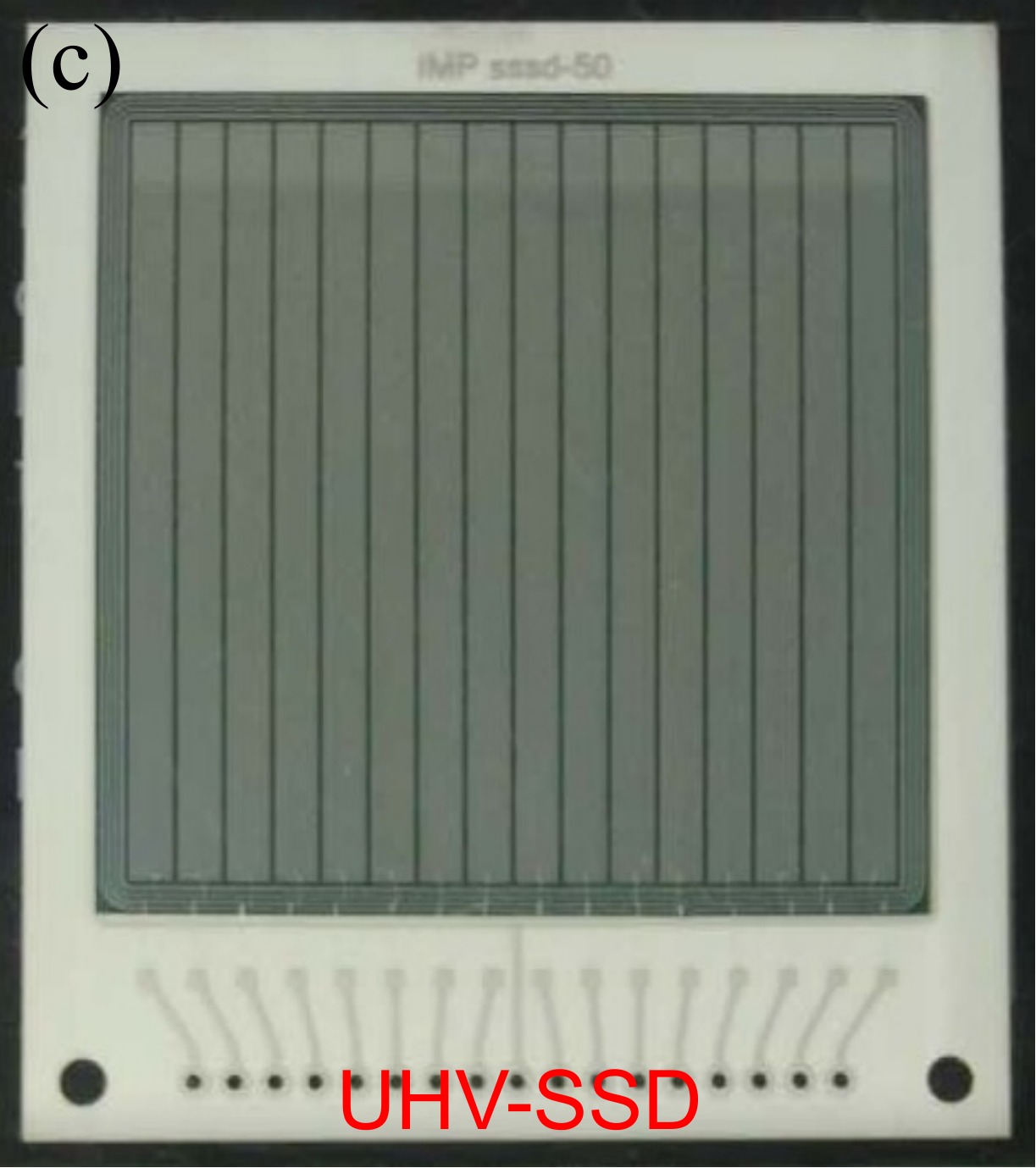}
\caption{(Colour online) 
The reaction setup at the CSRe in Lanzhou. 
(Left) The detailed sketch of the reaction setup at the gas-jet target. An additional SSD detector will be mounted in the free CF100 flange at 35$^\circ$ degrees. 
(Right) The photograph of the employed SSD detector which is placed directly into the ultra-high vacuum of the CSRe.
Adopted from \cite{Zhang-2019}.
}
\label{EXL_CSRe1}
\end{figure*}
Very recently, a dedicated setup for reaction studies has been developed at the CSRe \cite{Zhang-2019}.
There, a single-sided silicon strip detector, SSD, with a thickness of 300 $\mu$m was mounted at the internal gas jet chamber, see Figure~\ref{EXL_CSRe1}.
The present detector has merely 16 strips with about 3 mm width, the active size of $(48\times48)$ mm$^2$, 
and covers the angles in the laboratory frame from 85$^\circ$ to 90$^\circ$.
The SSD was placed at a distance of about 50 cm from the interaction point which somewhat compensated for the small number of strips.
The first reaction investigated with this setup was proton elastic scattering on stable medium-charged $^{58}$Ni$^{19+}$ ions stored and cooled at 95 MeV/u \cite{Yue-2019}.
At the flange at 35$^\circ$ in the laboratory frame, see Figure \ref{EXL_CSRe1}, a Si(Li) detector was mounted.
This detector was employed for measuring X-rays from the (atomic) ionization reaction in the target.
The measured X-rays corresponding to electron capture onto K-shell (K-REC) were taken for luminosity determination.
As a result, the measured differential cross-sections on the SSD are normalized by the measured K-REC X-rays.
Such atomic cross-sections can reliably be calculated with dedicated state-of-the-art codes.
In this particular case of medium-charged ions, the Relativistic Ionization CODE (RICODE) has been employed.
It is based on the relativistic Born approximation \cite{Shevelko-2011} and is a further development of the LOSS and LOSS-R codes \cite{Shevelko-2001}.
The RICODE has widely been applied to predict the single-electron loss cross sections for
collisions of heavy many-electron ions with neutral atoms in the relativistic energy region \cite{Shevelko-2012, Shevelko-2018}.
The new results on proton elastic scattering were found to be in a good agreement 
with theoretical calculations by using global phenomenological optical model potentials. 

In the future, the detection setup at the CSRe will be further extended by adding new detectors and by covering a larger solid angle.
For the next series of experiments planned at the ESR an upgraded detector setup including 10-20 telescope arms, 
and thus covering a considerably larger angular range, is presently under consideration by the EXL collaboration.
The full potential of this new technique applied to short-lived nuclei far off stability will only be available once the experimental conditions provided by the future 
FAIR facility with high intensity exotic beams from the new fragment separator Super-FRS \cite{Geissel-2003s} have been realised~\cite{Ilieva-2007,Kalantar-2009}. 

\begin{figure*}[t!]
\centering\includegraphics[angle=-0,width=0.8\textwidth]{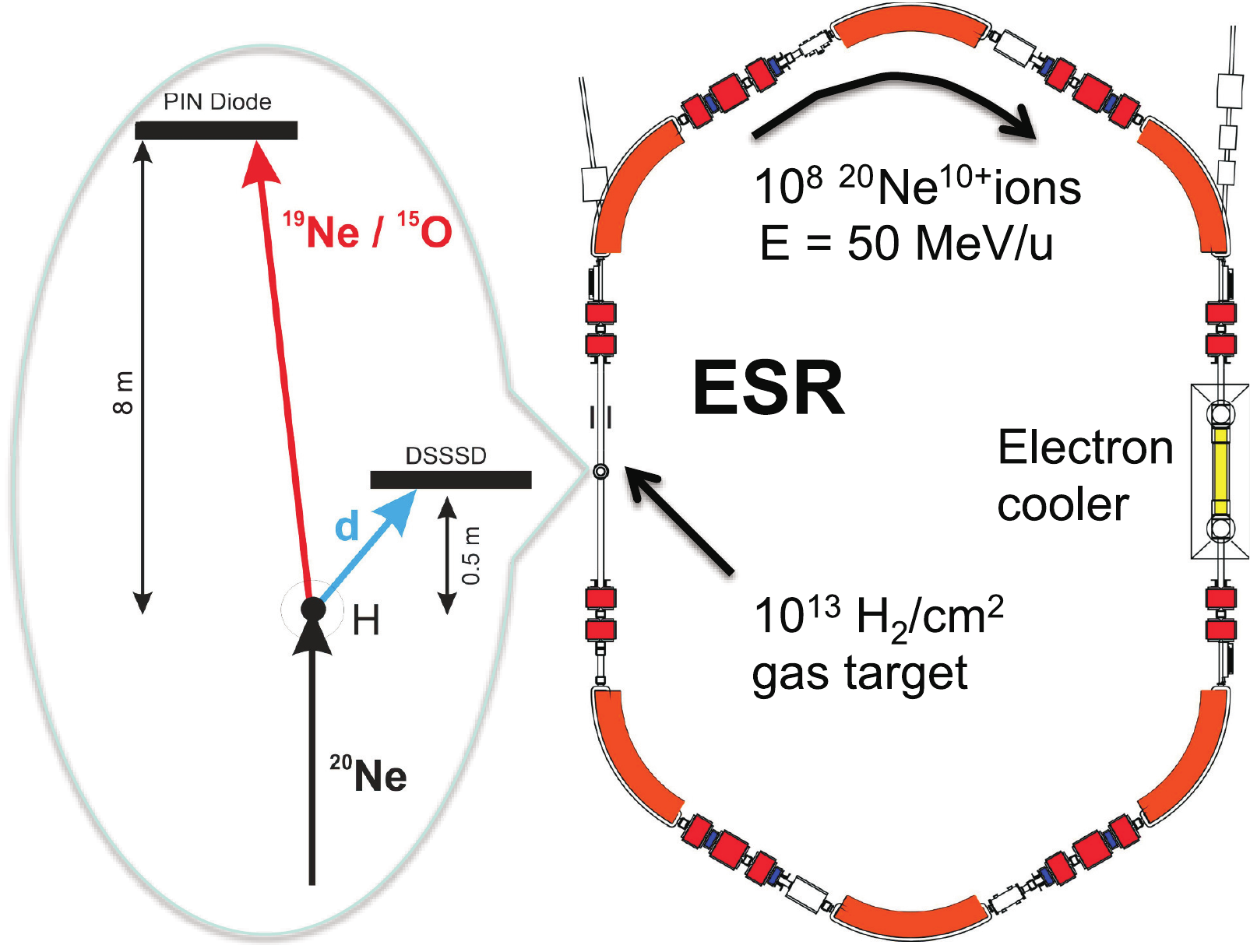}
\caption{(Colour online) The scheme of the experiment at the ESR to study $^{20}$Ne$(p,d)^{19}$Ne$^*$ reaction and the decay of the 4.033~MeV excited state in $^{19}$Ne.
The stored and cooled $^{20}$Ne$^{10+}$ beam interacts with H$_2$ gas target. The deuterons emitted in the $(p,d)$ reaction are detected by the DSSD telescope in the pocket placed about 50 cm downstream the target.
The excited states in $^{19}$Ne$^*$ decay either by emitting $\alpha$-particle producing $^{15}$O daughter ions or through $\gamma$-de-excitation resulting in $^{19}$Ne in the ground state. 
The daughter ions are detected by the PIN-diode detector placed about 8 m downstream the target on the opposite ring side of the DSSD telescope.
The PIN diode is the same as used in the scattering experiments on Ni ions \cite{Yue-2011}. 
Adopted from \cite{Bosch-2013}.
}
\label{Ne_ESR}
\end{figure*}
The last example of so far accomplished reaction studies at medium/high energies relates to the proof-of-principle experiment at the ESR to study the $^{15}$O$(\alpha,\gamma)^{19}$Ne reaction.
This reaction has an astrophysical application in X-ray busters.
Such stellar events are generated by thermonuclear explosions in the atmosphere of an accreting neutron star in a close binary system~\cite{Woosley-1976}. 
In between the bursts, energy is generated at a constant rate by the hot CNO 
cycle driven by the in-flow of hydrogen and helium material from the less evolved companion star. 
It is uncertain how the transition occurs from the hot CNO cycle to the rapid proton-capture nucleosynthesis process, $rp$-process, 
which produces elements possibly up to the tin-antimonium region~\cite{Schatz-2001}.
The $^{15}$O$(\alpha,\gamma)^{19}$Ne reaction is considered to be the most probable candidate for the breakout from the CNO cycle.
However, to address this reaction in a direct kinematics and in the relevant energy region turned out to be very complicated \cite{Davids-2003}.
This is a resonant reaction, in which the key resonance is at $E_r = 504$~keV, corresponding to a $3/2^+$ state at an excitation energy of 4.033~MeV in $^{19}$Ne.
This state in $^{19}$Ne decays predominantly by $\gamma$-decay.
A very weak $\alpha$-decay branch of about $10^{-4}$ was predicted to exist~\cite{Langanke-1986}.
It was the goal to verify this prediction in experiment.

Fully-ionized $^{20}$Ne$^{10+}$ atoms were stored and cooled in the ESR at 50~MeV/u \cite{Doherty-2015}.
The population of the key 4.033~MeV state in $^{19}$Ne was done via the $(p,d)$ direct reaction by employing hydrogen gas-jet target with $10^{13}$~atoms/cm$^2$ density.
Energetic deuterons emitted in the reaction were identified using two DSSDs, arranged as a $\Delta E-E$ telescope.
The detector was placed into a specially constructed detector pocket on the inner side of the ESR about 50 cm downstream the interaction point, see Figure~\ref{Ne_ESR}.
The segmentation of the DSSD allowed for the determination of the emission angle.
Excitation energy spectra were reconstructed by gating on deuteron events in the particle identification spectra. 
A good energy resolution of about 260~keV could be estimated from the FWHM of the 2.795~MeV peak, which is known to be a singlet.
The decay products of the $^{19}$Ne$^*$ ions, $^{19}$Ne$^{10+}$ or $^{15}$O$^{8+}$ ions, after $\alpha$-decay, 
were detected with a PIN-diode detector \cite{Yue-2011} at about 8~m downstream the interaction point.
The same PIN-diode detector as described above was used here.
In an ideal case, for each detected deuteron corresponding to the populated 4.033~MeV state in $^{19}$Ne$^*$, 
a $^{19}$Ne$^{10+}$ or $^{15}$O$^{8+}$ ion shall be detected in coincidence.
For higher energy states in $^{19}$Ne$^*$, $^{15}$O$^{8+}$ ions were successfully detected in coincidence with deuterons.
However, the duration of the experiment, of a few hours only, was too short to investigate the very weak branch of interest.
This was the first transfer reaction ever measured in the ESR.
The performed proof-of-principle experiment shows that the proposed approach to study the $^{15}$O$(\alpha,\gamma)^{19}$Ne is feasible.
In a real experiment, though, the more efficient reaction to produce the key astrophysical resonance in $^{19}$Ne, namely the $^{21}$Ne$(p,t)^{19}$Ne shall be used \cite{Woods-2015}.

\subsubsection{Nuclear reaction studies at beam energies of 10 MeV/u and below}
\label{s:pg}

Whereas the techniques for nuclear reaction studies at medium/high energies are meanwhile well developed, 
addressing nuclear reactions at low energies remains a challenge.

The main motivation driving such studies so far were astrophysical nucleosynthesis processes, 
where the relevant energies for nuclear reactions are determined by the corresponding Gamow windows.
Nucleosynthesis processes involve radioactive, often very short-lived, nuclides.
Various reactions are of interest for different processes and are being considered for future studies.
As of today, only two experiments were performed at the ESR, both aiming at proton capture reactions for the $p$-process nucleosynthesis.
The $p$-process is believed to happen in the explosive burning O/Ne layers in the type-II Supernovae and the explosive carbon burning in the type-Ia Supernovae,
which can be characterised by high temperatures of $2 - 3$~GK and densities of about $10^6$~g/cm$^3$~\cite{Arnould-2003, Rauscher-2013}.
The process involves about 2000 nuclei connected by more than 20000 reactions, mainly $(\gamma, n)$, $(\gamma, p)$ or $(\gamma, \alpha)$.

Both experiments were conducted on stable beams and can thus be seen as proof-of-principle experiments.
Beams of fully ionized $^{94}$Ru$^{44+}$ and $^{124}$Xe$^{54+}$ ions were produced at high energies of above 100 MeV/u 
by stripping all bound electrons in 11~mg/cm$^2$ carbon foil.
They were decelerated in the ESR, cooled and brought into interaction with the H$_2$ gas-jet target.
The reaction kinematics of proton-capture, $(p,\gamma)$, reactions on heavy ion beams
gives rise to a narrow, beam-like cone of heavy reaction products, caused
by the negligible momentum transfer to the prompt $\gamma$ ray \cite{Langer-2017}. 
In the nuclear reaction, the daughter recoils have the charge $(Q+1)$ as compared to the charge of the primary beam $(Q)$.
After separation of the heavy recoils from the primary beam in a
dipole magnet, an efficient interception of the
heavy recoils by using suitable particle detectors is possible. 
It can be shown, that due to momentum conservation, if a primary beam particle has a bound electron and looses it in an atomic reaction in the target, 
its trajectory is basically the same as the 
one of the nuclear recoils of interest. 
Therefore, a prerequisite of this experimental scheme is the absence of beam particles being
further ionized $(Q+1)$ in the target, which is ensured through using fully ionized primary beams.
This is done at the cost of a lengthy deceleration process but offers clean experimental conditions.

The deceleration of ion beams in the ESR, see Section \ref{s:deceleration}, is routinely done for fast extraction to the HITRAP setup.
However, in-ring experiments at lowest ESR energies were never done before and required dedicated machine studies.
In the first experiment in 2009 \cite{Zhong-2010, Mei-2015}, 
the cross-section of $^{94}$Ru$(p,\gamma)^{95}$Rh reaction could be measured for $^{94}$Ru$^{44+}$ ions stored at 11, 10 and 9 MeV/u~\cite{Zhong-2010}.
These energies correspond to center of mass energies of $E_{CM}=10.971$, 9.973 and 8.976 MeV, respectively.
About $5\cdot10^6$ $^{96}$Ru$^{44+}$ ions were stored and cooled at the final energy.
Taking into account the revolution frequency of about 500~kHz and the thickness of the H$_2$ target of about $10^{13}$ atoms/cm$^2$, 
luminosity of about $2.5\cdot 10^{25}$~cm$^{-2}\cdot$s$^{-1}$ was achieved.
The main reaction channel in the target was the atomic electron pick-up (REC) from the target atoms which is accompanied by an emission of an X-ray~\cite{Eichler-2007}.
As a result, the charge state of $^{96}$Ru ions is changed from $44+$ to $43+$ and the $^{96}$Ru$^{43+}$ ions 
can be detected by particle detectors inside the first dipole magnet downstream the gas-jet target (see Figure~\ref{esr-cry}).
Since the K-shell REC cross sections are known to a very few percent, 
the X-rays detected in coincidence with $^{96}$Ru$^{43+}$ ions provide an {\it in situ} measurement of the luminosity.
The covered energies are much higher than the Gamow window of the $p$-process. 
Also, these energies are above the $(p,n)$ reaction threshold.
The nuclear reaction products of $(p,\gamma)$, $(p,n)$ and $(p,\alpha)$ reactions, the $^{97}$Rh$^{45+}$, $^{96}$Rh$^{45+}$ 
and $^{93}$Tc$^{43+}$ ions, respectively, 
were bent to inside orbits by the dipole magnet and are detected by position sensitive DSSDs.
The DSSDs were installed in air in vacuum pockets separated from the ring vacuum by 25~$\mu$m stainless steel windows.
To extract the cross-sections of interest, deconvolution of several reaction channels in measured spectra was required.
The measured $^{96}$Ru$(p,\gamma)^{95}$Rh cross-sections at 8.976, 9.973, 10.971~MeV 
were determined to be $8.28^{+2.58}_{-2.76}$, $7.83(2.13)$, and $9.13^{+2.59}_{-2.94}$ mbar, respectively.
We note that the energy range below the neutron-emission threshold is probably most interesting for constraining theoretical calculations 
since there the  proton width is typically the most sensitive contribution. 

\begin{figure*}[t!]
\centering\includegraphics[angle=-0,width=0.8\textwidth]{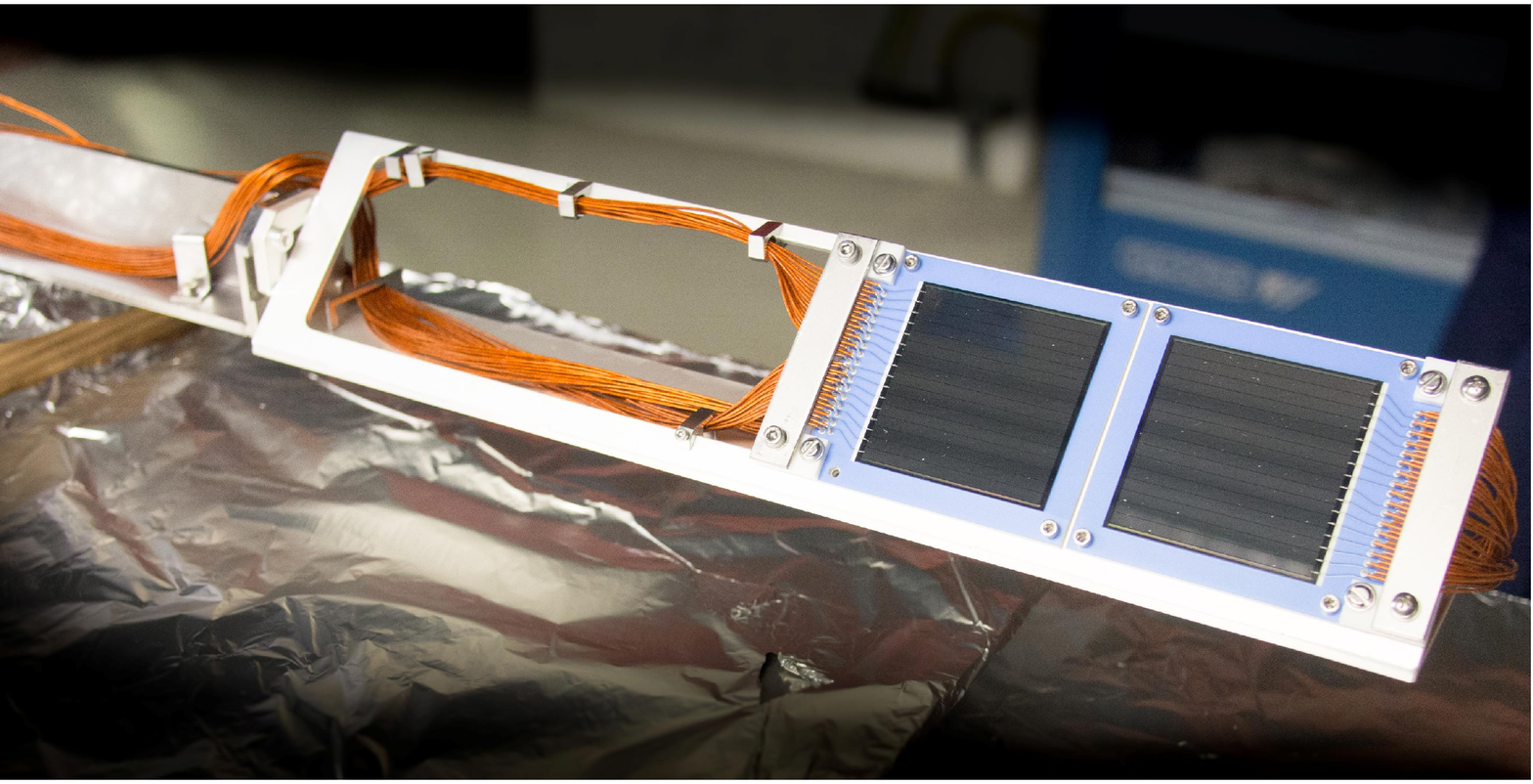}
\caption{(Colour online) Photograph of the ultra-high vacuum capable DSSD detectors employed for proton capture reaction studies in the ESR.
The detector is made of low outgassing materials, is bake-able and is placed directly into the ESR vacuum environment. 
In this configuration two DSSDs are mounted onto the detector frame. 
Please note the arrangement of the cables around the opening required for the stored beam to freely pass the detector.
In the $^{124}$Xe$(p,\gamma)$ experiment, the assembly with a single DSSD was used, see Figure \ref{PG_Setup}.
Adopted from \cite{Lestinsky-2016}.
}
\label{ESR_Detector}
\end{figure*}
On the one hand, the feasibility of such studies and a reliable operation of the ESR have been shown.
On the other hand, the DSSDs employed at that time were placed in air behind a stainless steel window.
Heavy recoils with energies below 9 MeV/u were stopped before reaching the detector. 
Therefore, to continue this research program, special ultra-high vacuum compatible DSSDs have been designed.
The detectors were made of materials that provide low outgassing rates.
The employed silicon wafer is 0.5 mm thick and its active area is
$50 \times 50$ mm$^{2}$ with both sides perpendicularly segmented into
16 strip. For heavy ion detection it provides 100\% efficiency, a spatial
resolution of about 3 mm as well as an energy resolution of better than 1\%.
In 2016, the detectors were mounted into the ultra-high vacuum of the ESR.
An ultimate vacuum pressure of $2 \cdot10^{-10}$ mbar has been achieved after the bake-out. 
In order to increase the flexibility of the operation of the ESR, a new position in the ring with a smaller dispersion was chosen.
The detector was inserted into the vacuum chamber of the C-type dipole magnet.
The complexity was that one needs to intercept orbits in the inner part of the ESR.
To reach these orbits, the detector was mounted on an about 2 m long arm and was moved from outside of the ring across the whole ring aperture into its most inner part.
The detector was designed such that the detector frame and cables allow the primary beam to pass through.
The photograph of the detector is shown in Figure \ref{ESR_Detector}.
 
\begin{figure*}[t!]
\centering\includegraphics[angle=-0,width=0.8\textwidth]{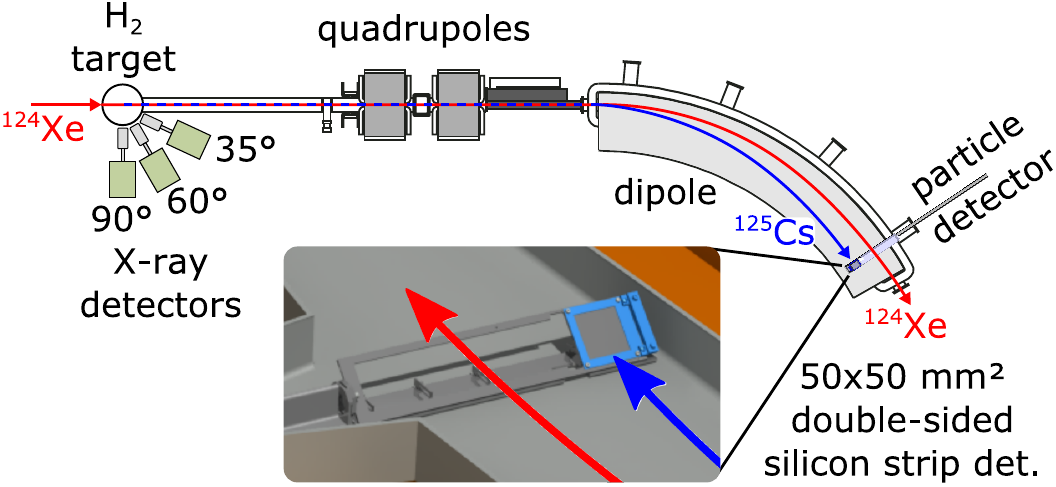}
\caption{(Colour online) The experimental arrangement for the $^{124}$Xe$(p,\gamma)^{125}$Cs experiment at the ESR.
The primary $^{124}$Xe particles circulate in the ring and interact with thin H$_2$ gas-jet target.
The reaction products $^{125}$Cs have the charge larger than the primary beam ions and are thus deflected to the inner orbits by the ESR dipole magnet.
These ions are intercepted by the DSSD.
The detector, see Figure \ref{ESR_Detector}, is moved from the outside of the ESR across its entire aperture.
Therefore, the detector has an opening to allow for the primary beam to freely circulate in the ESR.
The X-ray detectors placed around the target are used to measure K-REC radiation, which is then used to determine the reaction luminosity.
Adopted from \cite{Glorius-2019}.
}
\label{PG_Setup}
\end{figure*}
In the experiment on $^{124}$Xe$^{54+}$ \cite{Glorius-2019}, measurements at 5 beam energies 
starting from 8 MeV/u and reaching down as low as 5.5 MeV/u were performed.
An internal H$_2$ target was operated at an increased average density of $10^{14}$ atoms/cm$^2$.
Taking into account the revolution frequency of about 250~kHz, the peak luminosity of about $L=10^{26}$ cm$^{-2}\cdot$s$^{-1}$ was obtained.
Atomic interactions with the atoms of the target and the residual gas limit the beam storage time.
At 7 MeV/u a beam lifetime of about 2.5~s could be achieved, which indicates the complexity of the conducted experiments at such low beam energies.
Data recording period was about 12 s before the ring had to be refilled. 
A single fill cycle of the ESR including deceleration, cooling, and measurement was about 50 s, implying a duty cycle of about 25\%.

The low beam energies do not allow for using gas-filled detectors in pockets and 
thus disable calibration of the luminosity through detection of REC daughter $^{124}$Xe$^{53+}$ ions.
X-ray detectors were installed at $35^\circ$, $60^\circ$ and $90^\circ$ laboratory angles viewing the interaction region.
The measured K-REC X-rays were taken for luminosity determination.
The K-REC process for fully-ionized projectiles can be well described by theory within the framework of the relativistic distorted-wave approach \cite{Artemyev-2010}.
Differential K-REC cross-sections for the angles of interest can be predicted with an uncertainty of below 2\%,
where the main source of the uncertainty is due to the used molecular H$_2$ target instead of atomic H assumed in theory.

\begin{figure*}[t!]
\centering\includegraphics[angle=-0,width=0.8\textwidth]{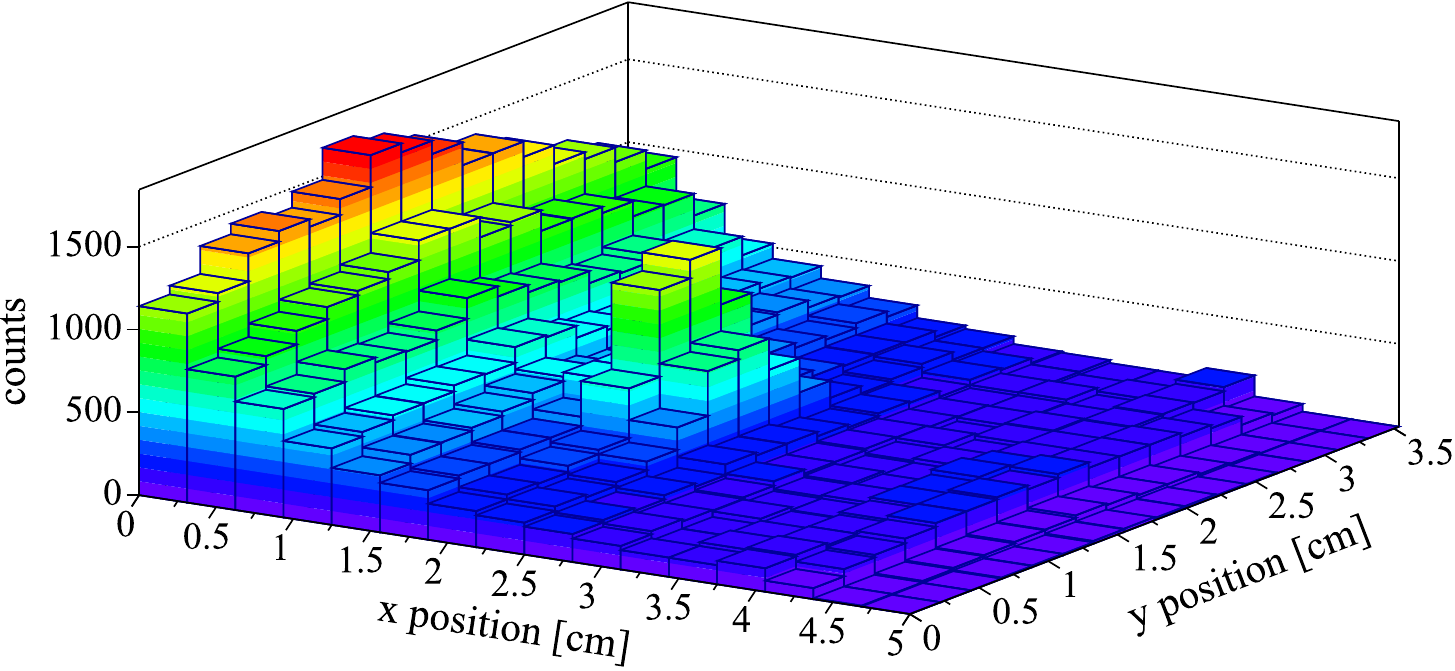}
\caption{(Colour online) The hit distribution of ions across the surface of the DSSD for the primary beam energy of 7 MeV/u. 
The narrow cluster of counts at about the center of the detector corresponds to the $(p,\gamma)$ reaction products, $^{125}$Cs$^{55+}$ ions.
The smooth background is due to elastic scattering of $^{124}$Xe on the protons of the target. 
The shape of this background can be well described in dedicated simulations.
At the largest $x$ values also the contribution due to $(p,n)$ reaction can be seen.
Adopted from \cite{Glorius-2019}.
}
\label{F:pg}
\end{figure*}
The distribution of hits of $^{125}$Cs ions on the DSSD detector is shown in Figure~\ref{F:pg} for the center of mass energy $E_{CM}=6.96$ MeV.
This energy and also $E_{CM}=7.92$ MeV are the two points measured above the $(p,n)$ threshold of $E_{CM}=6.71$ MeV.
Most of the events in Figure \ref{F:pg} are due to the elastic scattering of $^{124}$Xe$^{54+}$ on protons.
The shape of this background is well understood through Monte-Carlo simulations including the reaction kinematics and ion-optical properties of the ESR.
The simulations were performed with the MOCADI code \cite{Iwasa-1997, Iwasa-2011}.
At large $x$ positions, an excess of events can be seen. These events correspond to the $^{124}$Xe$(p,n)^{124}$Cs reaction products. 
The recoil energy due to neutron emission is larger than due to $\gamma$-emission, which leads to a different distribution on the detector.
The peak from the $(p,\gamma)$ reaction is clearly seen at about the center of the detector.
The reaction cross-section at lowest $E_{CM}=5.47$ MeV is deduced to be $\sigma(p,\gamma)=14.0\pm2.4_{stat}\pm0.9_{syst}$ mbarn.
For more details including the impact on the theoretical calculations, the reader is referred to \cite{Glorius-2019}.

After the successful experiment on $^{124}$Xe, the next step is to perform first studies on secondary beams.
The corresponding proposal has been approved at GSI and the experiment is pre-scheduled for early 2020 \cite{GSI_Beamtime}.
It is planned to use a dedicated metal scraper installed in front of the analyzing dipole magnet in the ESR to largely remove the background due to elastic scattering events.
The lowest energy reached in the past experiments lies at the higher edge of the Gamow window of the $p$-process.
In principle, the ESR is capable to decelerate the beams further down to a minimum energy of about 3 MeV/u.
However, the operation of the ESR for in-ring experiments at such low energies has not been demonstrated yet.
Furthermore, in order to achieve longer storage times, the vacuum of the ring needs to be improved.

Therefore, a dedicated low-energy storage ring, CRYRING@ESR, has been built behind the ESR, see Figure \ref{esr-cry}.
CRYRING@ESR will receive pre-decelerated, pre-cooled beams of highly-charged ions from the ESR.
It is aimed to achieve a lower rest gas pressure in the CRYRING@ESR as compared to the one in the ESR,
which shall allow for longer storage times for ion beams.
Although the main research program at CRYRING@ESR relates to high-precision atomic physics studies, there is a range of nuclear physics and astrophysics experiments planned \cite{Lestinsky-2016}.
In addition to lower energies, the experiments at CRYRING@ESR will profit from shorter circumference, which is exactly 1/2 of the ESR circumference.
For a beam with a given energy, this immediately transfers in an increase in the reaction luminosity by a factor of 2.
It is foreseen that the beams will be transported to CRYRING@ESR at energies of about 10-15 MeV/u.
This energy range is ideal to address nuclear structure through transfer reactions.
A dedicated chamber for reaction studies is being constructed at the University of Edinburgh.
Together with versatile in-vacuum DSSD detectors arranged to cover a large fraction of $4\pi$ around the interaction zone, this setup is named CARME.
The details of the design can be found in the approved Technical Design Report \cite{CARME}.
In addition to H$_2$ targets, it is planned to employ gases like D$_2$, $^3$He and $^4$He, etc.
Many of the envisioned experimental ideas are being prepared for publication.
Among the reported planned experiments there is a straightforward continuation of $(p,\gamma)$ reaction rate measurements for nucleosynthesis application,
which in the beginning will be possible directly in the Gamow window of the $p$-process and later, when the techniques and methods are further developed, possibly at even lower energies relevant to $rp$- and other processes.

The electron cooler in the CRYRING@ESR implements the expansion of the electrons into the decreasing magnetic guiding field and thus extremely cold electron beams can be produced.
It has been proposed to employ this feature to measure resonance strength of specific resonant astrophysical reactions.
This may be important in situations where within the corresponding Gamow window the energy level density is low and the amount of states to be considered is just a few.
Then, the calculations based on Hauser-Feschbach theory might be inadequate.
This holds, for instance, for the case of the $^{33}$Cl$(p,\gamma)^{34}$Ar and $^{34}$Cl$(p,\gamma)^{35}$Ar reactions in O-Ne novae, where only several states are relevant.
These reactions are needed to constrain the parameters for the S-Cl burning cycle and set the limits on the existence of the galactic $^{34m}$Cl $\gamma$-emitter.
Furthermore, the knowledge of these reaction rates will contribute to the understanding of S isotope abundances in pre-solar grains.
The details on the proposed investigations of these reactions can be found in \cite{Lestinsky-2016}.

\begin{figure*}[t!]
\centering\includegraphics[angle=-0,width=0.6\textwidth]{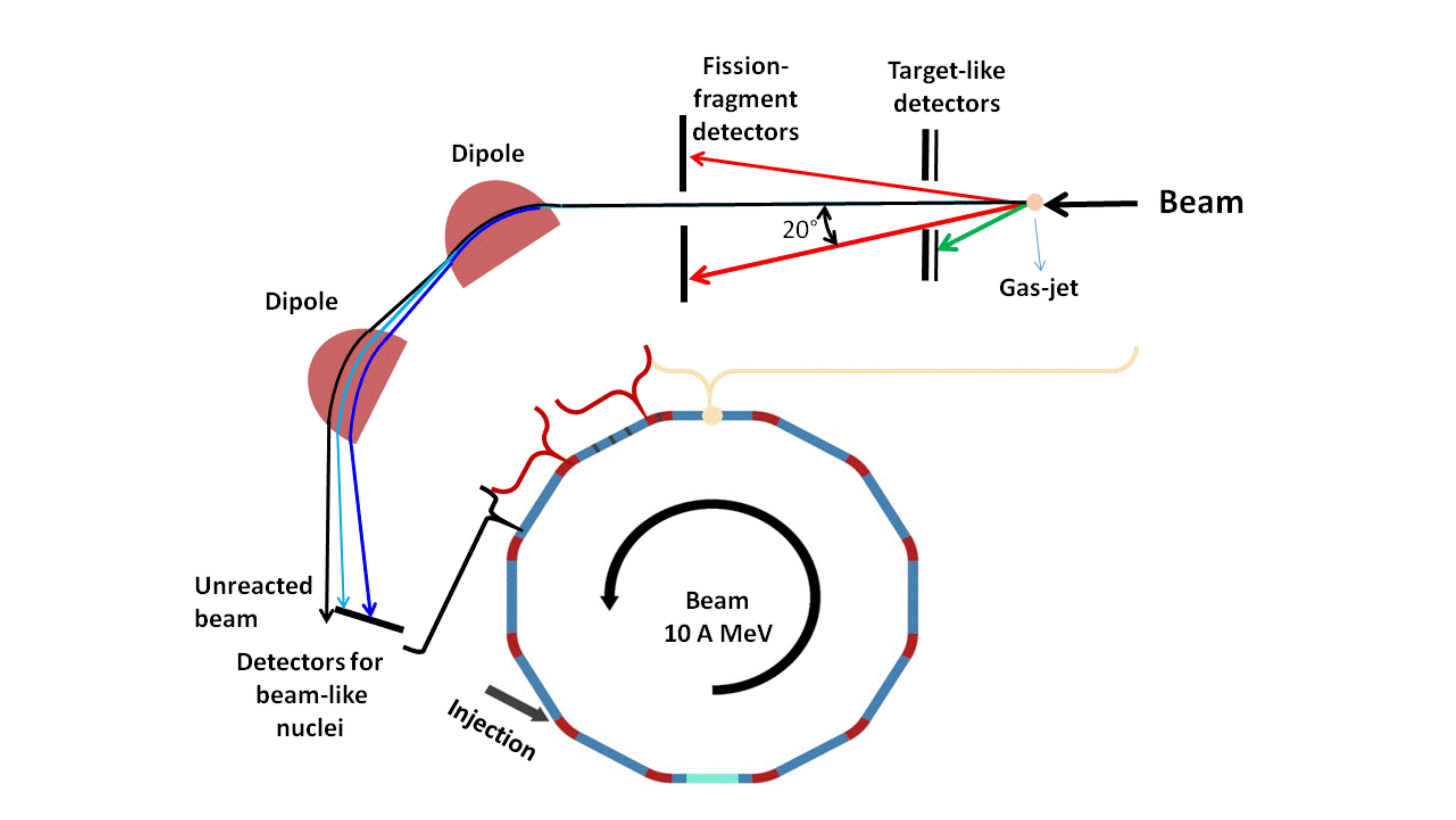}
\caption{(Colour online) Proposed setup for studying neutron-induced reactions through surrogate reaction method in the CRYRING@ESR.
The primary beam interacts with the gas-jet target producing the compound nucleus of interest, which promptly decays.
The target-like ejectiles and fission fragments are detected with the detectors mounted within the CARME setup \cite{CARME}.
The beam-like nuclei are detected downstream the target. The CRYRING@ESR dipoles are used as the mass separator.
In addition, $\gamma$ rays can be detected with conventional detectors placed outside of the vacuum chamber.
Adopted from \cite{Henriques-2019}.
}
\label{Surrogate}
\end{figure*}
Another proposal concerns the study of neutron-induced reactions by means of the surrogate method.
In this method the excited compound nucleus of interest with mass number $(A-1)+n=A^*$ is produced in a different reaction $X+Y=A^*$.
Though the production process is different, the probabilities for various decay channels shall be the same, though,
in the correct description of the $(A-1)+n=A^*$ reaction, one shall take into account that spin-parities of excited states populated via $X+Y=A^*$ reaction might be different \cite{Escher-2012, Escher-2018}.
In the last years the theoretical description of reaction data obtained with the surrogate method show very good agreement with available direct measurements \cite{Ratkiewicz-2019}.
It is envisioned to employ this surrogate reaction method to study neutron-induced reactions on heavy (long-lived) radioactive ions stored and cooled in the CRYRING@ESR.
In particular, the fissioning nuclei are in the focus. 
In such cases, the compound nucleus can either de-excite by emitting $\gamma$-rays, can emit a neutron or can fission.
The reaction setup is schematically shown in Figure \ref{Surrogate}.
In such an arrangement, the entire information on the fate of produced compound nuclei will be available.
The target-like ejectiles will be measured with DSSDs available in the CARME setup and will provide the excitation energy of the compound nucleus.
Projectile-like fragments will be measured with particle detectors by using the ring dipole magnets as a mass separator.
In addition, $\gamma$ rays may be detected with standard high efficiency detectors placed outside of the ring vacuum system.
An interesting development pursued within this project is to employ conventional solar cells as position sensitive, ultra-high vacuum compatible, radiation hard particle detectors \cite{Perez-2019}.
If successful, this will be a cost-efficient alternative to silicon detectors.
In this arrangement at CRYRING@ESR, solar cells can be used to detect fission fragments.
More details on the project the reader can find in \cite{Henriques-2019}.

The CRYRING@ESR is presently the only dedicated low-energy storage ring.
It offers unparalleled capabilities for nuclear reaction studies.
However, the lengthy process until the heavy-ion beam is stored and cooled in the CRYRING@ESR at the energy of interest,
which is mainly due to long deceleration time in the ESR, restricts the use of CRYRING@ESR for forefront nuclear physics research,
where the main interest lies in the reaction studies on short-lived nuclei.

\subsection{Atomic physics experiments with highly-charged ions} 
\label{s:atomic}
The capability of heavy-ion storage rings to provide cooled beams of heavy ions in a defined, high atomic charge state enables 
precision studies of atomic properties and tests of fundamental interactions and symmetries.
The TSR, ESR and since recently also CSRm and CSRe contributed to this research.
A huge discovery potential is expected from future experiments at CRYRING and HESR.
Many dedicated spectrometers, targets, detection methods, etc. were developed in the last decades.
To our knowledge, no review exists covering all aspects of atomic physics research at storage rings.
The reader is referred to proceedings of specialized conferences and special journal issues, like, e.g., \cite{Aumayr-2019}.
In this work a brief summary is given of actual research activities, 
which are based on the presently approved experimental programs.
A multitude of other experiments which are not backed by a running experimental proposal at the ESR or CSRe,
like, for instance, the exciting precision tests of general relativity conducted in the TSR \cite{Novotny-2006} and ESR \cite{Botermann-2014}, are left outside of the scope of this paper.

\subsubsection{Precision tests of bound-state quantum electrodynamics}
\label{S:QED}
The ESR has provided world-wide unique conditions for precision studies of bound-state quantum electrodynamics (QED) 
in the regime of strongest electromagnetic fields as present in high-Z one- and few-electron ions, up to hydrogen-like uranium. 
The high intensity beams of basically any element in any desirable charge state can be stored and cooled guaranteeing brilliant quality for precision experiments. 
In particular, the cooling is of high importance, since it guarantees long storage times as well as accurately defined beam properties such as position, size, and velocity. This is indispensable for reducing the uncertainties associated with relativistic Doppler effect when one deals with measurements on relativistic ion beams.  

Since the early days of its operation, various experiments addressing the bound-state QED 
effects in high-Z ions have been carried out at the ESR. 
The experiments explored different observables in different systems, thus providing comprehensive tests of the bound-state QED. 
These can be sorted in the following manner; the ground-state Lamb shift in hydrogen-like ions, 
the two-electron QED in helium-like ions, intra-shell transitions in lithium-like systems and hyperfine structure in hydrogen- and lithium-like systems.

Understandably, heavy hydrogen-like ions, being the most fundamental atomic systems, have attracted much of the experimental efforts. The main technique used in these experiments has been the X-ray spectroscopy of Lyman transitions produced by electron capture in the excited states of the hydrogen-like ions.  
The first measurement on hydrogen-like uranium at the gas-jet target of the ESR \cite{Stoehlker-1993} could already improve the accuracy of the first measurements at BEVALAC \cite{Briand-1990, Lupton-1994}. 
This has been followed by experiments at the electron cooler of the ESR, 
where the accuracy has been further enhanced \cite{Beyer-1995}. 
The next improvement of accuracy has been made possible by exploiting the deceleration mode of the ESR in experiments at the gas-jet target \cite{Stoehlker-2000} and at the electron cooler \cite{Gumberidze-2005}. 
The most precise value available currently for the ground-state Lamb shift in hydrogen-like uranium 
from \cite{Gumberidze-2005} is in good agreement with the state-of-the-art QED calculations \cite{Yerokhin-2015} 
and provides a test of the first-order QED contributions at the 2\% level. 
This represents the most stringent test of the bound-state QED for heavy hydrogen-like ions. 
However, the experimental accuracy still does not permit sensitivity to higher-order QED contributions, whose evaluation has been accomplished following decades of extensive theoretical work, see \cite{Yerokhin-2015, Indelicato-2019} and references therein. 
One of the limitations on the experimental side is due to the energy resolution 
of the semiconductor detectors which have been used in all these measurements. 
In order to overcome this barrier, dedicated high-resolution detection systems have been designed and developed; 
the FOCAL crystal spectrometer \cite{Beyer-2009, Beyer-2015} and 
microcalorimeters \cite{Pies-2012,Hengstler-2015,Kraft-Bermuth-2017}. 
These two devices differ by their detection principle, but both provide about one order of magnitude higher resolution than standard semiconductor detectors. 
Both devices have been fielded at the gas-jet target of the ESR to measure the ground state Lamb shift in hydrogen-like gold \cite{Gassner-2018, Kraft-Bermuth-2017} and both
have demonstrated the superb energy resolution. 
However, the accuracy of these results has been limited by relatively large systematic uncertainties, 
mainly due to the position of the gas-jet target. 
Reduction of this systematic uncertainty requires more studies including modification of the detection setup for the case of FOCAL. Here, we would like to mention that in the near future a measurement is planned 
with the microcalorimeters at the electron cooler of CRYRING \cite{Lestinsky-2016}. 

For high-Z helium-like systems, two experiments have been conducted at the ESR. 
In \cite{Gumberidze-2004}, the K-shell radiative recombination (K-RR) 
transitions in initially bare and hydrogen-like uranium ions have been measured 
at the electron cooler providing a direct access to the two-electron contribution 
to the ground-state binding energy in helium-like uranium. 
The obtained experimental result is in good agreement with the QED predictions \cite{Yerokhin-1997}. 
Moreover, the accuracy of this experimental result has already achieved the sensitivity to the two-electron self-energy contribution, i.e. to a second-order QED effect. In \cite{Trassinelli-2009}, the $1s2p~^3P_2 \rightarrow 1s2s~^3S_1$ transition has been measured at the gas-jet target in high-resolution using a dedicated Bragg crystal spectrometer. 
In this experiment, the accuracy was limited by low counting statistics and the observed peak asymmetry due to imperfections of the used crystal. 
With better crystals, which are already available, and longer measurement time this accuracy can be readily improved by about an order of magnitude. This would be enough to test the two-electron and two-loop QED contributions. 
We would like to note, that to the best of our knowledge, these two measurements are the only ones available up to date for these particular observables in helium-like uranium providing sensitive tests of electron-electron interaction in the heaviest two-electron system.

\begin{figure*}[t!]
\centering\includegraphics[angle=-0,width=0.8\textwidth]{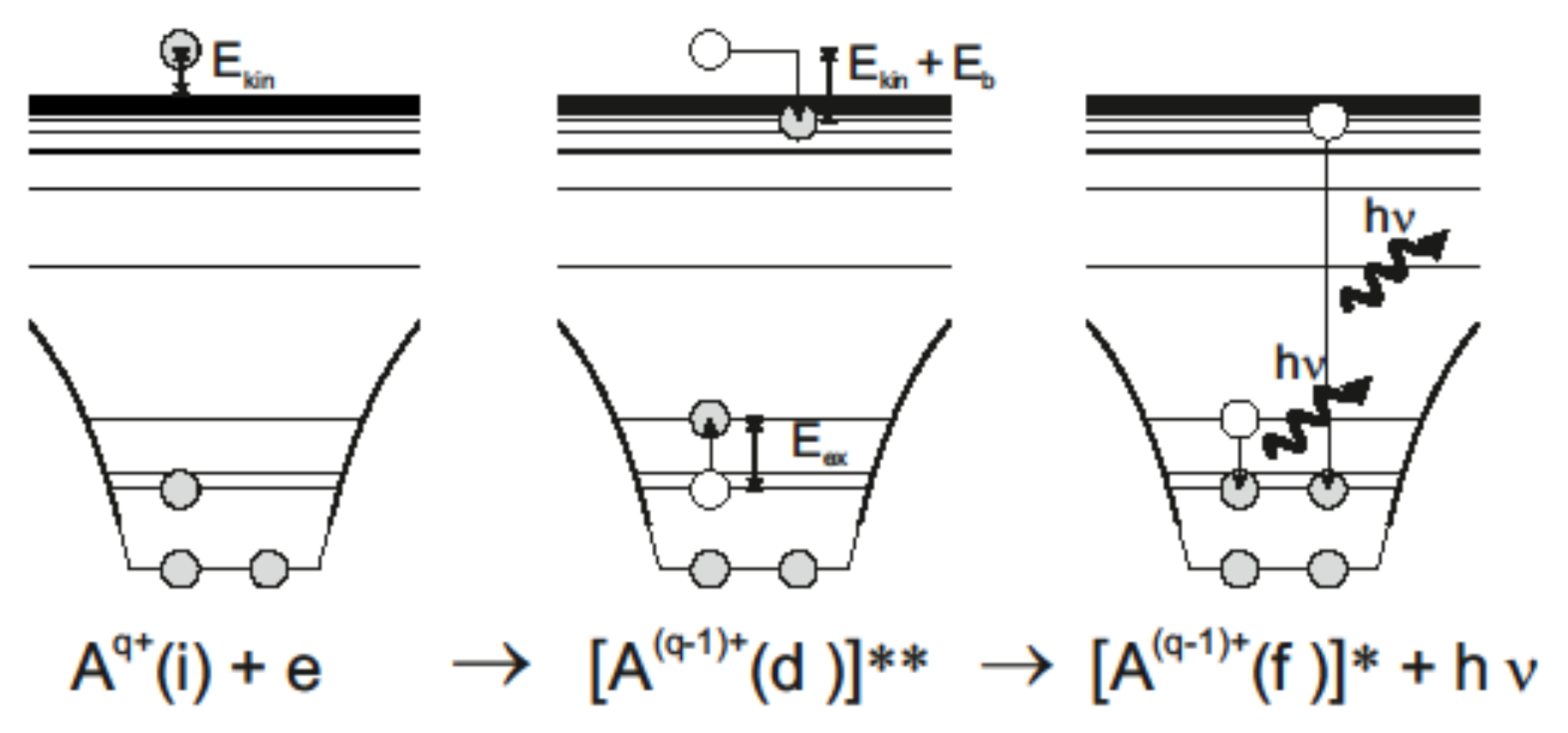}
\caption{(Colour online) 
Scheme of dielectronic recombination. (Left) The initial situation (i) with bound electrons of an ion in the charge state $q+$, 
and a free electron with kinetic energy $E_{kin}$ above threshold. 
(Middle) The free electron is captured into a bound state with binding energy $E_b$ and, 
simultaneously, a bound electron will be excited with the resonance condition $E_{\rm ex} = E_{\rm kin} + E_b$ thus forming a doubly excited state (d). 
(Right) The final ionic state (f) is obtained via the stabilisation of the doubly excited state (d) by the emission of photons or Auger electrons. Taken from \cite{Bosch-2013}. (Courtesy C. Brandau).}
\label{F:DR}
\end{figure*}

To study lithium-like ions, dielectronic recombination (DR) at the electron cooler of the ESR has been employed.
In the DR process, free electrons with matching kinetic energy are captured resonantly with simultaneous excitation 
of an already bound electron resulting in a doubly excited autoionizing atomic configuration \cite{Mueller-2008, Brandau-2010, Lestinsky-2008, Graham-2012, Brandau-2012}, see Figure \ref{F:DR}. 
The DR can be regarded as Auger spectroscopy in inverse kinematics since autoionization and the initial dielectronic capture 
are in statistical equilibrium. 
The electron cooler in a storage ring provides a dedicated co-propagating (merged-beams) free-electron target. 
Fast changing of the electron velocity in the cooler allows for (short time) tuning of the relative velocity between ion and electron beams, see Section \ref{s:targets}.
Such tuning can be done in very fine energy steps. 
If the resulting relative collision energy matches the resonance condition the recombination process producing $(Q-1)$ ions is enhanced.
The down-charged ions can be registered by particle detectors in analogy to reaction experiments described in Section~\ref{S:reactions}. 
For QED studies, DR has been exploited as a tool to measure $2s_{1/2}-2p_{1/2}$ excitation energy in Au, Pb, and U \cite{Brandau-2003} as well as $2s_{1/2}-2p_{1/2}$ and $2s_{1/2}-2p_{3/2}$ energies in Xe \cite{Bernhardt-2015}. 
A good agreement with QED predictions has been observed for all cases. For the heaviest lithium-like ions, the accuracy of theoretical values is lower than the experimental one due to the uncertainties of nuclear parameters. 
Here, we would like to note that currently the most precise measurement of the $2s_{1/2}-2p_{1/2}$ 
excitation energy in lithium-like uranium has been performed at the SuperEBIT facility in Lawrence Livermore National Laboratory \cite{Beiersdorfer-2005}.

The hyperfine structure reflects the interaction between the nuclear and electron magnetic moments and thus it provides an opportunity to test the quantum electrodynamics in the regime of extremely strong magnetic fields. 
In contrast to the Lyman and intra-shell transitions, the energy range of the hyperfine transitions in high-Z one- and few-electron ions is accessible with conventional laser systems and accordingly the laser spectroscopy has been the tool for corresponding experimental studies at the ESR. The first observation and measurements of the hyperfine transitions have been performed at the ESR for hydrogen-like bismuth ($^{209}$Bi$^{82+}$) \cite{Klaft-1994} and lead ($^{207}$Pb$^{81+}$) \cite{Seelig-1998}. 
A discrepancy with the QED predictions has been observed.
However the comparison has been hampered by uncertainties of the nuclear parameters; the nuclear magnetic moment and the magnetization distribution. In order to circumvent this problem, a measurement of the specific difference between the hyperfine splittings of hydrogen- and lithium-like ions with the same nucleus has been proposed which would eliminate the uncertainties due to the nuclear magnetization distribution and provide a direct test of the QED predictions \cite{Shabaev-2001, Volotka-2012}. 
After quite an extensive experimental effort, the first observation of the hyperfine transition in lithium-like bismuth ($^{209}$Bi$^{82+}$) \cite{Lochmann-2014} and the first measurement has been accomplished \cite{Ullmann-2017}. 
The obtained experimental result for the specific difference deviated from the QED predictions by about 7$\sigma$ which has been termed as a ``hyperfine puzzle'' \cite{Karr-2017}. 
Although the specific difference mostly eliminates the dependence on the nuclear magnetization distribution, 
it is still affected by the value of the nuclear magnetic moment. 
Therefore, to explore this aspect, a new measurement of the nuclear magnetic moment  of $^{209}$Bi has been performed combined with new chemical shift calculations \cite{Skripnikov-2018}. 
The result of this measurement deviates strongly from the previously used literature value of the nuclear magnetic moment for $^{209}$Bi and, when combined with the QED calculations, it resolves the discrepancy with the experiment for the specific difference. Nevertheless, the uncertainty of this new measurement of the $^{209}$Bi nuclear magnetic moment is still relatively high and does not yet permit a stringent test of the QED in the regime of very strong magnetic fields. Therefore, further investigations are needed to fully clarify the issue. These include more precise measurements of the $^{209}$Bi nuclear magnetic moment as well as measurements of the hyperfine structure for different isotopes of bismuth ($^{207}$Bi and $^{208}$Bi) as well as measurements in Penning traps, i.e. ions at rest. 
Since $^{207}$Bi and $^{208}$Bi are radioactive, their intensities in the ESR will inevitably be much lower than that of stable $^{209}$Bi. 
Whereas the excitation of the upper hyperfine level through a suitable laser will be done in a conventional way as all other laser-spectroscopic studies at the ESR \cite{laserspec}, the optical detection of the de-excitation will not be feasible.
It is therefore proposed to investigate whether the state-selective application of the resonant DR process can be employed for this purpose.

\subsubsection{Electron spectroscopic methods to study few-body quantum dynamics at heavy-ion storage rings}
Storage rings offer excellent conditions to perform precision studies of quantum dynamics in ion-atom collisions.
Again, these conditions are due to the brilliant quality of cooled stored beams and clean experimental conditions.
Such studies were extensively pursued in the past in the TSR, where ground-breaking experiments were conducted, like, for example, implementation of a reaction microscope to investigate break-up processes 
of laser-cooled atoms in a magneto-optical trap used as a target in the TSR \cite{Fischer-2012}.
Today the relevant experiments play a central role in the ESR program as well as there is a bright research program envisioned at CRYRING, CSRe and future HESR and the rings of HIAF. 

Electrons are arguably the most immediate messengers from ion-atom collisions 
and thus serve as supremely efficient tools to study few-body quantum dynamical processes for atomic collision systems. 
This is particularly valid in the high $(Z_1+Z_2)$ region, i.e. in the relativistic regime, 
where additionally considerable contributions from QED add to expose otherwise invisible details of quantum dynamics.

The central device of the present experimental campaign at the ESR is the forward electron spectrometer located at the internal target. 
The imaging forward electron spectrometer has the aim to reconstruct the vector momenta of electrons emitted in collisions of the stored ion beam with atoms of the gas target \cite{Hagmann-2007}.
The spectrometer consists of a dipole magnet with 60$^\circ$ bending angle 
followed by a quadrupole triplet, another 60$^\circ$ dipole and a set of slits. 
Its design allows for accurate momentum analysis of electrons represented by the images on dedicated 2D position sensitive detector \cite{Hillenbrand-2014a, Hillenbrand-2015a}.
The detector conventionally used for this purpose is a micro-channel plate detector implemented into the ultra-high vacuum of the ESR.

The main quantity to be measured in collisions are the electrons emitted into the continuum. 
The X-rays and up or down-charged projectile-like recoils are detected in coincidence.
The setup enables investigations of several fundamental processes ranging from kinematically complete studies of multiple ionization and electron-impact ionization, $(e, 2e)$, on ions to radiative and non-radiative electron transfer processes to the projectile continuum and kinematically complete measurements of the high-energy limit of the electron-nucleus Bremsstrahlung \cite{Hagmann-2007, Hillenbrand-2013a}.

In very slow collisions, when the ion velocities $v_{ion}$ are 
much smaller than the velocity of the K-shell electron of the target $v_K$, $v_{ion}<<v_K$,
the ion and the target atoms can form a superheavy quasi-molecule.
In transient superheavy $(Z_1+Z_2>173)$ quasimolecules formed in adiabatic collisions the innermost quasimolecular orbital $1s\sigma$ has been predicted by theory to dive into the negative Dirac continuum for suitable internuclear distances of the collision partners \cite{Gershtein-1970,Tupitsyn-2010, Tupitsyn-2012}. 
When a vacancy in one of the collision partners, e.g. a hydrogen-like or bare ion, is brought into the $1s\sigma$ orbital during the collision, this vacancy can be filled by an electron from the negative continuum,  appearing as a spontaneously emitted positron.  
The spectroscopy of spontaneous positron emission in superheavy quasimolecules is one of the great challenges and promises for storage rings.
It is important to stress that the duration of the collision and thus the lifetime of the $1s\sigma$ molecular orbital (MO) decreases rapidly with the ion velocity.
Here, the recent capability of the ESR to store cooled beams at very low energies, see Section \ref{S:reactions},
is a crucial achievement for performing spectroscopy of inner orbitals in transient superheavy quasi-atoms $Z>100$.
Furthermore, even more research opportunities open up with the commissioning of the CRYRING@ESR.
An extended study has been performed of the dynamical behavior of the innermost quasimolecular orbitals via electron-, photon- and positron spectroscopy in CRYRING/ESR. 
This entails investigation of predicted strong oscillations of the KK vacancy transfer in 
symmetric collisions of bare and hydrogen-like projectiles with atoms, e.g. $Xe^{54+} + Xe^{0+}$, 
with oscillation patterns as function of impact parameter giving directly access to the instantaneous energy 
of the $1s\sigma$ MO \cite{Tupitsyn-2010, Tupitsyn-2012}.
High-resolution measurement on electron-, positron- and the X-ray emission from the $1s\sigma$ MO 
during the collision will pave the way to understanding of whether the empty or singly occupied $1s\sigma$ 
orbital might be diving into the Dirac sea \cite{Sepp-1981}.

Another fundamental process of coupling radiation to the matter field is electron nucleus Bremsstrahlung (eNBS), 
which is a fundamental process whose experimental \cite{Nofal-2007} and theoretical \cite{Jakubassa-2006} 
studies have been for decades landmarks in atomic physics. 
The need for stringent tests of its claimed relationship at the high energy end of the bremsstrahlung (BS) spectrum with other fundamental processes like photoionization, radiative recombination and pair production has been emphasized for decades by theory. 
Only now first experiments directed at this claim have been possible with the appearance 
of heavy-ion storage rings and coincidence experiments in inverse kinematics of the high energy BS photon 
with the associated electron from the BS elementary process \cite{Hillenbrand-2014b, Hillenbrand-2015a, Hillenbrand-2020}. 
More extensive studies are anticipated for the near future.

The  supreme  quality of ion beams in heavy-ion storage rings has enabled in inverse kinematic the investigation of the fundamental process of electron impact ionization of ions in complete kinematics, $(e, 2e)$ \cite{Kollmus-2002, Hillenbrand-2013a}. 
Such theoretically very important studies of the full differential ionization cross section of the ion is hindered in standard kinematics  
by luminosity problems of several orders of magnitude. 
The same technique in the storage ring can be investigated to study fully differential cross sections  for $1s$ photoionization 
of H-like very heavy ions like U$^{91+}$  via electron impact ionization in the limit of vanishing momentum transfer. 
This is a significant test of relativistic theories predicting a fundamentally different emission pattern in  the relativistic cases compared to the non-relativistic ions.         

In the future, new experimental capabilities will be enabled at the HESR \cite{Hagmann-2013a, Hillenbrand-2015c}, where highly relativistic cooled beams will be available.
In near relativistic collisions free-free electron-positron pair production for very heavy collision systems faces great challenges 
as it cannot be treated  as a first order process and does not possess restrictions on the available volume in the phase space 
of the coincident electron-positron pair. 
This resulting challenge on coincidence efficiencies has so far inhibited any benchmark test of the predictions of theory. 
A new toroidal lepton spectrometer for simultaneous cover of the entire electron and positron phase space 
suitable for implementation into the future HESR storage ring has been recently proposed \cite{Hagmann-2019}. 
This should open a new window on stringent tests for the most advanced pair production theories. 

The advance of strong femtosecond lasers has brought the study of field-assisted electron-ion fundamental processes into the reach of experiments \cite{Voitkiv-2001a, Voitkiv-2001b, WangZ-2019}. 
A very promising  case is the fundamental $0^\circ$ binary encounter collision in ion-atom collisions. 
This, in the emitter frame a 180$^\circ$ elastic collision, 
generates a prominent electron peak which is spectacularly split into a doublet 
with the energy spilt given by $\Delta E=E_{laser}(p_e/\omega_l)$ 
resulting in very sensitive features for relativistic beams.

Zero-degree lepton spectroscopy has also application in accelerator design.
For the FAIR project, it is essential to achieve the highest intensities of uranium primary beams.
For this purpose, in order to minimize space charge effects, acceleration of the uranium beam in charge state $28+$ is crucial.
However, it has been calculated that due to a large number of weakly bound electrons (the ionization potential of U$^{28+}$ is $E_B = 930$ eV) 
the losses of the beam through the projectile ionization process U$^{28+} + A \rightarrow $U$^{(28+n)+} + {A^{+?}} + xe$ 
will be large \cite{Shevelko-2009a, Shevelko-2010, Shevelko-2011}.
The complication of the theoretical description of the ionization of U$^{28+}$ 
arises from dense unoccupied states above the least bound electron ([Kr]$4f^{14}5s^25p^2$) \cite{Voitkiv-2010}.
Differential cross sections for the electron projectile ionization cusp in U$^{28+}$ were measured in coincidence 
with the outgoing charge state of the projectile \cite{Hillenbrand-2016a}. 
Such measurements are essential to unambiguously distinguish between direct
single and multiple ionization contributions which help to critically benchmark the theory \cite{Hillenbrand-2016a}.

\subsubsection{Dielectronic recombination as a tool to study nuclear properties}
\label{s:dr}
Apart from its importance in atomic physics and plasma physics \cite{dr, Mueller-2012}, the DR mechanism can successfully be used to address nuclear properties such as nuclear charge radii, nuclear spins and nuclear magnetic moments. 
For a current review about the application of DR as a spectroscopic tool see~\cite{Brandau-2012}.

The sensitivity of DR to atomic structure, which in turn depends on the spin-parity of the nucleus, 
gives access to measurements of isotope shift (IS) of stable and radioactive ions \cite{Schuch-2005, Brandau-2008, Brandau-2013}.
On the one hand, the precision of IS measurements is worse than in dedicated optical and/or laser experiments.
On the other hand, the DR measurements are performed on few-electron, typically lithium-like, systems, which allow for accurate theoretical description of electron configurations.
Therefore, a clear interpretation of the obtained spectroscopic information, unbiased from atomic many-body effects, is possible~\cite{Kozhedub-2008}.
First DR-IS experiments were performed in low-energy storage rings like CRYRING and TSR on stable ions of rather complex electronic structure ($^{207}$Pb$^{53+}$~/~$^{208}$Pb$^{53+}$)~\cite{Schuch-2005} and later on stable three-electron isotopes 
by determining the hyperfine splitting of stable $^{45}$Sc$^{21+}$~\cite{Lestinsky-2008}.
With the availability of the ESR, the experiments are routinely performed with lithium-like ions. 
For instance, the measurements of the nuclear volume shift studies in stable Nd-isotope pair $^{142}$Nd$^{57+}$~/~$^{150}$Nd$^{57+}$~\cite{Brandau-2008}. 

In the last years, the sensitivity of DR measurements in the ESR was improved such that measurements on low-intensity beams of merely $10^2$ to $10^3$ stored particles became possible.
Such intensities are now suitable for the experiments on radioactive secondary ions.
To date, DR studies of radioactive ions were conducted only in the ESR.
A major breakthrough was achieved with the first studies on 
$^{237}$U$^{89+}$ and $^{234}$Pa$^{88+}$~\cite{Brandau-2010, Brandau-2012, Brandau-2013, Brandau-2009}.  
The latter nuclide has a low-energy isomeric level at $73.92+x$~keV with a half-life of $T_{1/2}=1.17$ min \cite{NNDC}, 
which decays predominantly by beta-decay~\cite{AME-2016}. 
This condition renders $^{234}$Pa$^{88+}$ an 
ideal candidate for pilot experiments on DR-IS experiments with nuclear isomers 
since the presence of the isomer can be independently monitored via 
counting of its decay daughters with a particle detector blocking the inner orbits of the storage ring~\cite{Brandau-2012, Brandau-2013}.
The production of $^{234}$Pa$^{88+}$, the verification of its isomeric state in the stored samples and the measurement of the 
DR spectra were successfully performed in the ESR. 
DR spectra for different isomeric population have been acquired.
Thanks to the short half-life of the isomer, this is possible by simply taking the spectra at different times after the production of radioactive ions.
In this case, the corresponding spectrum was taken 5 min after the injection of the ions into the ESR when the isomeric population should be zero.
\begin{figure*}[t!]
\centering\includegraphics[angle=-0,width=0.8\textwidth]{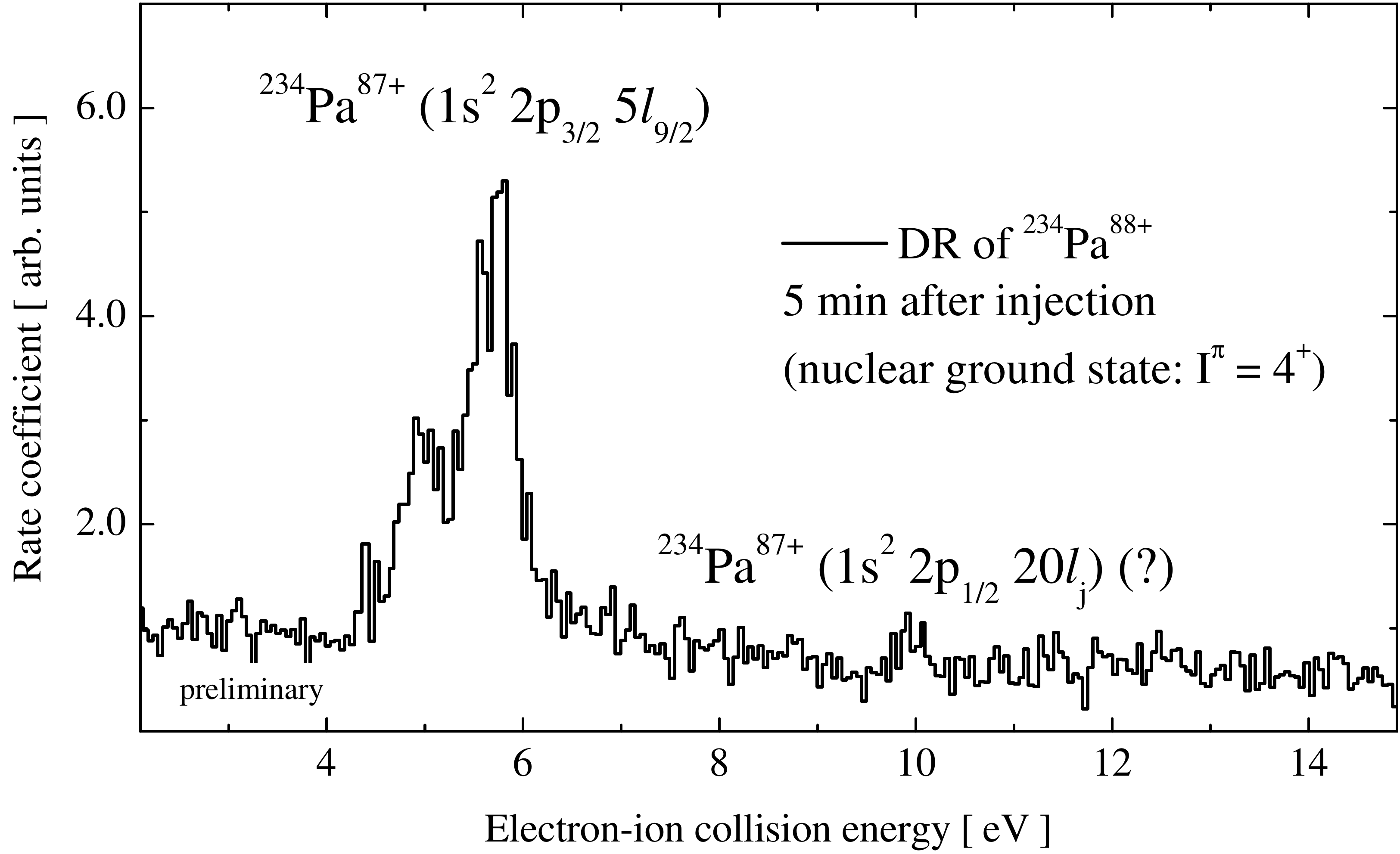}
\caption{(Colour online) 
Dielectronic recombination spectrum of the radioactive $^{234}$Pa$^{88+}$ taken 5 min after injection of the ions into the ESR. 
The isomeric state of $^{234}$Pa has a half-life of $T_{1/2}=1.17$ min \cite{NNDC}.
At this time the isomers should be completely $\beta$-decayed. 
The remaining $^{234}$Pa ions are in their nuclear and hyperfine ground states. 
The spectrum was obtained with a few $10^4$ stored ions. 
Taken from \cite{Brandau-2013}.}
\label{F:DR}
\end{figure*}
The corresponding DR spectrum is illustrated in Figure~\ref{F:DR}. 

The potential of this measurement is manyfold.
For instance, through measurement of the DR resonances, the beam composition 
can be determined or monitored, if performed (semi)continuously ~\cite{Brandau-2012}. 
Following the application of the DR to study atomic metastable states and hyperfine induced decays~\cite{Schmidt-1994, Schippers-2007},
it is proposed to use DR for detecting the population 
of the hyperfine states in laser spectroscopy on radioactive $^{207}$Bi and $^{208}$Bi, see Section \ref{S:QED}.

Of particular interest are the DR-assisted studies of long-lived nuclear isomers 
with nuclear excitation energies which are too low to be investigated with the majority of other approaches. 
The most interesting case is the so-called 'nuclear clock' isomer in $^{229}$Th with 8.28(17)~eV \cite{Seiferle-2019}
nuclear excitation energy connected by an M1 transition to the nuclear ground-state. 
The wealth of $^{229}$Th physics cases and applications is enormous: 
amongst others it can be employed as a clock with 19 decimal places precision~\cite{Campbell-2012}, 
as an ideal test-bed of fundamental symmetries~\cite{Berengut-2009}, or as a primary candidate for a nuclear laser~\cite{Tkalya-2011}. 
Although the excitation energy of the isomer has been measured and the laser spectroscopy experiments are being prepared \cite{Thielking-2018, Thirolf-2019}, 
storage ring based investigations on highly charged ions allow for addressing effects which are inaccessible otherwise.
Extensive studies of $^{229}$Th are envisaged at the present ESR and CRYRING as well as at the upcoming new facilities of the FAIR project~\cite{Brandau-2013}.


	\section{Conclusion}
	
	Storage rings are facilities for precision experiments in atomic and nuclear physics. Various manipulation methods employing beam cooling were developed in order to support the specific needs of the experiments. It is still an active research field which aims at experiments with specific beam requirements, particularly more exotic secondary beams but also the extension of beam parameters to very low energy and ultimate beam quality.  

	On the one hand, the developments on the machine side are complemented by the implementation of novel instrumentation, like high resolution lepton spectrometers, X-ray and $\gamma$-ray detector setups with a large solid-angle coverage, high sensitivity non-destructive diagnostics, etc.
On the other hand, the development of experimental instruments goes hand in hand with the developments of new operation modes of storage rings, like the efficient deceleration of exotic beams resulted in the need of placing position sensitive detectors directly into the ultra-high vacuum of the ring. New technological solutions, like cryogenic micro-calorimeters or current comparators, find their place in state-of-the-art experiments and enable new experimental capabilities.

	In this review we made an attempt to illustrate the deep connection of the machine, the heavy-ion storage ring, and the precision experiment. Indeed, it is hard to separate between the accelerator and the physics experiment and we are therefore used to say that the storage ring itself is an essential indispensable part of a successful experiment. In this work we could focus only on selected highlight research programs in atomic and nuclear physics leaving aside many interesting experimental studies.

	The success of storage ring based experiments in the last decades attracted an increased attention from the cross-discipline community. This resulted in a multitude of new heavy-ion storage ring projects for addressing questions in modern atomic physics, nuclear structure and astrophysics, aiming at specific solutions for various energies, different driver accelerators and challenging physics cases. Some of these projects are approved and are being constructed like FAIR and HIAF. Other projects, especially on the low-energy frontier, are in the planning stage. 

	To conclude this review, the research with stored and cooled secondary beams has a history full of successes as well as the new developments in the field point to an exciting future.
\section*{Acknowledgements}
The examples of research results presented here are due to the common effort of our colleagues working with us over many years in different collaborations. We appreciate very much their essential contributions and help concerning practical issues as well as enlightening discussions about physics which we enjoy. 
In particular we warmly acknowledge help in preparing this review from Carsten Brandau, Siegbert Hagmann, Jan Glorius, Alexandre Gumberidze, Nikos Petridis, Shahab Sanjari, and Takayuki Yamaguchi.
YAL acknowledges the support by the European Research Council (ERC) under the European Union's Horizon 2020 research and innovation programme (grant agreement No 682841 ``ASTRUm'').

\addcontentsline{toc}{section}{References}
\bibliography{storage-ring-master}
\end{document}